\documentclass[amssym,useAMS,usenatbib,times]{mn2e}

\usepackage{graphicx}
\usepackage{subfigure}
\usepackage{amssymb}
 at 10truept

\def\and  {\it {et al.} \rm}

\def\spose#1{\hbox to 0pt{#1\hss}}
\def\simlt{\mathrel{\spose{\lower 3pt\hbox{$\mathchar"218$}}
     \raise 2.0pt\hbox{$\mathchar"13C$}}}
\def\simgt{\mathrel{\spose{\lower 3pt\hbox{$\mathchar"218$}}
     \raise 2.0pt\hbox{$\mathchar"13E$}}}
\def\be{\begin{equation}}
\def\ee{\end{equation}}
\def\bce{\begin{center}}
\def\ece{\end{center}}
\def\bea{\begin{eqnarray}}
\def\eea{\end{eqnarray}}
\def\ben{\begin{enumerate}}
\def\een{\end{enumerate}}

\def\ni{\noindent}

\def\brr{\begin{array}}
\def\err{\end{array}}

\def\nh1{n_{\rm HI}}

\def \p1dk {P_{\rm 1D}(k)}
\def \simlt {\mathrel{\spose{\lower 3pt\hbox{$\mathchar"218$}}
     \raise 2.0pt\hbox{$\mathchar"13C$}}}
\def \simgt {\mathrel{\spose{\lower 3pt\hbox{$\mathchar"218$}}
     \raise 2.0pt\hbox{$\mathchar"13E$}}}

\def \vm {{\bf m}}

\def \vx {{\bf x}}
\def \vq {{\bf q}}

\def \be {\begin{equation}}
\def \en {\end{equation}}
\def \bea {\begin{eqnarray}}
\def \ena {\end{eqnarray}}

\def \bi {\begin{itemize}}
\def \ei {\end{itemize}}

\def \apj {{\it ApJ}}
\def \apjl {{\it ApJ Let.}}
\def \apjs {{\it ApJ Sup.}}

\def \aap {{\it A\&A}}
\def \aj {{\it AJ}}
\def \mnras {{\it MNRAS}}
\def \nat {{\it Nature}}
\def \procspie {{\it Proc.~SPIE}}
\usepackage[usenames]{color}
\definecolor{Blue}{rgb}{0,0.08,0.65}
\definecolor{Red}{rgb}{0.65,0.08,0.05}
\definecolor{Green}{rgb}{0.15,0.45,0.25}

\newcommand{\distance}[2]{{\cal D}\!\left({\rm #1},{\rm #2}\right)}

\begin{document}
\title[The fully connected N-D Skeleton]{The fully connected {\it N} dimensional Skeleton:\\ 
probing the evolution of the cosmic  web 
}

\author[T. Sousbie, S. Colombi \&  C. Pichon]
{\ni T. Sousbie$^{1,2}$, S.  Colombi$^1$  \& C. Pichon$^{1,3}$ \\
$^1$ Institut d'Astrophysique de Paris, CNRS UMR 7095 \& UPMC,  98 bis boulevard Arago, 75014 Paris, France \\
$^2$ Department of Physics, The University of Tokyo, Tokyo 113-0033, Japan \\
$^3$ Institut de Recherches sur les lois Fondamentales de l'Univers, DSM,  l'Orme des Merisiers, 91198 Gif-sur-Yvette, France\\
{sousbie@iap.fr, pichon@iap.fr, colombi@iap.fr} 
}

\maketitle

\begin{abstract}

A method to compute the full hierarchy of the critical subsets of  a
density field  is presented. It is based on a watershed technique
and uses a probability propagation scheme to improve the quality of
the segmentation by circumventing  the discreteness  of the sampling.
It can be applied within spaces of arbitrary dimensions and geometry.
This recursive segmentation of space yields, for  a $d$-dimensional
space, a  $d-1$ succession of  $n$-dimensional subspaces that fully
characterize the topology of the density field.  The final  1D
manifold  of the hierarchy is the fully connected network of  the
primary critical lines of the field : the skeleton. It corresponds to
the subset of lines  linking maxima to saddle points, and provides  a
definition of the filaments that compose the cosmic web as a precise
physical object, which makes it possible to compute any of its
properties such as its length, curvature, connectivity etc...

When the skeleton extraction is applied to initial conditions of
cosmological N-body simulations and their present day non linear
counterparts, it is shown that the time evolution of the cosmic web,
as traced by the skeleton, is well accounted for by the  Zel'dovich
approximation. Comparing this skeleton to the initial
skeleton undergoing the Zel'dovich mapping shows that  two effects
are competing during the formation of  the cosmic web: a general
dilation of the larger filaments that is captured by a simple
deformation of the skeleton of the initial conditions on the one
hand, and the shrinking, fusion and disappearance of the more
numerous smaller filaments on the other hand. Other applications 
of the N dimensional skeleton and its peak patch hierarchy are discussed.\\
\end{abstract}
\begin{keywords}
Cosmology: simulations, statistics, observations, Galaxies: formation, dynamics.
\end{keywords}

\section{Introduction}

The web-like pattern certainly is the most striking feature of matter
distribution on megaparsecs scale in the Universe. The existence of
the ``cosmic web'' \citep{zeldo70} \citep{bbks} has been confirmed more than twenty
years ago by the first CfA catalog \citep{delap86} and the more recent
catalogs such as SDSS \citep{sdss} or 2dFGRS \citep{2df}. These observations, together with
the dramatic improvement of computer simulations (e.g. \cite{teyssier} \cite{ocvirk}) have largely improved the
picture  of a Universe formed by an intricate network of voids ({\it
  i.e.} globular under-dense regions) embedded  in a complex
filamentary web which nodes are the location of denser halos. The
traditional way of understanding large scale structures (LSS) formation and
evolution relies on Friedman equations and assumes  that LSS are the
outcome of the growth of very small primordial quantum fluctuations by
gravitational  instability (see {\it e.g.} \cite{peebles80} or
\cite{peebles93} and references therein). In this theory, the solution for structure formation
is described in terms of a mass distribution that one needs to grasp
({\it i.e.} by  following the evolution of its most important
features) and compare these to observations. Comprehending the mass
distribution  as a whole, especially at non-linear stages, is a very
difficult task. A possible solution therefore  consists in extracting and
studying simple characteristic features of matter distribution such as
voids, halos and filaments as individual physical objects. So far,
mainly because of  the relatively higher complexity of the filaments,
most theoretical and computational researches have focused on the
voids and halos. 

The dark matter halos have arguably been the most studied component of
the cosmic web. Their density profiles for instance are very well
described by so-called NFW profiles \citep{NFW97} and non-parametric
models are still under investigation \citep{merritt06}. The dependence
of these density profiles on the halos mass ({\it e.g.} \cite{bbks},
\cite{lacey93}) has also been investigated thoroughly and its
relationship with redshift and environmental properties  are a very
active topics ({\it e.g.}  \cite{harker06}, \cite{aubert06},
\cite{wang07}, \cite{aragon-calvo07}, \cite{3dskel} or \cite{hahn07}). From a computational
point of view,  much effort has been put into the development of
various algorithms to identify halos in simulations and galaxies  in spectroscopic
redshift galaxy surveys. The friend-of-friend algorithm \citep{FOF} is
now widely spread, as well as more complex hierarchical sub-structures
identifiers such as HFOF \citep{gottlober98}, SUBFIND
\citep{springel01}, VOBOZ \citep{neyrinck05} or  ADAPTAHOP
\citep{ADHOP}. 

Voids are another feature of cosmological matter
distribution that also have a long history of theoretical and
computational modeling. The first voids were observed by
\cite{kirshner81} and are in some sense the counterpart of halos: the
initial quantum perturbations collapsing into halos at non-linear
stages leave room to voids in the under-dense regions. The first
theoretical voids models where developed by \cite{hoffman82},
\cite{icke84} or  \cite{bertschinger85} among others, while numerical
void finders exist, such as the one described in \cite{el-ad97},
ZOBOV \citep{neyrinck08}, based on Voronoi tessellation, or the
recent Watershed Void Finder, based on the Watershed transform ({\it
  e.g.} \cite{beucher79}, \cite{beucher93}), by \cite{platen07} (see
the introduction and references therein for a more complete review of
the subject).
The improvements in our understanding of voids and halos
properties led to the formulation  of powerful theories such as the
patches theory \citep{bbks} the extended Press-Schechter theory ({\it
  e.g.}  \cite{bond91} and \cite{sheth98}) or the skeleton-tree
formalism \citep{hanami01}.

 But our investigation of the filaments as
individual objects is not yet  as thorough as for the halos and voids:  the
definition of a well established mathematical framework  for their
study could therefore lead to significant improvements in our
understanding of matter distribution in the Universe. The first
attempts date from \cite{barrow85}, who used a graph-theory construction: 
the minimal spanning tree (MST). This method defines the cosmic web
as the network linking galaxies (or particles from a numerical
simulation), having the property of being loop-free and of minimal
total length. This technique was later developed in order to try
quantifying in an objective way the properties of the cosmic web (see
{\it e.g.}  \cite{graham95}, \cite{colberg07} and a review on the
subject can be found in \cite{martinez02}). The so-called shape finders (\cite{sathyaprakash96}, \cite{sahni98} or
\cite{bharadwaj00}) allow a statistical study of the filaments and another method, based on
the CANDY model, commonly used to detect road networks, uses a marked
point process and a simulated annealing algorithm to trace the
filaments \citep{stoica05}. More recently, the skeleton formalism and
its local approximation, that describe the filaments as particular
field lines of the density field, was introduced by \cite{2dskel} and
\cite{3dskel} with the advantage of framing a well-defined
mathematical ground for theoretical predictions of the filaments
properties as well as an efficient numerical identification
algorithm. Finally, an interesting first attempt to unify  halos,
voids and filaments identification using the Multiscale Morphology
Filter (MMF) technique was also proposed by \cite{aragon-calvo07a}.

In this paper, we introduce a framework and algorithm to
identify the full hierarchy of critical lines, surfaces, volumes... of density distribution in the general case of $d$-dimensional
spaces. For 3D space, these critical
subspaces can be identified to the void and peak patches, as well as
filaments and other primary critical lines of the distribution. The algorithm
extracts the filaments as a differentiable
and, by definition, fully connected networks that traces the backbone
of the cosmic web. This method is closely related to the skeleton
formalism presented in \cite{2dskel} and \cite{3dskel} and is also based on both
Morse theory (see {\it e.g.} \cite{colombi00}, \cite{milnor63} or \cite{jost}) and
an improved Watershed segmentation  algorithm that uses a
probability propagation scheme.\\

 This paper is organized as follows. In section \ref{sec:algo}, we present
a general definition of the critical sub-spaces that we use as well as
a method to extract them from sampled density  field with a sub-pixel
precision (focusing more specifically on the filaments in the 2D and
3D case). In section \ref{sec:applications}, we use  this formalism to
study the time  evolution of the cosmic web,  and understand
the change of its properties as a specific
object via the truncated Zel'dovich approximation
\citep{zeldo70}. Finally, in section \ref{sec:conclusion}, we
summarize our findings and discuss a few possible applications to
N-body simulations and observational spectroscopic galaxy surveys. The
details of  a general simplex minimization algorithm used in section
\ref{sec:algo} are presented in Appendix A while the general behavior
of the inter-skeleton pseudo-distance as defined in section
\ref{sec:applications} is given in appendix B.

\section{Method}
\label{sec:algo}
The main goal of the algorithm presented here is to allow a robust extraction
of the non-local primary critical lines (among which the skeleton) as introduced in \cite{2dskel} and
\cite{3dskel}. In these papers, the skeleton was defined as the set
of points that can be reached by following the gradient of the field,
starting from the filament type saddle points (i.e.  those where only
one eigenvalue of the Hessian is positive). Let $\rho(\mathbf{x})$ be
the density field, and $\nabla\rho$ its gradient at position $\mathbf
x$, the skeleton can be retrieved by solving the following
differential Equation:

\begin{equation}
\frac{d \mathbf{x}}{dt}\equiv \mathbf{ v}=\nabla\rho\,,\label{eq:skel}
\end{equation}
using the ``filament" type saddle points as initial boundary conditions.
Because of the difficulty of designing a robust algorithm to solve
this equation, it was achieved only in $2D$ in \cite{2dskel}
and a solution to a local approximation in $3D$ was proposed
in \cite{3dskel}. This local approximation allowed the extraction of a
more general set of critical lines linking critical points together,
the subset of this lines linking saddle points and maxima together
corresponding to the skeleton (i.e. the ``filaments'' in the large
scale distribution of matter in the universe). See \cite{pogo}  for 
a discussion of these various sets.  \\

This method works in a very general
framework and allows the extraction of a fully connected {\em
  non-local} skeleton as well as an extension of the primary critical lines
introduced in \cite{2dskel} and \cite{3dskel} to a hierarchy
of critical surface.  Following the idea, already present in
Equation (\ref{eq:skel}), that the topology of a field can be
expressed in terms of  the properties its field lines, it takes ground
in Morse theory {\citep{jost}} and is roughly based on an extension of
the patches theory \citep{bbks}. For a sufficiently smooth and non-degenerate field \footnote{It would be beyond the scope of this paper to define such a field
from the mathematical point of view, but it certainly has to be a Morse function that obeys the Morse-Smale-Floer condition, {\it e.g.} the discussion in \cite{2dskel}. }

of dimension $d$, the peak patches -- PP hereafter -- ({\sl resp.} void patches
-- VP hereafter) are defined as the set of points from which the field
lines solution of Equation  (\ref{eq:skel}) all converge to the same
maximum ({\sl resp.} minimum) of the field. Within this framework, we shall qualitatively show that in
a $d$-dimensional space, the skeleton can be thought of as the result
of $d-1$ successive identifications of VPs or, equivalently, as the
one dimensional interface between at least $d$ VPs (an actual rigorous demonstration can be found in {\it e.g.} \cite{jost}). 
Using this definition, extracting the skeleton of a distribution thus simply
amounts to finding a way of robustly and consistently identifying the
patches.\\

 Whether considering a particle distribution obtained from a numerical
 simulation  or a density field sampled on a grid, the major
 difficulty arises from the discrete nature of the data. In fact, even
 if the underlying  density field is supposedly smooth and continuous,
 the discreteness of the sampling implies a relatively large
 uncertainty on the precise location of the patches boundaries, as
 sampling is limited by computational power, which is even more true
 when considering higher dimensions space. The algorithm we use is an
 improved version of the Watershed transform method \citep{beucher79}, based on a
 probability propagation scheme and aims at attributing  a probability of belonging to a
 given patch to every
 sampled point of the  density field. This scheme is very general and efficient as it allows
 dealing with discrete dataset in a naturally continuous fashion and
 on manifolds of arbitrary dimensions. 


\subsection{Probabilistic patches extraction}
\label{sec:patch}
\begin{figure*}
\centering
\includegraphics[width=13cm]{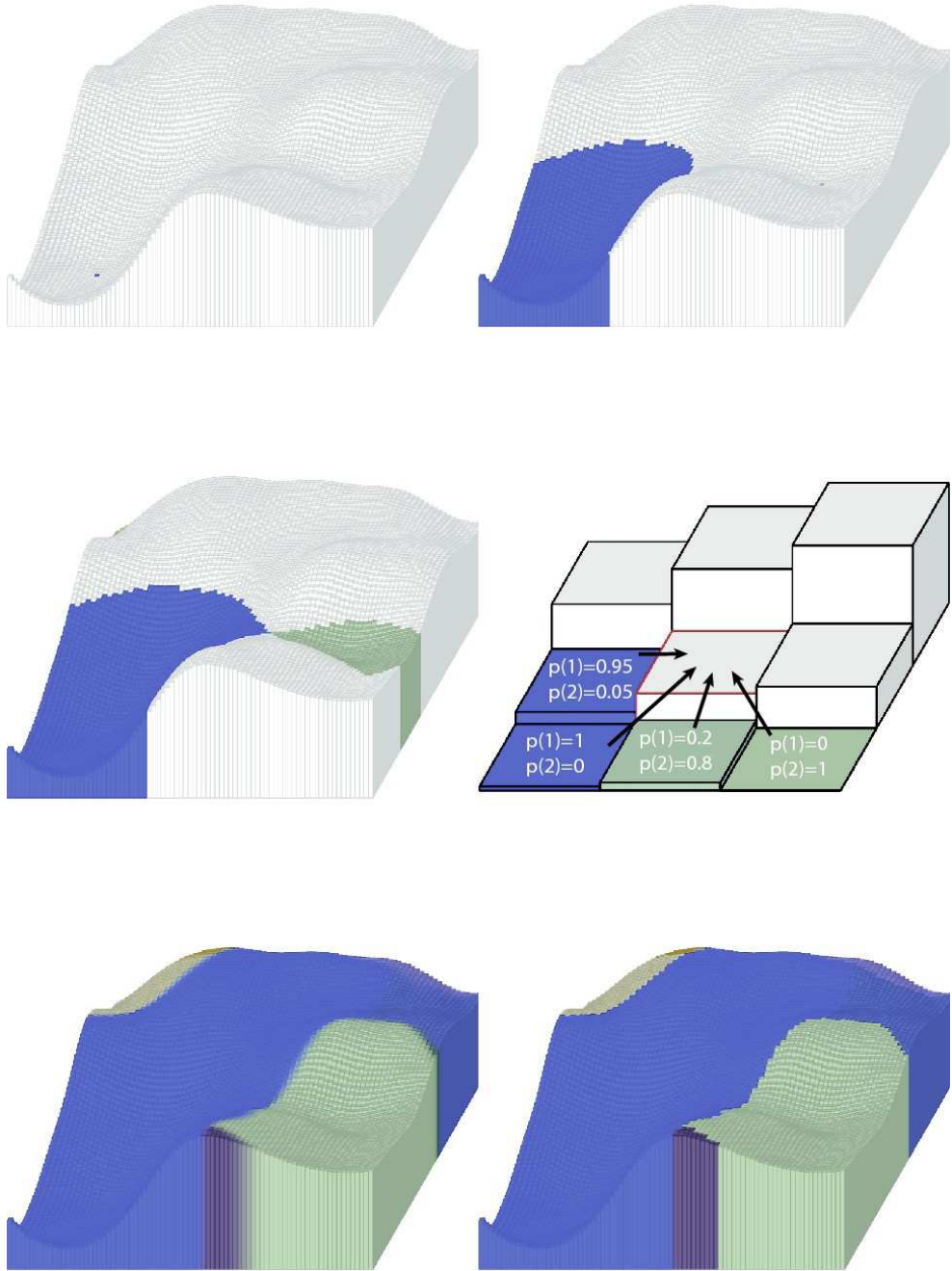}\label{fig_algo_tile}
\caption{The different steps of the probabilistic algorithm for
  finding the patches. The height of the histograms is proportional to
  the density at each pixel of a $2D$ random field. {\em Top-left}:
  the pixel with lowest density is identified and tagged as belonging
  to the void-patch number $1$ (blue colour) with probability
  $P(1)=1$. {\em Top-right}: the pixels are then considered in
  ascending order and are tagged according to the tag of their already
  visited surrounding pixels. This is repeated until the level of the
  minimum with second lowest density is reached. As this pixel does
  not have any tagged  neighbour, a new void-patch index is added and
  the pixel is tagged as belonging to it (green colour) with
  probability $P(2)=1$. {\em Middle-left and middle-right}: the process
  is repeated until one reaches the saddle point with lowest density,
  located at the border of two patches (middle-left). Above this
  threshold, a pixel can have several  neighbours, each tagged with
  different patch indexes (middle-right). A list of probabilities
  associated to the different patch index of the  neighbouring pixels is
  attributed to the current pixel by computing the density difference
  weighted average of the respective patches probabilities of the
  surrounding pixels. {\em Bottom-left}: repeating the process until
  all pixels have been visited, one obtains for each pixel a list of
  possible patches index together with their respective probabilities
  (hence the blurred borders between patches on the picture). {\em
    Bottom-right}: a clean border between the patches can be found
  by defining the index of the patch a pixel belongs to as the one
  with highest probability. It is very straightforward to extend this
  method to spaces with arbitrary number of
  dimensions.\label{fig_algo_steps}}
\end{figure*}

The initial idea beyond our patches identification algorithm is that a
patch can be defined as the set of field lines (i.e. curves that
follow  the gradient of a field) that originate from a given minimum
(VP) or maximum (PP) of a field. Considering a sampled field, being
able to identify the patches thus amounts to being able to decide, for
any given pixel $p$, from which extremum all field lines that cross
$p$ originate. It is therefore easy to understand that the discrete nature
of the sampling rapidly  plagues such a task: for each pixel,
considering the measured gradient, one has to decide from which, in
the fixed number of  neighbouring pixels, the field line comes
from. Within a $d$-dimensional space, having to select between only
$3^d$ possibly different direction for field lines is a crude
approximation that leads, because of accumulation, to a largely wrong
answer for pixels located far away from the extrema.\\     

Although we present the algorithm in the general case here, the reader
can refer to figure~\ref{fig_algo_steps} and its legend for a simpler
and more visual explanation of the algorithm in the $2D$ case. More
generally, our algorithm involves considering each pixel of a sampled
field in the order of their increasing ({\sl resp.} decreasing) value,
depending on whether we want to compute the VPs or PPs and, for each
of them, computing the probability that it belongs to a given VP
({\sl resp.} PP). This probability map is simply computed by scanning the
probability distribution of its $3^d-1$ neighbours (within a
$d$-dimensional space, here $d=2$) and deducing the current pixel
patch probability distribution from it. Two cases are possible:
\begin{enumerate}
\item none of the  neighbours has already been considered (i.e their
  respective densities are all higher -{\sl resp.} lower- than that of the
  current pixel). This means that the pixel is a local minimum
  ({\sl resp.} maximum) of the field: a new VP ({\sl resp.} PP) index is created
  and the probability that the current pixel belongs to it is set to
  $100\%$.
\item At least one  neighbour has already been considered (i.e its
  density is lower -{\sl resp.} higher- than that of the current pixel). The
  current pixel probability distribution is computed as a
  gradient weighted average of its lower -{\sl resp.} higher- density
  neighbours' probability distributions.
\end{enumerate}

\begin{figure}
\centering
\includegraphics[width=7cm]{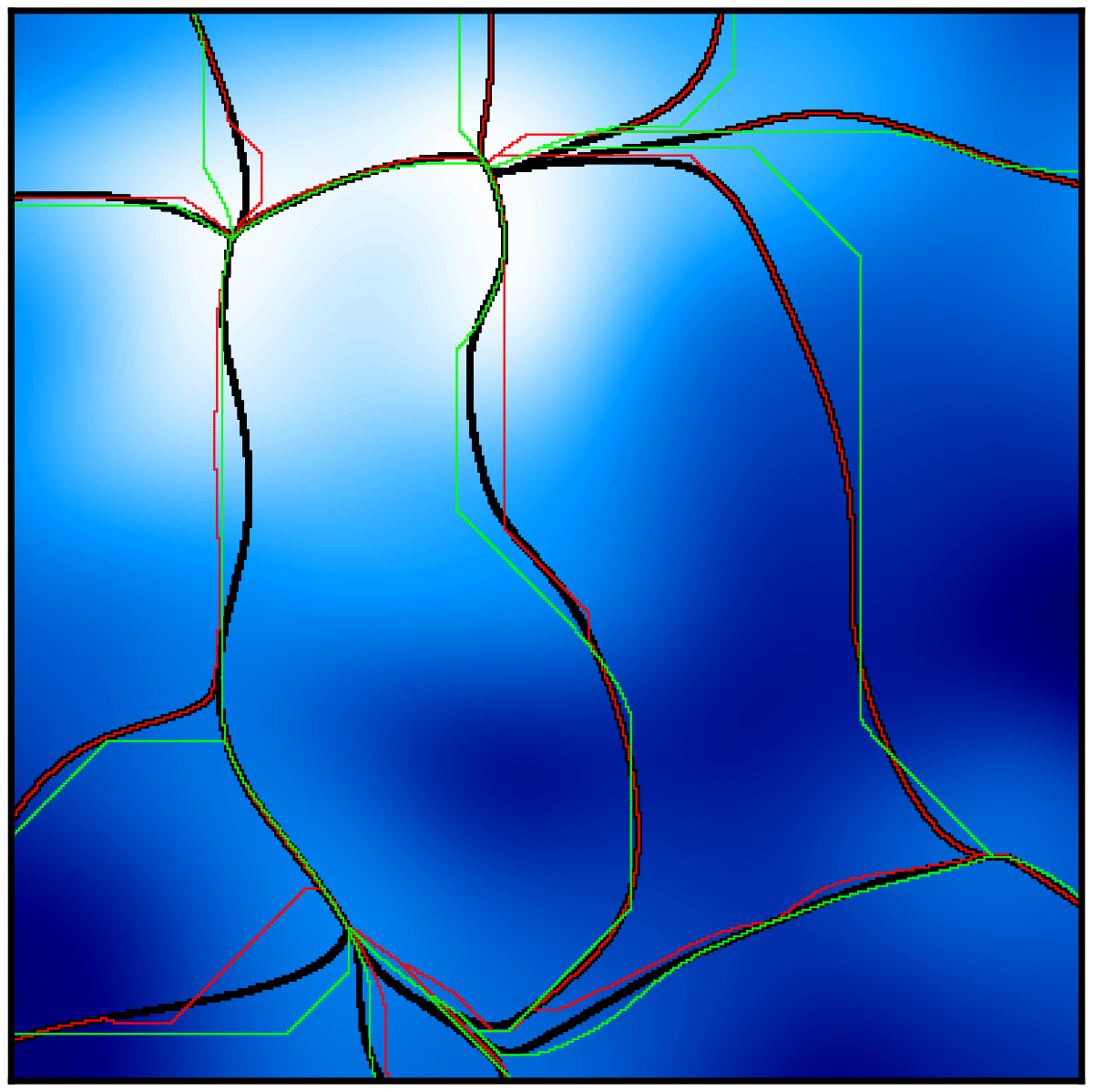}
\caption{Illustration of the virtue of the probabilistic
  algorithm. These three curves represent the borders of the void
  patches obtained with the probabilistic algorithm, by limiting
  the maximum number of probabilities recorded for each pixel. The
  black line was derived without any limitation, while for the red
  one two probabilities were kept and only one for the green one. This
  last case is equivalent to not using any probability list, as only
  the values of neighbouring pixels is taken into account. Also note
  the tendency of the borders to be aligned with the major directions of the sampling grid
  (namely, the sides and diagonals) when not taking advantage of the probabilistic
  algorithm.\label{fig_proba_check}}
\end{figure}

Once all pixels have been visited, a number $N$ of patches have thus
been created and a list of $N$ probabilities $P_i^k$,$k\in
\lbrace1,..,N\rbrace$, has been computed for each pixel, $i$. These
probabilities quantify the odds that a given pixel $i$ belongs to a
given patch $k$. Figure~\ref{fig_proba_check} illustrates the
advantages of our  probability list scheme compared to the naive
approach: without it, the patches borders have a strong tendency to be
aligned with the sampling grid and the problem tend to get much worse
when considering lower sampling and of course higher dimensions.\\

\begin{figure*}
\centering  \subfigure[density
  field]{\includegraphics[width=8.5cm]{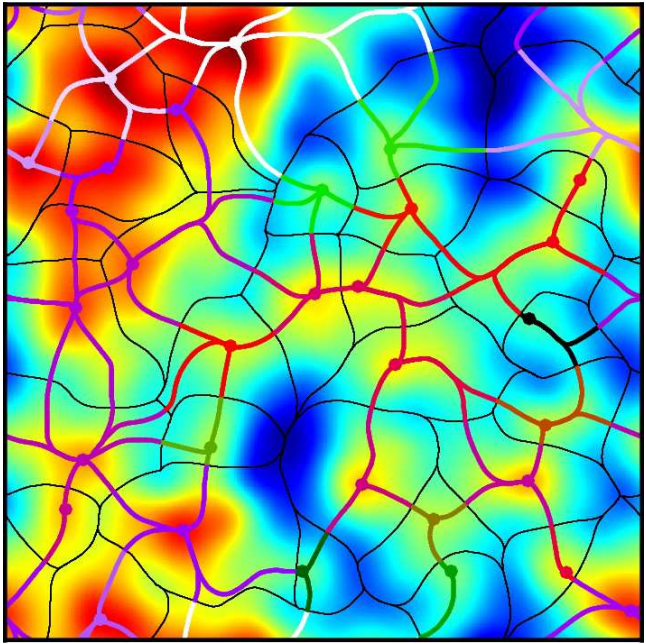}\label{fig_dfield}}
\hfill  \subfigure[skeleton presence
  probability]{\includegraphics[width=8.5cm]{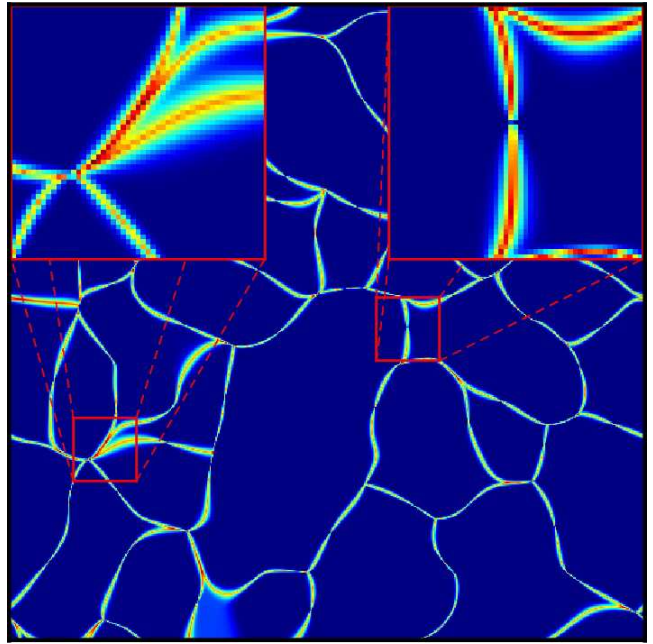}\label{fig_proba}}\\ 
\subfigure[void
  patches]{\includegraphics[width=8.5cm]{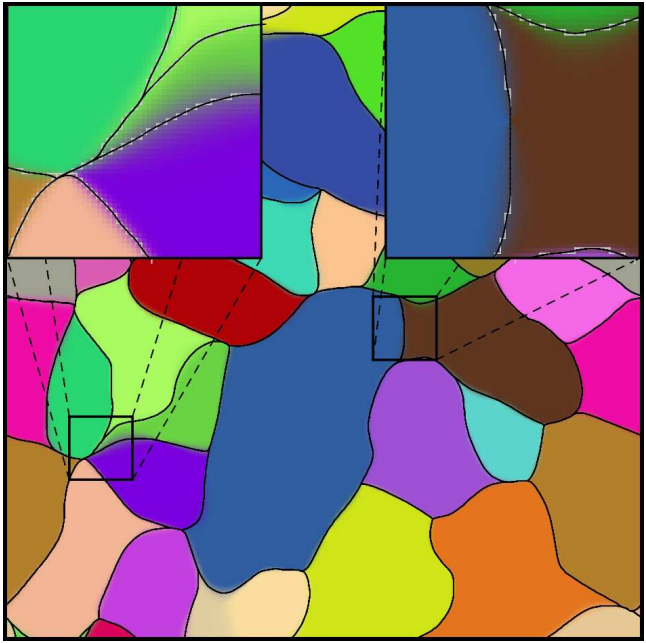}\label{fig_vpatch}}
\hfill  \subfigure[peak
  patches]{\includegraphics[width=8.5cm]{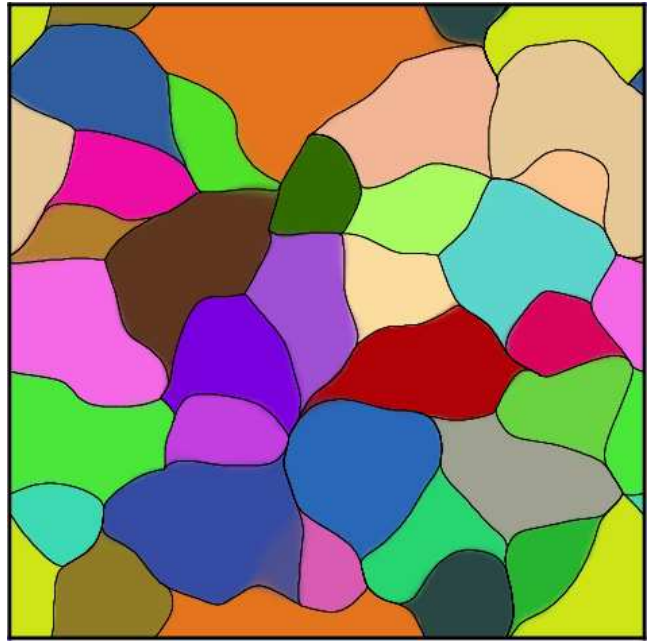}\label{fig_ppatch}}
\caption{Figure \ref{fig_dfield} represents a 2D density field
  together with its anti-skeleton (black curve) and skeleton (thick
  coloured curve). The skeleton is coloured according to the value of
  the index of the underlying void patch, which allows the detection
  of the saddle points (intersection of the skeleton and void patch
  borders). A skeleton branch starts from a field maximum (large dots)
  and goes through one saddle point before reaching another
  maximum. Figure~\ref{fig_proba} represents, for each pixel, the
  value of the  probability that it belongs to its most probable
  patch. By definition, the skeleton is the set of points that do not
  belong to any patch so the lower this value, the more probable the
  pixel belongs to the skeleton.  Figure~\ref{fig_vpatch} was obtained
  by attributing a given random colour to each patch index and
  representing each field with the colour resulting from the probability
  weighted blend of all patches colours. The zoomed parts show patches
  borders where the uncertainty on the index of the most probable
  patch index is maximal. The skeleton is represented in white,
  together with its smoothed counterpart (black). Figure~\ref{fig_ppatch} represents the peak patches of the same
  field.\label{fig_algo}}
\end{figure*}

Figure~\ref{fig_vpatch} presents the results obtained by applying this
algorithm to the $2D$ Gaussian random field of Figure~\ref{fig_dfield}. On this picture, each patch is assigned a different
shade, and the colour of each pixel is the probability weighted average
colour of its possible patches. As expected, a majority of pixels seems
to belong to a definite void patch with high probability (close to
$100\%$). In fact, considering two  neighbouring void patches A and B,
all the pixels that belong to one of these patches and have a value
lower than that of the first kind saddle point(s) on their border
(i.e. where the Hessian only has one positive eigenvalue) have a
$100\%$ probability of belonging to either A or B. Hence, the
probabilities of belonging to different patches only starts mixing
above first kind saddle points. This can be seen on the top right
zoomed panel of Figure~\ref{fig_vpatch} where probabilities only start
blending mildly for densities above this threshold (the saddle point
are represented by the probability ``nodes'' on the picture). This
results in a complex distribution of patch index probabilities in the
vicinity of higher density borders (see upper left panel of Figure~\ref{fig_vpatch}), and thus a higher uncertainty of the location of
the void patches border. This uncertainty on the precise patch index
is directly linked  to the location of the skeleton. In fact, as
explained in the next section, the skeleton can also be defined as the
set of field lines that do not belong to any patch, or in other terms,
where sampled pixels have an equal probability of belonging to several
distinct patches.\\

\subsection{The d-dimensional skeleton}
\label{sec:dskel}
As one can easily see, the major strengths of this simple patch
extraction algorithm are that it is robust and can be trivially
extended to spaces of any dimensions and topology,  the only
requirement being that one needs to be able to define  neighbouring
relationships between pixels and measure distances between them. So we
now have a robust algorithm for extracting the VP and PP of, in practice, nearly arbitrary
scalar fields. In this subsection, we show that it is  possible to
generalize the definition of  the skeleton \citep{2dskel} to
spaces of arbitrary dimension and present a simple method to compute
the skeleton, as well as critical lines and surfaces, based on our
patches extraction algorithm.\\  

\subsubsection{Definition} 

Let us first present important results of the Morse theory without
demonstrating them. The more thorough reader can refer  to \cite{jost} for
a mathematical demonstration.\\

Let us consider the general case of a sufficiently smooth and non-degenerate $d$-dimensional scalar
field $\Phi_d(\vx)$, with $\vx\in M_d$ and $M_d$ a manifold (i.e
$\mathbb{R}^d$, the sphere $S^2$, ...)\footnote{ It will be assumed
  throughout this paper that the field satisfies the Morse-Smale-Floer condition
  \citep{jost}.}. Following \cite{jost}, the field lines of
$\Phi_d(\vx)$ fill $M_d$ and  a VP can be defined as the set of points
that can be reached by following the field lines originating from a
given minimum of $\Phi_d(\vx)$. The VPs of $\Phi_d(\vx)$ thus segment
a set of $d$-dimensional volumes that completely fill $M_d$, each of
them encompassing exactly one minimum of $\Phi_d(\vx)$. The interface
of the VPs, $M_{d-1}$, defines a $(d-1)$-surface (i.e. a surface of
dimension $d-1$ embedded in $M_d$). It is therefore possible to apply
our probabilistic algorithm to $\Phi_{d-1}(\vx)$, the restriction of
$\Phi_d(\vx)$ to $M_{d-1}$, in order to extract the VPs on this
interface. For clarity, we will call the VPs of $\Phi_{d-1}(\vx)$ the
first order VPs of $\Phi_d(\vx)$, noted $1$-VPs
hereafter. Recursively, the $1$-VPs define $(d-1)$-dimensional volumes
that pave $M_{d-1}$, each of them encompassing, by definition of a VP,
exactly one minimum of $\Phi_{d-1}(\vx)$, with coordinates $\vm\in
M_{d-1} \subset M_d$, and the reasoning can be applied to the whole
hierarchy of $\alpha$-VPs, $\alpha\in\lbrace 0,..,d-1\rbrace$.\\

Starting from a $d$-dimensional $C^2$ scalar field $\Phi_d(\vx)$, it
is thus possible to define a complete hierarchy of sets of
$\alpha$-VPs, $\alpha \in \lbrace 0,..,d-1\rbrace$. These $\alpha$-VPs
are $(d-\alpha)$-dimensional volumes that partition $M_{d-\alpha}$,
where $M_{d-\alpha}$ is defined as the $(d-\alpha)$-dimensional
interface of the $(d-\alpha+1)$-patches. Each set of $\alpha$-VPs is
defined as the set of void patches of $\Phi_{d-\alpha}(\vx)$, the
restriction of $\Phi_d(\vx)$ to $M_{d-\alpha}$. Let us call a critical
point, $\vx$, of kind $n$ a critical point with Morse index
$\mu\left(\vx\right)=n$ ({\it i.e.} where the Hessian ${\cal
  H}\left(\vx\right)$  has exactly $n$ positive eigenvalues). Then,
$M_{d-\alpha}$ encompasses the whole set of saddle points of kind
$n\leq d-\alpha$, of $\Phi_d(\vx)$,  the minima of
$\Phi_{d-\alpha}(\vx)$ associated to each $\alpha$-patch being the
saddle points of $\Phi_d(\vx)$ of kind $d-\alpha$. The interface $M_1$
is thus a curve embedded in $M_d$ that links the maxima of
$\Phi_d(\vx)$ to its saddle points of kind $1$: the skeleton of
$\Phi_d(\vx)$. It is interesting to note that  this approach also
allows a rigorous definition of the whole set of critical lines
similar to the one introduced with the  local approximation of the
skeleton in \cite{2dskel} (see also \cite{3dskel}), as well as their extension to critical
hyper-surfaces of any number of dimensions.\\

 Although we have only addressed the $\alpha$-VPs case so far, the
 exact same argumentation holds for the whole hierarchy of
 $\alpha$-PPs, which leads to $M_d$ being the skeleton of the voids
 that links minima to  saddle points of kind $d-1$. Moreover,
 alternating a selection of $n_v$ $\alpha_v$-VPs and $n_p$
 $\alpha_p$-VPs, $n_v+n_p=d$, leads to $M_d$ being the curve that
 links saddle points of kind $n_p$ to saddle points of kind $n_p+1$: a
 peculiar set of critical lines of the field. One can note that, as
 rigorously demonstrated in Morse theory \citep{jost}, critical
 lines defined in such a way can only link critical points whose Morse
 index only differ by unity.

\subsubsection{Implementation}

The representation of the critical lines of a given scalar field as a
peculiar limit of a peak or void patches hierarchy certainly has some
mathematical appeal. From a practical point of view, although
apparently straightforward, its direct numerical implementation can
nevertheless be somewhat problematic.  Let $G$ be an
initial sampling grid and $\bar G$ its reciprocal (i.e. $G$ shifted by
half the size of the pixels in every direction). Using our patch
computation algorithm on a scalar field $\Phi_d(\vx)$ sampled over
$G$, we obtain for every pixel, $i$, of $G$ a probability $P_i^k$ that
it belongs to a given patch, $k$. Those sets of probability
distributions could be used to define a border between the patches
and thus to compute the 1-PPs and 1-VPs.  Nevertheless, this is in
general not an easy task: one in fact first needs a very precise localization  of
the 1-PPs and 1-VPs (those living on the (hyper-)surface of the initial
VPs or PPs) to be able to compute the following segmentation  of the
hierarchy (as opposed to a density probability). In order to overcome this issue, we chose first to base our
implementation on a subset  only of the different patches
probabilities and only keep for every pixel the index of its most
probable patch. This way, we are able to simply define the borders
between patches as the set of pixels of $\bar G$ that overlap at least
$2$ pixels of $G$ with different most probable patch index. The
patches extraction algorithm can then be applied again over that
border, restraining pixels examination to the ones that lie on its
surface. Identifying pixels of $G$ that overlap at least $2$ pixels of
$\bar G$ with different most probable patch index, one can thus
identify the 2-PP or 2-VP and, repeating this procedure, all orders of
the patches hierarchy.\\

\begin{figure*}
\centering  \subfigure[The 1-VPs of a 3D Gaussian random field]{\includegraphics[width=14cm]{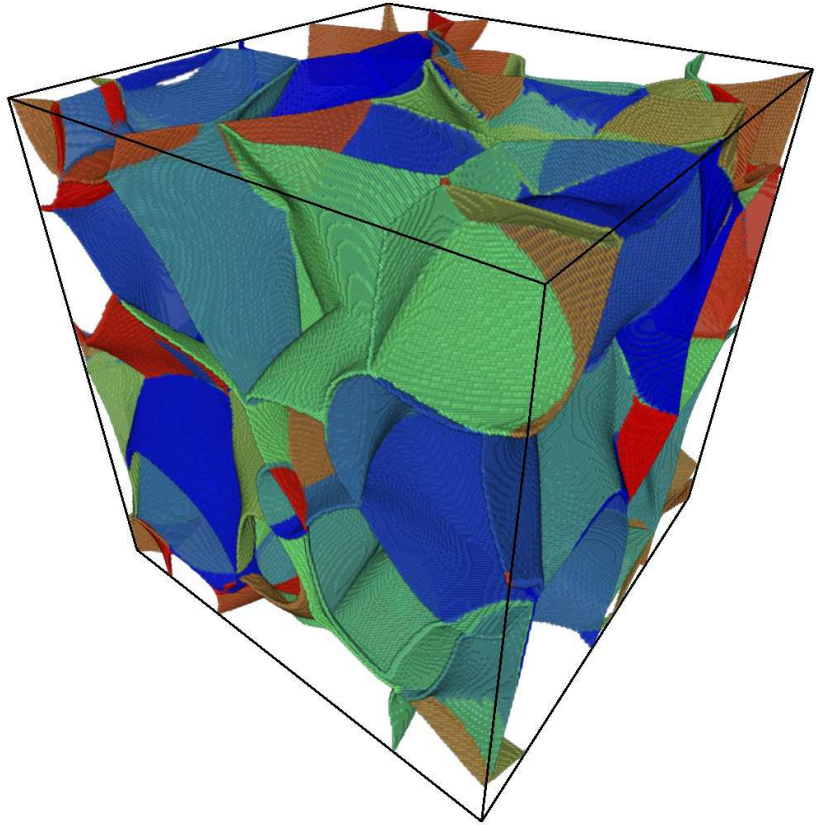}\label{fig_1vp}}\\
\subfigure[Recursive algorithm]{\includegraphics[width=8.5cm]{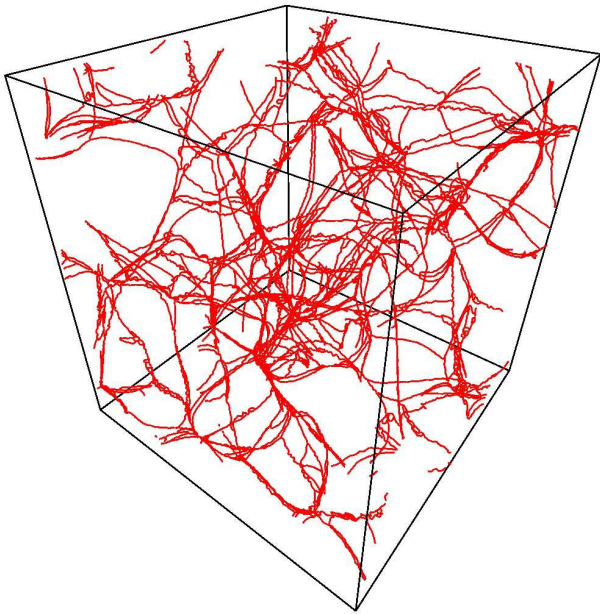}\label{fig_sklrecurse}}
\hfill\subfigure[Direct algorithm]{\includegraphics[width=8.5cm]{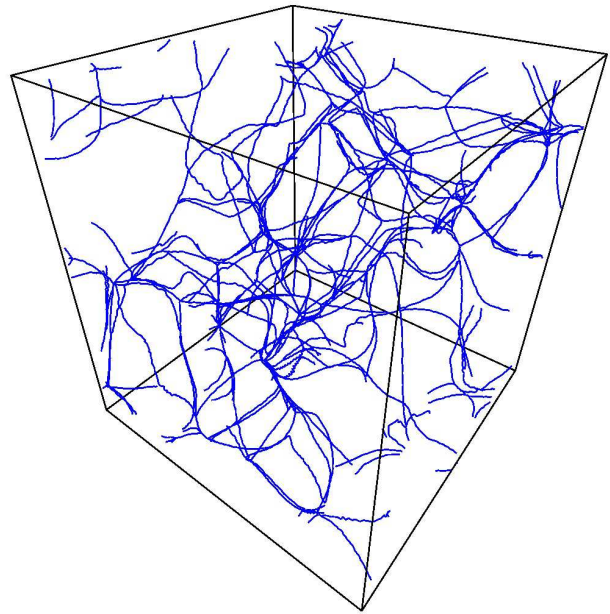}\label{fig_sklnormal}}
\caption{Illustration of the computation of the skeleton as the 1D
  interface of the 1-VPs for a 3D field,
  $\Phi_{3}(\vx)$. Figure~\ref{fig_1vp} presents the $1$-VPs of
  $\Phi_{3}(\vx)$. The 2D surface, $M_2$, is computed as the interface
  of the VPs of $\Phi_{3}(\vx)$. The 1-VPs are the void patches of
  the restriction, $\Phi_{2}(\vx)$, of $\Phi_{3}(\vx)$ to
  $M_2$. Similarly to picture \ref{fig_algo}, each colour corresponds
  to a given 1-VP of $\Phi_{2}(\vx)$, associated to a given minimum of
  $\Phi_{2}(\vx)$ (which is also a saddle point of kind $1$ of
  $\Phi_{3}(\vx)$). The rough appearance of $M_2$ is due to the fact
  that it is approximated by the set of pixels of the  sampling grid
  crossed by the interface of the VPs.  The skeleton of Figure~\ref{fig_sklrecurse} is defined as the interface of the 1-VPs of
  Figure~\ref{fig_1vp}: its location is not very precise and it seems
  to oscillate around its ``true'' location, mainly because only a
  locally flat approximation of $M_2$ is computed. Conversely, the
  skeleton of picture \ref{fig_sklnormal} is computed as the border
  between at least $3$ PPs of $\Phi_{3}(\vx)$ or equivalently as the
  set of points of the surface $M_2$ (pictured on  Figure~\ref{fig_1vp}) which are multiply connected (i.e. where $M_2$ folds
  onto itself). This algorithmically simpler definition leads to a
  much better defined skeleton.\label{fig_3dskl}}
\end{figure*}

For 2D Gaussian random fields, as pictured on figure~\ref{fig_ppatch} and \ref{fig_vpatch}, the skeleton
({\sl resp.} anti-skeleton) are identical to the VP ({\sl resp.} PP) borders and
the direct implementation of this algorithm leads to a very precise and
smooth skeleton. But the implementation in spaces of higher dimensions
raises a critical issue with this simplified method, due to the fact
that the borders of the $\alpha$-PPs and $\alpha$-VPs are only defined
by the index of the pixels they cross: thus they are jagged and
considered locally flat (on the scale of one pixel and its direct neighbourhood). Figure~\ref{fig_1vp} presents the 1-VPs
obtained by applying this algorithm to a 3D Gaussian random field,
each colour corresponding to a different 1-VP index. The 1-VPs live on
the 2D surface which is the border between the  cells formed by the
void patches of the field, each of this cell encompassing exactly one
minimum of the field. This surface is complex: it can be multiply
connected at the interface of more than two different void patches and
its curvature is locally significant. Although  neighbouring relationships
between pixels are easily obtained even where the surface is multiply
connected, only a rough approximation of the actual distances along
the surface can be computed, as the local curvature is not taken into account. Figure~\ref{fig_sklrecurse} shows the
corresponding skeleton, computed as the border of the 1-VPs of
Figure~\ref{fig_1vp}. This skeleton is clearly not very well defined,
the uncertainty in distance computation leading to errors in the
probability propagation algorithm. This bias results in multiple
skeleton branches that seem to oscillate and cross each other along
the true skeleton location.\\

In the end, it appears that dropping the full probability distribution
and approximating borders between patches is too coarse an
approximation. One solution would involve trying to compute the
precise location of the $\alpha$-VPs and $\alpha$-PPs using the full
set of probabilities, but, as it will be discussed in section
\ref{sec:smooth}, this raises complex issues. As it is the patches
interface computation  that seems to be difficult, the alternative we
chose to implement involves  computing directly the skeleton from the
$0-$VPs and $0-$PPs of the field, without having to consider the full
hierarchy of $\alpha$-VPs and $\alpha$-VPs. A close examination of
Figure~\ref{fig_1vp} led us to formulate the conjecture  that the
$(d\!-\!1)$-VPs or $(d\!-\!1)$-PPs interface corresponds in fact to the subspace
of $M_{d-1}$ where the manifold is sufficiently multiply connected
(i.e. where the $(d-1)$-surface defined by $M_{d-1}$ folds onto
itself). Equivalently, this locus can be defined in $3D$ as the
interface of at least $3$ different PPs or VPs (see Figure~\ref{fig_1vp}). This is  formally  demonstrated in
\cite{jost}. In the general case of $M_{d>3}$, the
skeleton should thus be the 1D interface between at least $d$ VPs or
PPs of $\Phi_d(\vx)$. Figure~\ref{fig_sklnormal} presents the skeleton
obtained using this method on the same Gaussian random field as the
one used for Figures~\ref{fig_1vp} and~\ref{fig_sklrecurse}. As
expected, as   there is no need to recursively compute the full
hierarchy of VPs, the resulting skeleton is much more precise and
well defined. Moreover, a quick comparison to Figure~\ref{fig_sklrecurse}  confirms that it is in fact the approximation of
the $\alpha$-patches interfaces by individual pixels that plagues the
algorithm, each recursive step exponentially increasing the error.

\subsubsection{The skeleton as a set of individual filaments}
\label{sec:filaments}
The concepts introduced above allow the definition and extraction
of the skeleton as a {\em fully connected} network that continuously
link maxima and saddle-points of a scalar field together. It is
certainly of interest to try understanding the topological and
geometrical properties of this scalar field through  the connectivity
and hierarchy relationship that it introduces between the critical
points. Applied to cosmology, it also  allows a formal definition of
the concept of individual filaments. Considering matter distribution
on large scales in the Universe, a natural definition of a {\em
  single} filament would be a subset of the cosmic web that directly
links two halos together. The transposition of such a definition to
the skeleton would  allow the introduction of useful concepts such as
 neighbouring relationship between halos in the cosmic web sense. It
would also make possible the study  of filaments as individual physical
objects, similarly to what has been done for years in the literature
with the halos and voids.\\ 

On Figure~\ref{fig_dfield}, the skeleton (coloured thick network, where
the colour corresponds to the underlying PP index) and anti-skeleton
(black network) are superimposed on the density field from which they
where extracted. 
 Let us define a filament as
a subset of the skeleton continuously linking two maxima together
while going through one - and only one - first kind saddle point.
These saddle points can be easily extracted as they are located on the
skeleton, at the border between the peak ({\sl resp.} void) patches (i.e
where the patch index along the skeleton changes, this definition
being valid for any number of dimensions). This way, all the filaments
of an N-dimensional distribution can be extracted individually by
starting from  each maximum of the field, following all the branches
of the skeleton, and storing only the paths that cross one saddle
point before reaching another maximum.  This algorithm thus allows the
individual extraction of filaments as well as a continuous wander of
the filamentary structure of a distribution, which should be very
useful in a wide range of applications in cosmology.

\subsection{Sub-pixel resolution and skeleton smoothing}
\label{sec:smooth} Let us first consider for simplicity a Cartesian sampling
grid (even though this sub pixel smoothing does not critically depend on this
geometry, see below).  The implementation of the procedure  of
Section~\ref{sec:dskel} naturally leads to a skeleton that lives along pixel
edges and is thus jagged at the pixels scale. The differentiability of the
skeleton is nonetheless a feature which may be critical  for a number of  its
characteristics: its length,  curvature, general connectivity ... In order to
enforce this differentiability,  we developed two smoothing methods which we
use in practice in turn. The first one  is based on a multi-linear
interpolation of the patches probability distribution which flows naturally
from  the original  algorithm used to create the skeleton.  It provides
sub-pixel resolution consistently with the probabilistic framework, thus allowing a precise extraction of the skeleton
even when the sampling is low.  The other
is used to control the level of smoothness away from fixed points (the maxima
or the bifurcation points) and can be used to enforce sufficient differentiability.

\subsubsection{Multi-linear sub-pixel skeleton }

\begin{figure*}
\centering  
\hfill\subfigure[Intersection of $2$ patches]{\includegraphics[width=5.5cm]{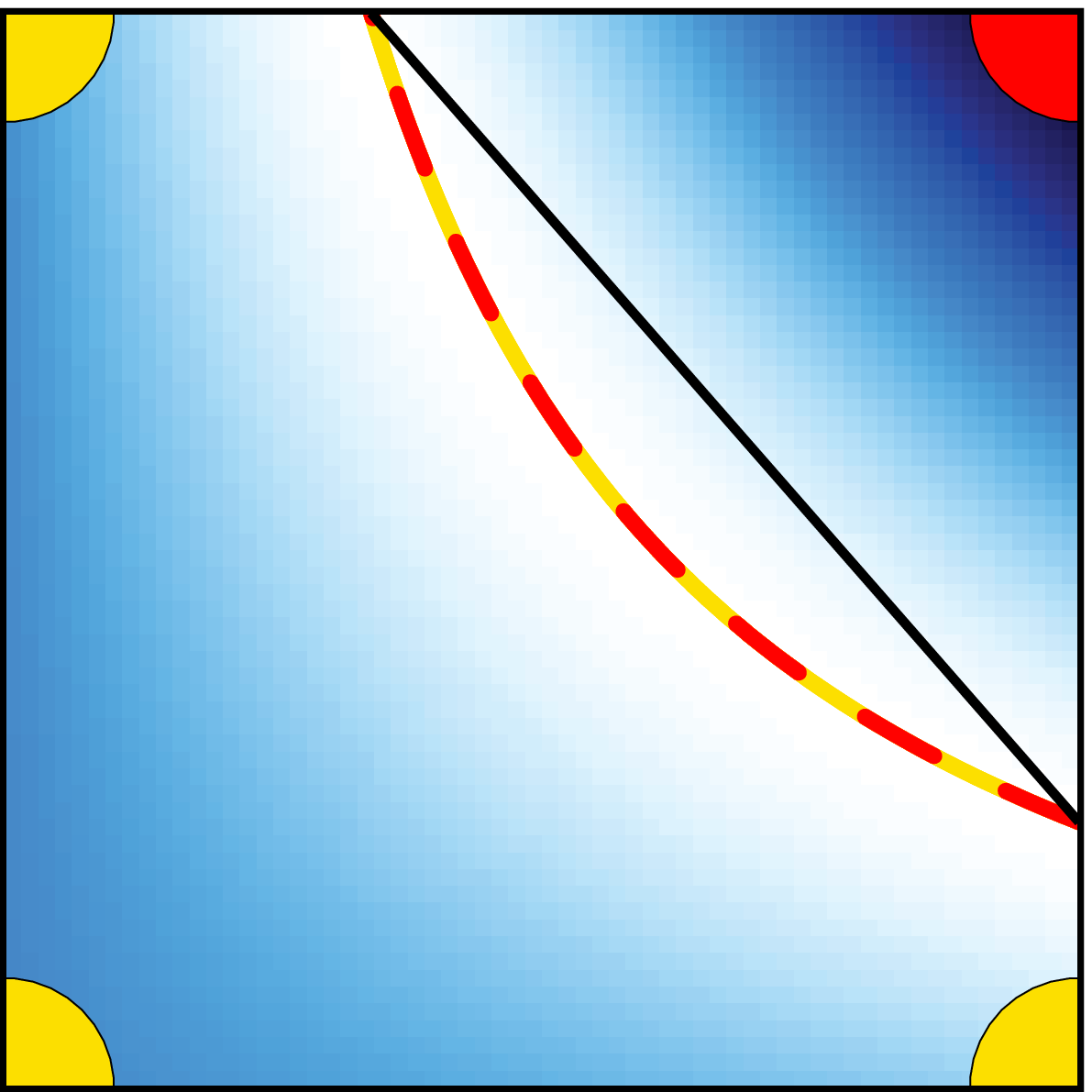}\label{fig_smooth2p}}
\hfill\subfigure[Intersection of $3$ patches]{\includegraphics[width=5.5cm]{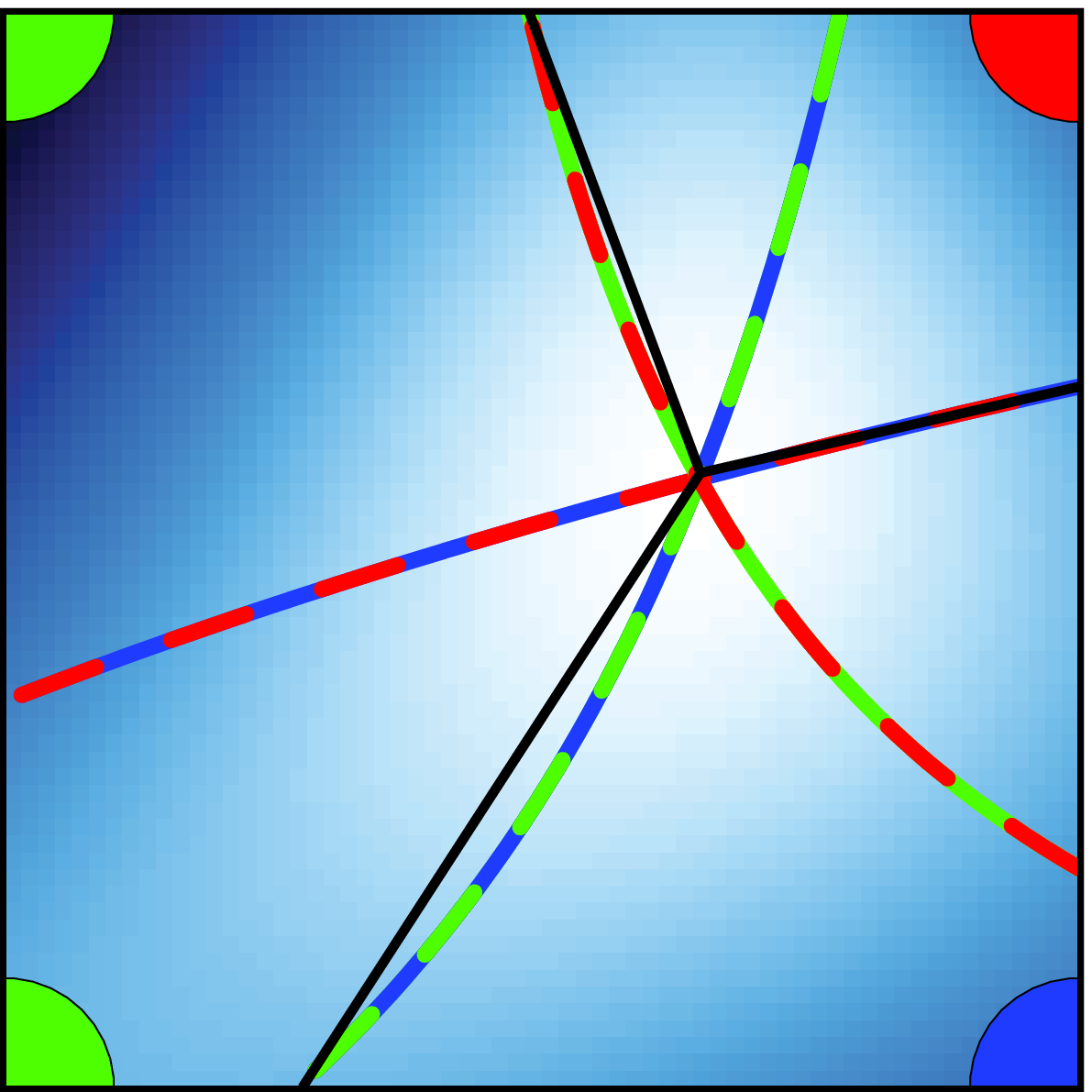}\label{fig_smooth3p}}
\hfill\subfigure[Intersection of $4$ patches]{\includegraphics[width=5.5cm]{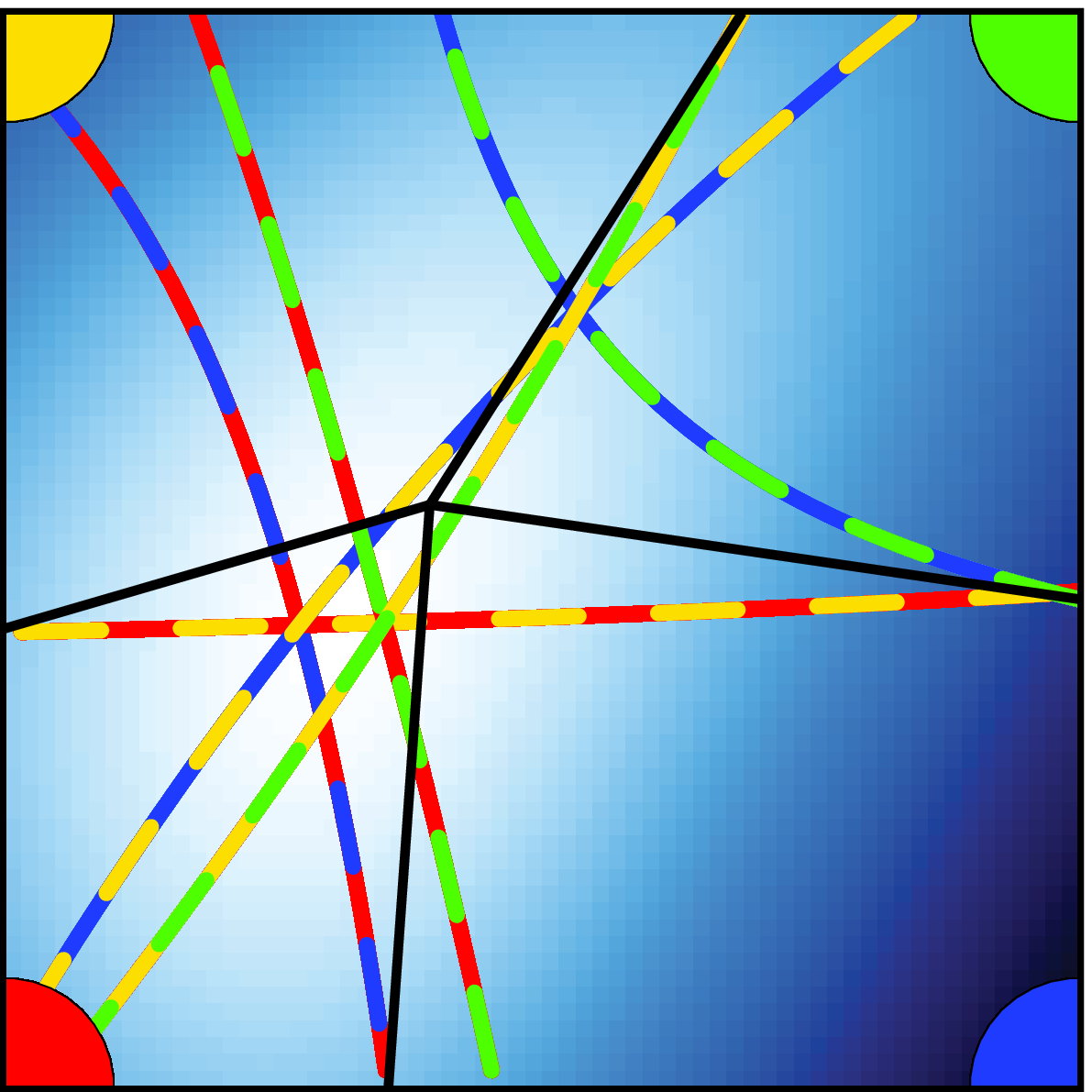}\label{fig_smooth4p}}
\caption{ Illustration of the computation of the sub-pixel skeleton in the
case of a 2D bi-linearly interpolated pixel located at the border of $2$, $3$
and $4$ different patches. The colour of the pixels vertices represent the
index of the dominant patch, while the two-coloured dotted lines are the
quadrics, solutions of Equation (\ref{eq_equalprob}). These lines are the
regions where the probabilities corresponding to the patches with similar
colours are equal. The underlying blue gradient corresponds to the value of
$\tilde P(\vx)$ (Equation (\ref{eq_pvar})), light colours encoding 
lower value. Finally, the black lines represent our approximation of the
smoothed intersection of  the skeleton with the
pixel. \label{fig_propersmoothing}}
\end{figure*}

Let us first find a way to obtain a sub-pixel resolution on the
skeleton position based on the patches probability distribution of
each pixel. The raw skeleton is made  of individual segments located
on the edges of the pixels of a Cartesian grid $G$. Each segment is
defined by its two end points, and each of them is  surrounded by
$2^d$ pixels with a full list of possible patches index, together with
their respective probabilities.  Recall that the probabilistic
algorithm we use works on individual pixels so the resulting skeleton
position, defined as the position of the border between several
patches, is computed with a precision of one pixel.  This implies that
the smoothing procedures may not move the skeleton on more than half
the size of a pixel. In other words, if we consider the dual sampling
grid, $\bar G$, of $G$,  the skeleton can be freely moved within the
pixels of $\bar G$ that its jagged approximation crosses. So it is
sufficient to consider individually each of these pixels.  Let $\bar
p$ be one of these pixels. We then know for each of its vertices,
$p_i$ with $i\in{1..2^d}$, the probability distribution of the
different VPs, $P_i^k$, where k is the index of a VP. In order to
obtain sub-pixel resolution, these probabilities can be interpolated
within $\bar p$. 

For simplicity, we will only  use a multi-linear interpolation and
define $P^k(\vx)$, the probability distribution of patch $k$,
interpolated at point $\vx=(x_1,..,x_d)\in[0,1]^d$ within $\bar p$ as:
\begin{equation}
P^k(\vx) = \sum_{i=1}^{2^d} P_i^k \prod_{j=1}^d \epsilon_j^i(x_j),
\end{equation} where $\epsilon_j^i(x)=x$ if the $j^{th}$ coordinate of $p_i$
within $\bar p$ is $1$ and $\epsilon_j^i(x)=(1-x)$ if it is
$0$. Ideally, the skeleton should not belong to any VP, so it should
be located where all the non null values of $P^k(\vx)$ are equal. Let
us define the arithmetic mean of the probability (over the VPs with
index $k$) over the pixel
\begin{equation}
\left<P(\vx)\right> = 1/N\sum_k^N P^k(\vx),
\end{equation}
and its root mean square,
\begin{equation}
\label{eq_pvar}
  \tilde P(\vx) = \sqrt{\sum_k^N \left(P^k(\vx) -
    \left<P(\vx)\right>\right)^2},
\end{equation}
where the sum is over all the $N$ subscripts $k$ such that there exist
a pixel $p_i$ where $\forall l\neq k, P_i^k>P_i^l$.  Clearly, all
patches with dominating probabilities $P^k(\vx)$ in $\bar p$ are equal
when 
\begin{equation}
\label{eq_pvar0}
\tilde P(\vx)=0.
\end{equation}
Equation (\ref{eq_pvar0}) is of fourth order and is thus difficult to
solve in general. 

\subsubsection{Approximate  quadrics sub-pixel smoothing}

Insights into the solution of Equation (\ref{eq_pvar0}) can be found while 
considering the intersection sets of points where pairs of
probabilities are equal instead of equating them all at the same
time. These sets are solution of the set of equations
\begin{equation}
\label{eq_equalprob}
P^k(\vx) = P^{k\prime}(\vx),\quad k \neq k\prime
\end{equation}
where $k $ and $ k\prime$ are subscripts of the patches that dominate on at
least one vertex of $\bar p$.

For clarity, let us consider the $d=2$ case first. With a proper indexing of  the
four pixels $p_i$, 
\begin{eqnarray}
P^k(\vx)& = &P_1^k (1-x_1)(1-x_2)+P_2^k (x_1)(1-x_2)\nonumber\\
 &&+P_3^k (1-x_1)(x_2)+P_4^k (x_1)(x_2),
\end{eqnarray}
Equation (\ref{eq_equalprob}) writes in this case:
\begin{equation}
\label{eq_quadric2d}
A\, x_1x_2 + B\, x_1 +C\,x_2 +D = 0\,,
\end{equation}
 where $A$, $B$, $C$ and $D$ only depend on the values of
$P_i^k$. Equation (\ref{eq_quadric2d}) is quadratic and its solutions are well
known curves of dimension $d-1=1$ called quadrics. Figure~\ref{fig_propersmoothing} illustrates solutions of Equation
(\ref{eq_quadric2d})  when $\bar p$ is located at the intersection of $N_p=2$,
$N_p=3$ or $N_p=4$ different VPs. In the most frequent configuration where
$\bar p$ is at the intersection of $2$ VPs, Equation (\ref{eq_quadric2d})
directly gives the first order approximation of the intersection of the
skeleton and $\bar p$, and we may approximate it by a straight
segment. Finding the end points of  this segment is easily achieved by
computing the location of equal probability along the two sides of $\bar p$
that link vertices with different patches (Figure~\ref{fig_smooth2p}). The
$N_p=3$  configuration is  rarer, and concerns only the maxima of the field as
well as all  bifurcation points of the skeleton. In this case, we know that
three different branches of the skeleton merge within the pixel, at a point
where all probabilities are equal. So, we may set the bifurcation point as the
locus where all the $C_2^3=3$ quadrics of Equation (\ref{eq_equalprob})
intersect (note that the three of them always intersect in a single point as
$P^1(\vx)=P^2(\vx)$ and $P^1(\vx)=P^3(\vx)$ implies $P^2(\vx)=P^3(\vx)$). The
three branches of the skeleton in $\bar p$ are thus obtained by linking the
bifurcation  point to the iso-probability along the three sides of $\bar p$
that link vertices associated to different patches (Figure~\ref{fig_smooth3p}).  Finally, the $N_p=4$ configuration  is very
rare\footnote{note that the scarcity of these points is directly related to
resolution, i.e. whether or not the skeleton is featureless at the sub-pixel
scale. Hence  these points may occur more often in higher dimensions,  which
for computational reasons may be relatively under-sampled.}  and also more
problematic. As previously, we know that there exist a bifurcation point
within $\bar p$, but this time with $4$ different skeleton branches. Since
there are now $C_2^3=6$ different Equations (\ref{eq_equalprob}), and given
that the solution of each of them is a 1D quadric, this system is clearly over
constrained to find the precise location of the bifurcation point. A solution
may well be to use a higher order interpolation, allowing more complex curves
than quadrics for equal probabilities regions, or to try solving directly
Equation (\ref{eq_pvar0}). As this case is clearly rare, it would  also be
possible to approximate the bifurcation point as the barycenter of the three
points of intersection  of the subsets of  Equations (\ref{eq_equalprob})
taken in pairs.  Again, the smoothed skeleton would  therefore be derived by
linking the bifurcation point to the four iso-probability points along the
four sides of $\bar p$ (Figure~\ref{fig_smooth4p}).

\subsubsection{Actual recursive implementation}
\label{sec:smooth-algo}

\begin{figure*}
\centering  
\subfigure[]{\includegraphics[width=4cm]{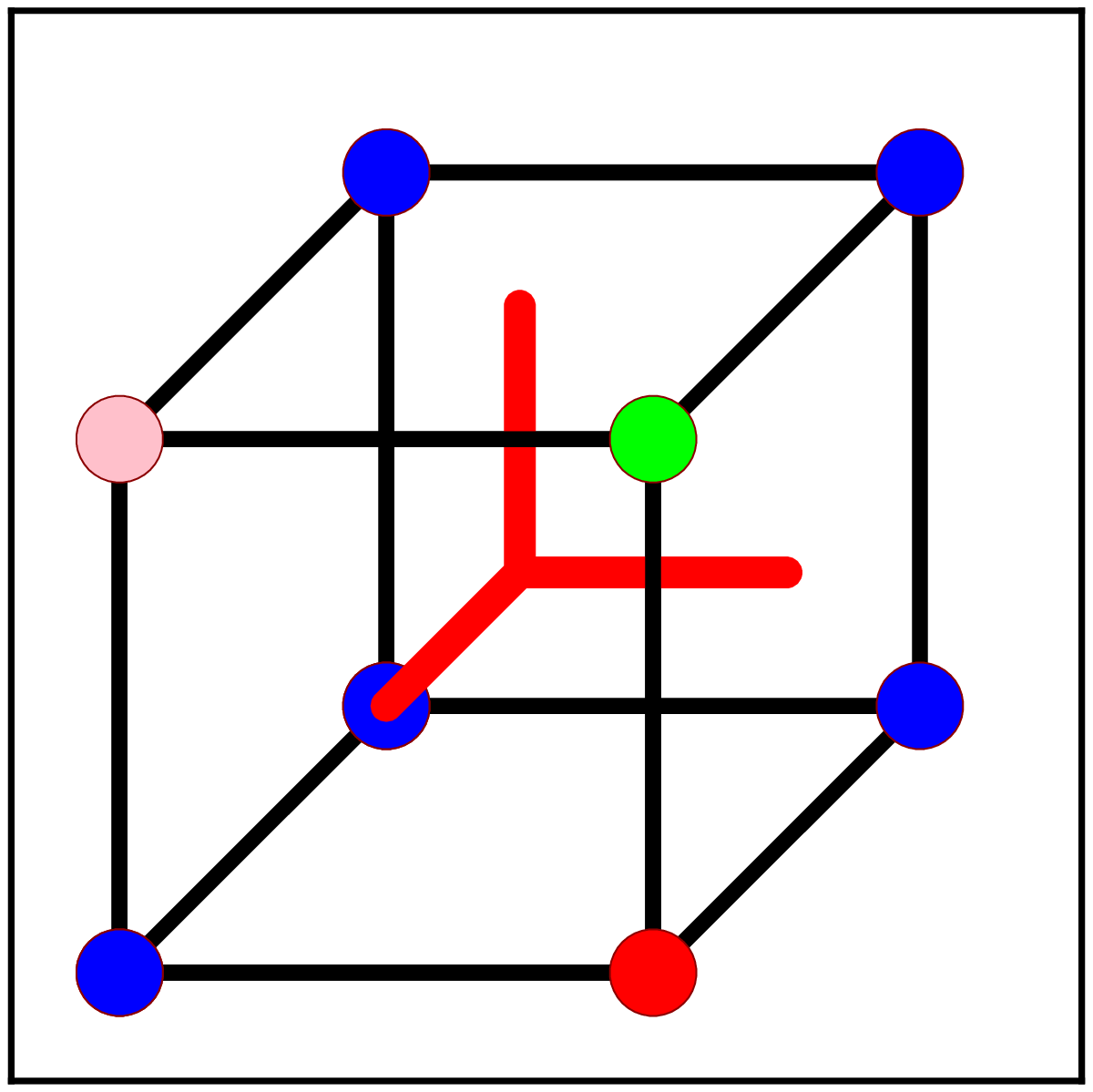}\label{fig_subp1}}
\hfill\subfigure[]{\includegraphics[width=4cm]{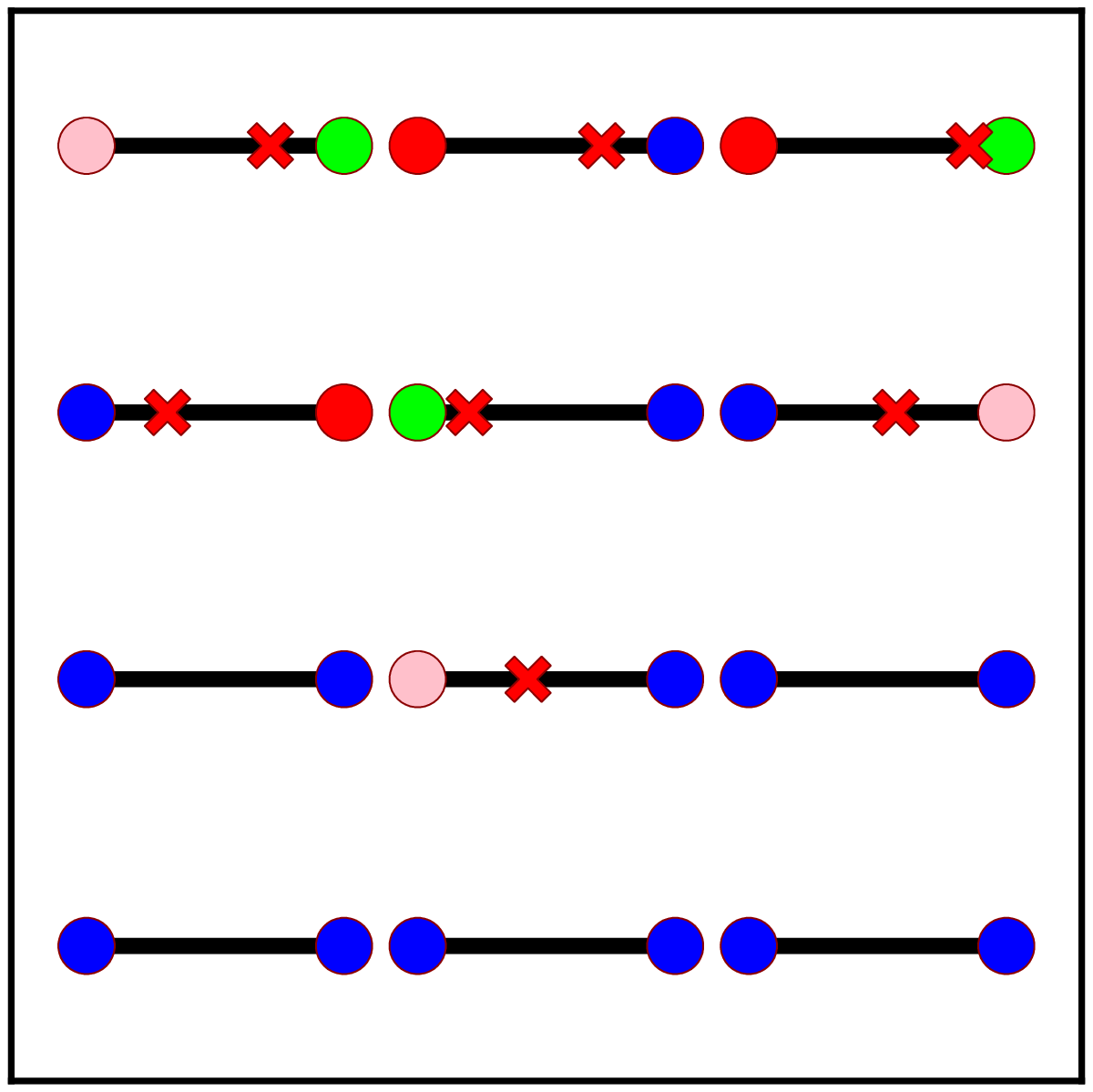}\label{fig_subp2}}
\hfill\subfigure[]{\includegraphics[width=4cm]{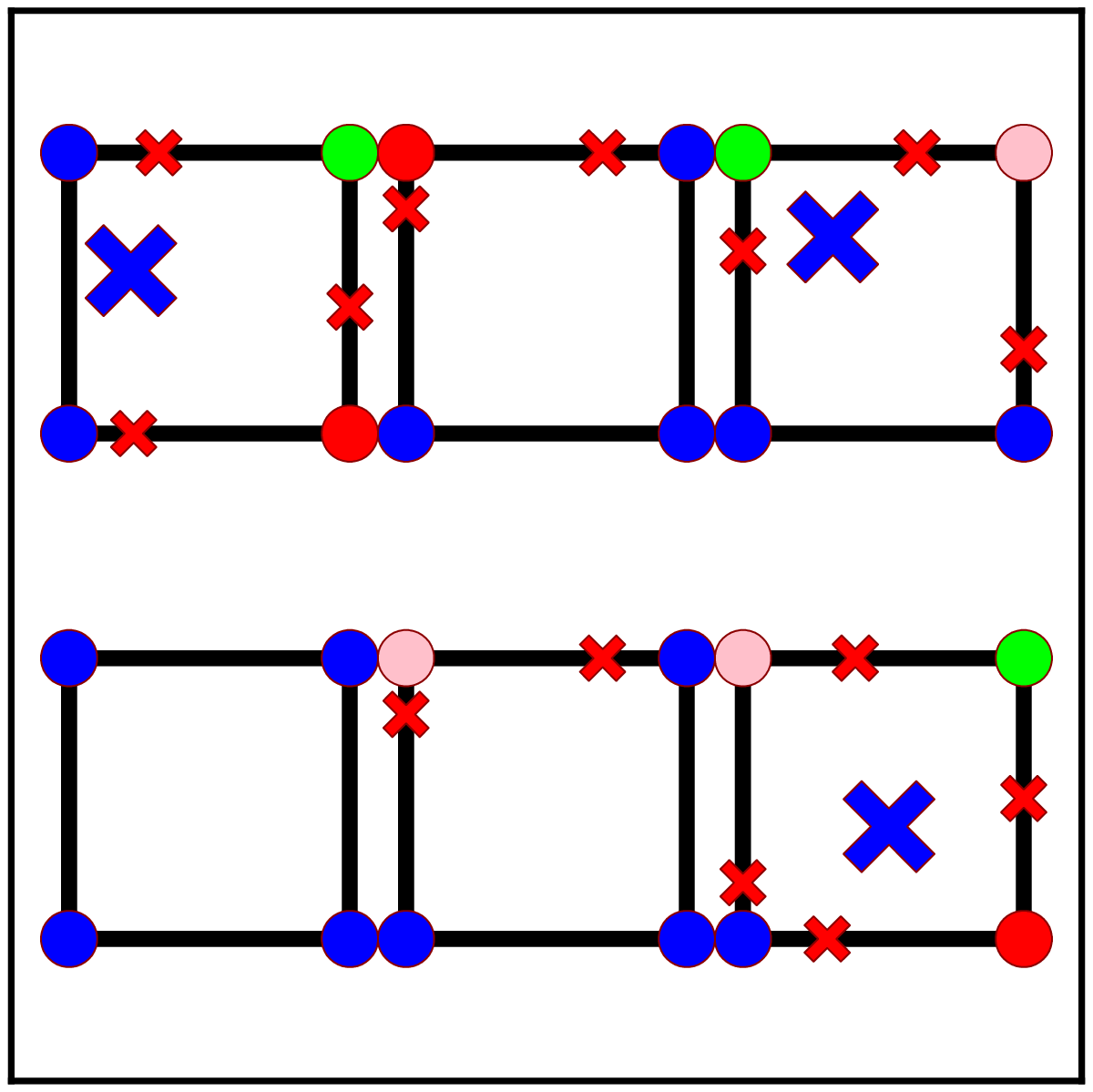}\label{fig_subp3}}
\hfill\subfigure[]{\includegraphics[width=4cm]{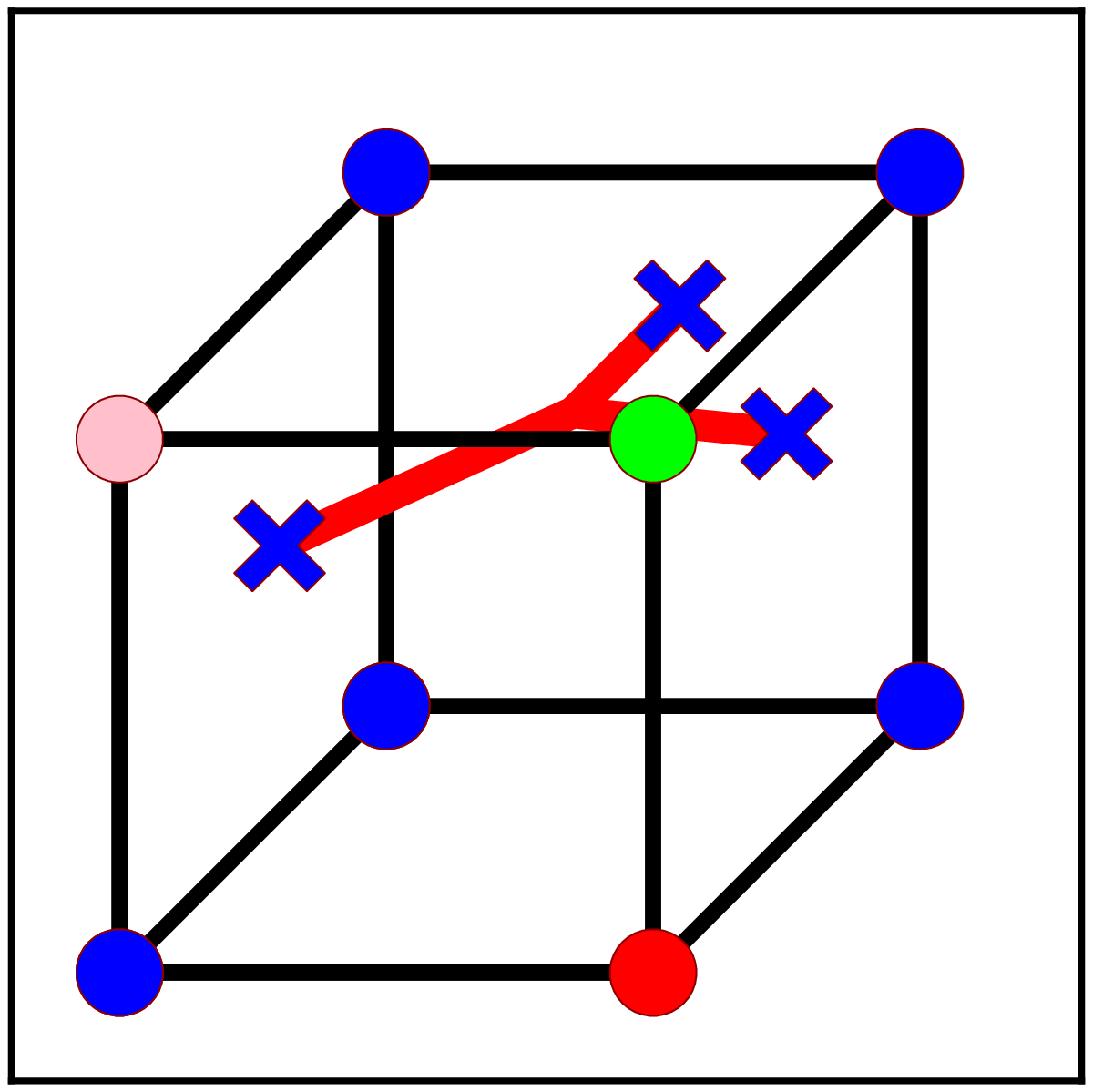}\label{fig_subp4}}
\caption{ Illustration of different steps of the  recursive algorithm
  used to obtain sub-pixel resolution for the skeleton, in 3D. The
  colour of the balls represent the index of the patch with maximal
  probability while the intersection of the skeleton and the cell is
  displayed in red. The algorithm consists in recursively considering
  the $n$-dimensional faces  of the sampling unit volume (here an
  hypercube). For a 3D Cartesian sampling grid, one starts equating
  dominant probabilities on the vertices of edges, then faces and
  finally the cube. \label{fig_subp}}
\end{figure*}
Having discussed the underlying geometry of the sub-pixel multi linear
interpolation, let us now turn to our actual sub-pixel smoothing
algorithm.  Indeed, in $d$-dimensions, Equation (\ref{eq_equalprob})
is of order $d$ and is linear  in each of the $d$ space coordinates
$x_i$. Its solutions are thus $d-1$ dimensional quadrics whose
intersections, as in $2D$, can be used to recover the skeleton
position down to a sub-pixel precision. Finding intersections of
quadrics in general remains nonetheless a highly difficult (or even
untractable!)  problem and even state-of-the-art solvers can only
achieve such a performance for $d=3$ at most. To circumvent this
difficulty, we thus opted in practice for a different solution that
consists in a recursive numerical minimization of the value of $\tilde
P(\vx)$ over the hierarchy of n-cubes (i.e hypercubes of dimension
$n$), $n\in \lbrace1,..,d\rbrace$, that are the faces of each cell of
the sampling grid. The trick is to always reduce the problem to a 1D
minimization of a polynomial of order $d$ (see appendix A).  Figure~\ref{fig_subp} illustrates the full process in 3D.  Let us consider
the grid cell of Figure~\ref{fig_subp1}, located at the interface of
$4$ different patches. The skeleton extraction algorithm produces  the
jagged skeleton represented in red. In order to improve its
resolution, we first consider each of the $12$ edges individually (see
Figure~\ref{fig_subp2}) and determine for each of them the point of
equal probability for the two patches that dominate at the  end points
of segment.  Of course this point only exists if different patches
dominate at the end points of a segment and we thus obtain at most
$12$ new points ($7$ in this instance, represented by the red
crosses). The edges of a cube can be considered as its ``one
dimensional faces'' or $1$-faces. The following step consists in
examining the configuration of its $2$-faces, usually called faces for
a 3D cube. Figure~\ref{fig_subp3} illustrates the configuration of
these  $6$ faces together with the iso-probability points computed
over their edges. We know that at least $3$ different patches have to
dominate on at least one of the $4$ vertices of each face for a
skeleton branch to enter the cell through this face. Using the
minimization algorithm presented in Appendix A and the iso-probability
points on the edges, it is thus possible to compute, over these faces,
the location of the minimum of $\tilde P(\vx)$ (represented as blue
crosses on Figure~\ref{fig_subp3}). Finally, considering the $3$-face
of the cell (i.e. the cube itself), one can determine the point of
minimal value of $\tilde P(\vx)$ over the cube, which is the point
where the skeleton branches connect  (see figure~\ref{fig_subp4}).\\

\begin{figure}
\centering
\includegraphics[width=7cm]{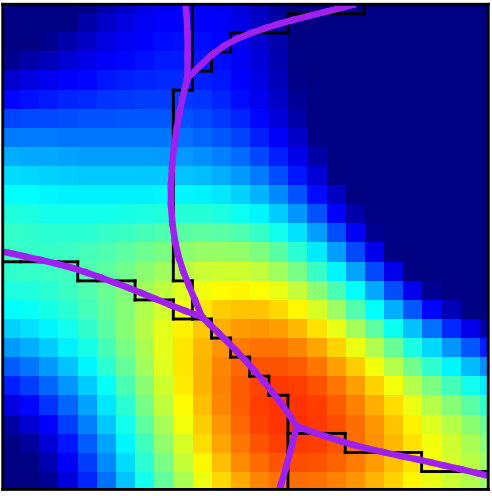}
\caption{Illustration of the skeleton with a sub-pixel resolution in
  2D. The background pixels colour represent the sampled density
  field while the black skeleton was obtained using our probabilistic
  algorithm. The purple skeleton is the post-treated version of the
  black one. Note how any sampling grid influence disappeared,
  especially in the originally vertical segment located in the
  upper-left corner of the image. \label{fig_sklpost_2D}}
\end{figure}

The generalization of this algorithm is relatively
straightforward. Let us again consider a cell  that is a hypercube of
dimension $d$. We know that the skeleton intersects this cell if at
least $d$ of its vertices have different maximal probability patches
index. In that case, the sub-pixel resolved skeleton can be recovered
by considering all the $n$-faces of the hypercube, $n\in\lbrace
1,..,d-1 \rbrace$, in ascending order of their dimension $n$. When
considering a $p$-face, we minimize the value of $\tilde P(\vx)$ in
order to obtain the point where its vertices respective patches have
equal probability, using the points obtained from the
$(p-1)$-faces. This point only exist for a $p$-face if at least
$(p+1)$ vertices most probably belong to different patches. In the
end, one thus obtains a number  of points from the $(d-1)$-faces that
are the points where the skeleton enters the cell and one point for
the $d$-face (i.e. the cell itself), which gives the location where
different branches of the skeleton connect. Figure~\ref{fig_sklpost_2D} illustrates  the result of applying this
algorithm in the $2D$ case.

\subsubsection{Artifacts correction and differentiability}
\label{sec:artifact}

\begin{figure}
\centering
\includegraphics[width=7cm]{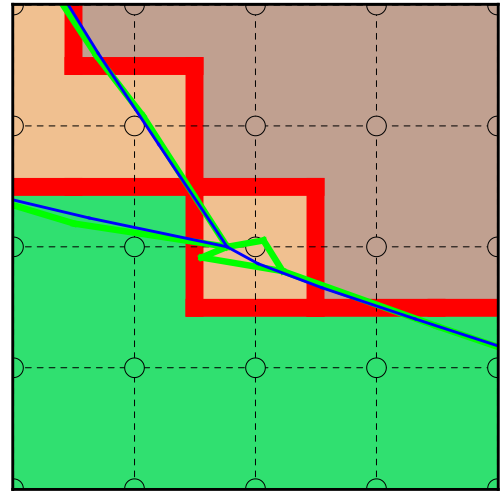}
\caption{A failure of the skeleton sub-pixel algorithm due to
the lack of sampling resolution. The dotted grid represents the reciprocal
sampling grid, $\bar G$, while the pixels colour represents their dominating
patches and the initial raw skeleton is represented in red. The green skeleton
is the result of applying the sub-pixel resolution algorithm while the blue one
was obtained from the green one, after removing one pixel sized loops and
smoothing.\label{fig_subp_fail}}
\end{figure}

Though the method presented above to obtain sub-pixel resolution works most of
the time, there nonetheless exist situations where it can fail due to sampling
effects. Figure~\ref{fig_subp_fail} illustrates such a situation, which can
sometimes occur when the sampling grid pixel size is not totally negligible
compared to the average extension of the patches. When the thickness of a peak
or void patch is smaller than a pixel size, it can in fact lead to mistakingly
isolated subregions of size one pixel, implying the creation of spurious loops
in the skeleton (in red). This phenomenon, although rare, occurs in spaces of
arbitrary dimension and triggers artifacts when applying our sub-pixel
resolution algorithm. The green skeleton on Figure~\ref{fig_subp_fail}
presents such an example of a spurious skeleton loop.\\

In order to fix these anomalous segments, we chose to post-treat the skeletons
by opening-up all one-pixel sized loops and
smooth the resulting skeleton to enforce a desired level of
differentiability in the skeleton trajectory (see the blue skeleton of Figure~\ref{fig_subp_fail}). The smoothing method that we use presents the advantage
of being quite robust, and involve fixing some specific points of the
skeleton, and averaging the position of each non-fixed segments end points with
the position of its closest  neighbouring end points a number of times. Let
$x^i_j$  be the $j^{th}$ coordinates of the $i^{th}$ sampled skeleton location
(among $N$) between two fixed points. Before smoothing, all $x^i_j$ are
located on the edges of $G$ and we can define their smoothed counterparts as
$y^i_j$, computed as:
\begin{equation}\label{eq_smoothing}
y^i_k = A^{ij} x^j_k,
\end{equation}
with
\begin{equation}
A^{ij}= \left\lbrace
\begin{array}{l}
3/4\quad{\rm if }\quad i=j=0\quad{\rm or }\quad i=j=N, \\
1/2\quad{\rm if }\quad i=j,\\
1/4\quad{\rm if }\quad i=j+1\quad{\rm or }\quad i=j-1, \\
0\quad{\rm elsewhere,}\\
\end{array}
\right.
\end{equation}
where Equation (\ref{eq_smoothing}) is applied $s$ times in order to
smooth over $s$ segments. Basically, Equation (\ref{eq_smoothing}) is
used to compute smoothed coordinates $y^i_j$ as a weighted average of
the original coordinates $x^i_j$ together with the coordinates of its
$2$ direct neighbours, $x^{i-1}_j$ and $x^{i+1}_j$. Applying this
scheme $s$ times thus produces  the final smoothed coordinates
$y^i_j$ to be a weighted average of $x^i_j$ and its $s$ closest
neighbours along the skeleton.

This smoothing technique introduces two parameters of importance: the
skeleton smoothing length $s$, and the type of fixed points.  In order
to determine the optimal value of $s$, it is possible  to minimize the
reduced $\chi^2$ corresponding to the discrepancy between  $y^i_j$ and
$x^i_j$ supplemented by a penalty corresponding to  the total length
of the skeleton (over-smoothing will increases the discrepancy,
under-smoothing will increase the total length). In practice, though,
as a post treatment to an already smooth skeleton (using the sub pixel
probabilities),  this choice is not critical.

\begin{figure}
\centering
\includegraphics[width=7cm]{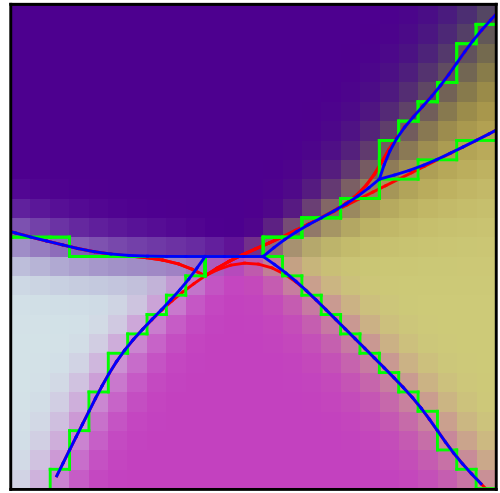}
\caption{Influence of the choice of fixed points on the shape of the
  smoothed skeleton. The original skeleton is represented in green,
  while the red and blue skeletons are smoothed $s=6$ times, while
  fixing the field maxima and the bifurcation (i.e multiply connected)
  points respectively. In both cases, the smoothed versions always
  stay within half the size of a pixel distance from the original
  non-smoothed skeleton. On this illustration, the smoothed skeleton
  was computed {\sl directly} from its raw jagged version  to
  emphasize the effect of the choice of different fixed points. This
  discrepancy between the two options is considerably weakened if the
  skeleton is previously post-treated. The background colour
  corresponds to the weighted probability each pixel has to belong to
  a definite patch.\label{fig_smoothing}}
\end{figure}

The choice of the skeleton points that should be fixed before smoothing
depends of the planned application;  in practice, we implemented two
possibilities: (i) fixing the field extrema, or (ii) the  bifurcation points of  the
skeleton  (i.e the points of the skeleton where two filaments merge into
one). Figure~\ref{fig_smoothing} illustrates the influence of this choice on
the shape of the smoothed skeleton. By fixing the extrema of the field, one
ensures that the skeleton subsets that link these extrema are treated
independently: this is the solution used to study the properties of individual
filaments in the dark matter distribution on cosmological scales. One should
note that, in this case, the parts of the skeleton that belong to several
individual filaments are duplicated (see the red skeleton on Figure~\ref{fig_smoothing}), affecting global  properties of the skeleton such as its
total skeleton length. In contrast, fixing bifurcation points enforces  the smoothing of the skeleton while conserving its global properties.

\begin{figure*}
\centering   \includegraphics[width=12cm]{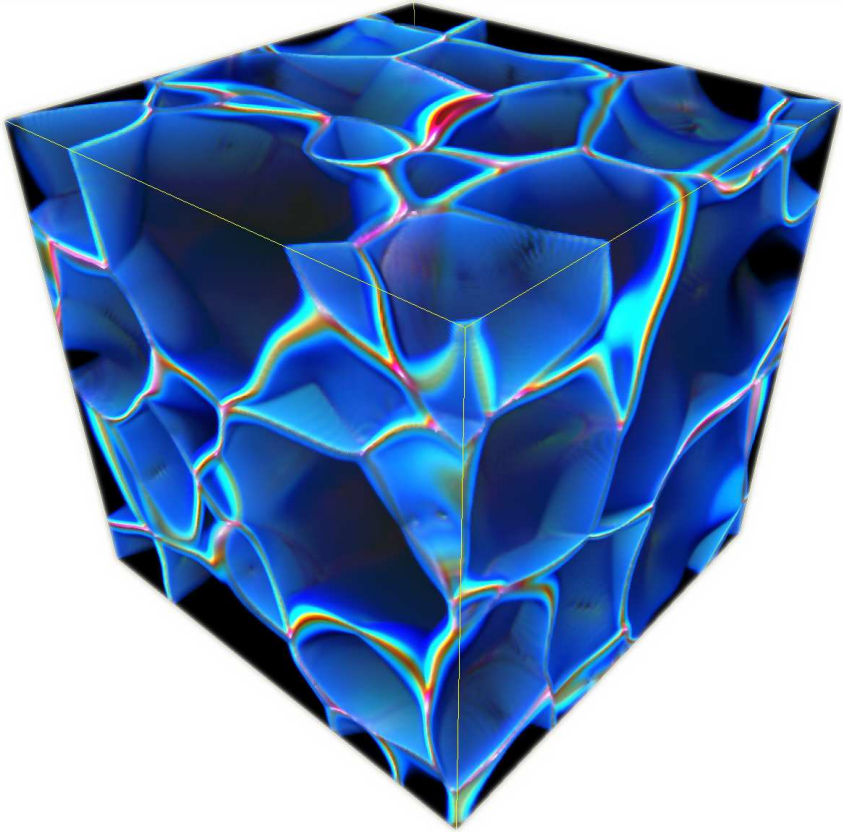}
\caption{ the 3D peakpatch colour coded probability function: warm colours correspond to equiprobability regions, 
dark colours to regions where one probability dominates (see also Figure~\ref{fig_proba}). 
This supplementary map complements the peakpatch map in the present algorithm
and allow for a precise sub pixel segmentation and
skeleton extraction.
Note the extended equiprobability sheets corresponding to places where the exact position of the filament will be more uncertain.
 \label{proba3d}}
\end{figure*}

\subsection{Summary}
\label{sec:algo_sum}

\begin{figure}
\centering   \includegraphics[width=8.5cm]{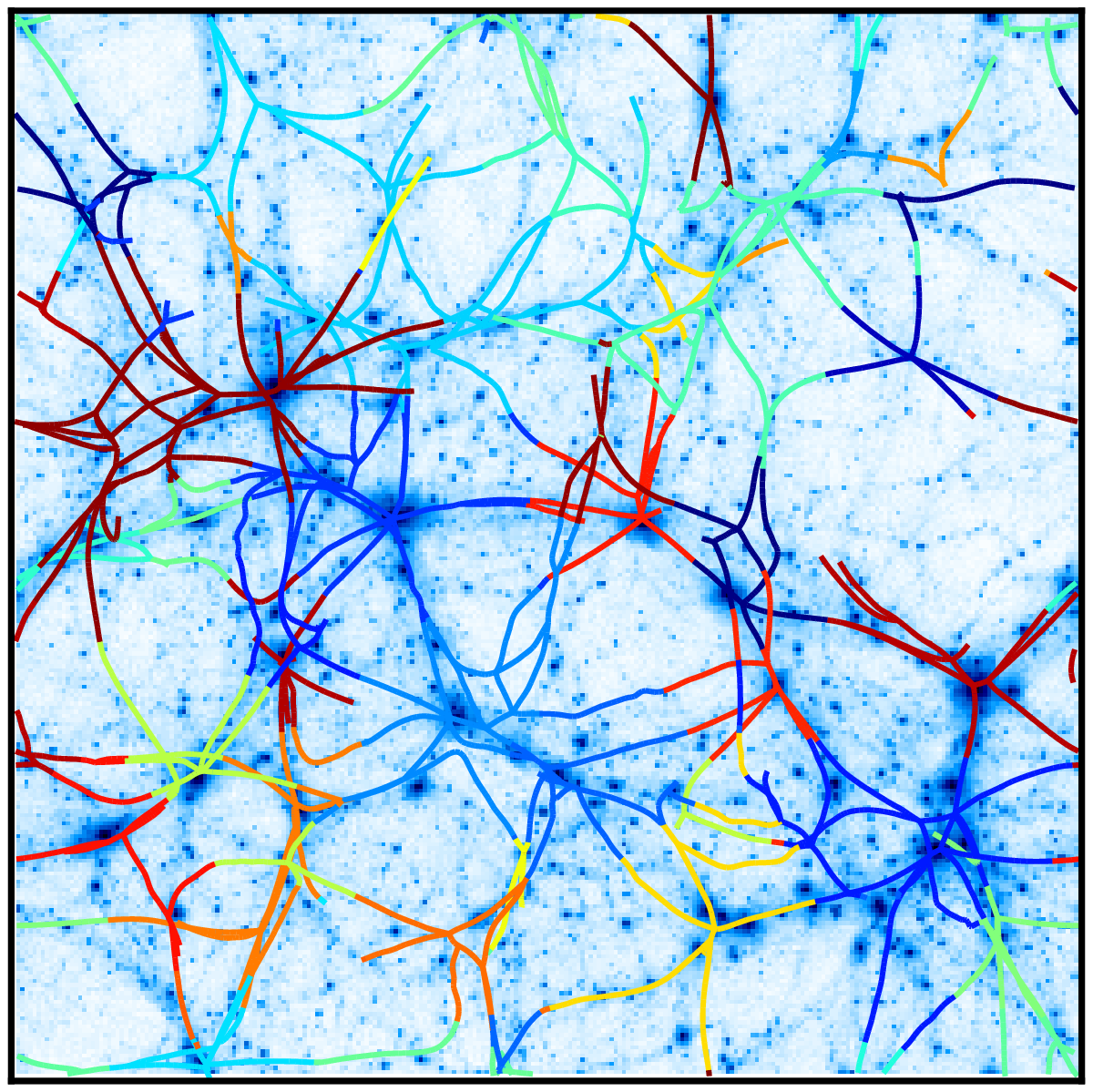}
\caption{ The 2D projection of a 3D skeleton computed on a simulation of the cosmological density
density field in a $50h^{-1}$ Mpc box with {\tt gadget-2}. This $20 h^{-1}$ Mpc thick section of skeleton was computed from a $128^3$ pixels sampling grid smoothed over $5$ pixels ($\approx 2 h^{-1}$ Mpc). The skeleton colour represents the index of the peak patch. Note that the 2D projection of a 3D skeleton differs from the skeleton of the 2D projection, hence the discrepancy between the skeleton and apparent filaments. 
\label{fig_3dskl_proj}}
\end{figure}

\begin{figure*}
\centering   \includegraphics[width=14cm]{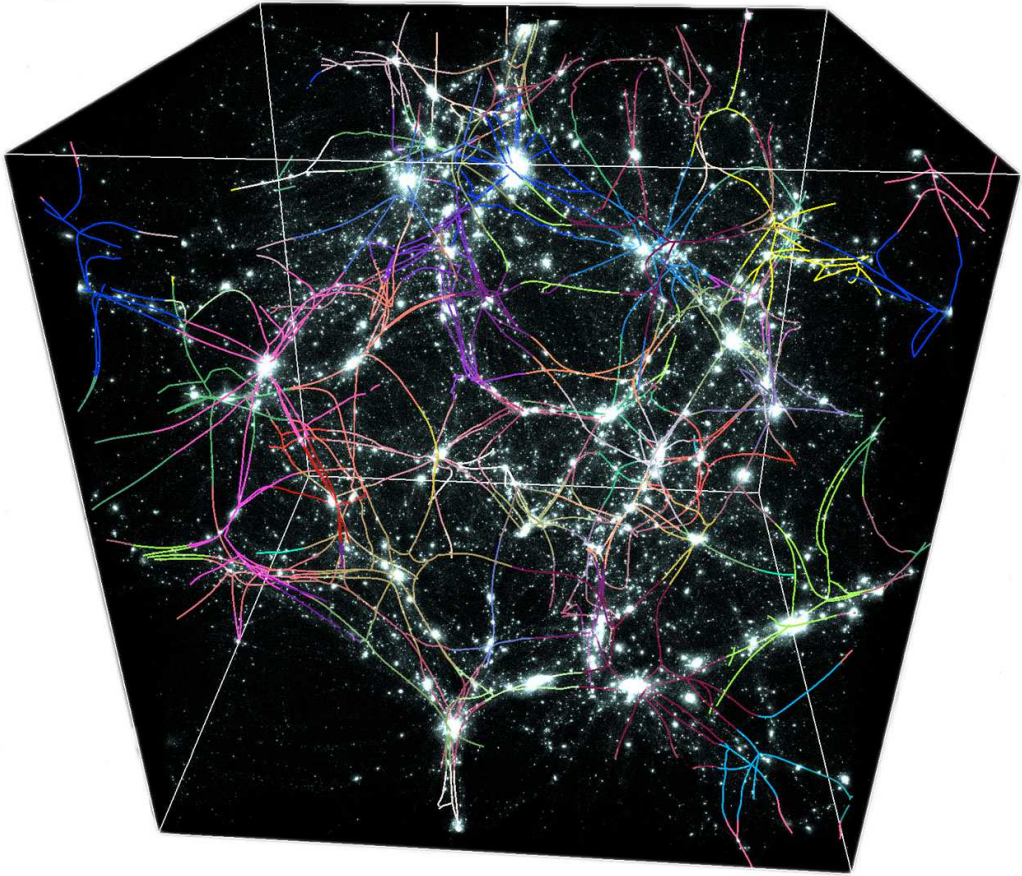}
\caption{ The 3D skeleton of the simulation of the cosmological
  density density field in a $50h^{-1}$ Mpc box with {\tt gadget-2}
  (see also Figure~\ref{fig_3dskl_proj}). This skeleton was computed
  from a $128^3$ pixels sampling grid smoothed over $5$ pixels
  ($\approx 2 h^{-1}$ Mpc). The skeleton colour represents the index
  of the peak patch (which provide by construction
    the natural segmentation of filaments attached to the different
    clusters.). Movies of 3D skeletons can be downloaded at {\it http://www2.iap.fr/users/sousbie/}\label{fig_3dskl_simu}}
\end{figure*}

Let us finally recap the main steps involved in the extraction
of the fully connected skeleton in  a
$d$-dimensional space.
\begin{enumerate}
\item The density field is sampled and smoothed in order to ensure
  sufficient differentiability. A smoothing scale of at least $5$
  pixels is  recommended when using a Cartesian grid.
\item All pixels are considered in the order of their ascending (or
  descending) density. Depending on their neighbours, they are
  labelled as minima (or maxima), or  assigned a list of probability to belong
  to a given VP (or PP) following the algorithm of section
  \ref{sec:patch}.
\item Considering only the patch index with highest probability for
  each pixel, skeleton segments are created on pixel edges when at
  least $d$ surrounding pixels  among $2^d$ have a different most
  probable patch index.
\item Calling a vertex connected to more than two segments a node
  of the skeleton and considering each node, the sets of connected
  segments that link them to other  nodes are recorded in order to later
 recover the information on the skeleton connectivity (and allow a continuous
  wander along the fully connected skeleton).
\item The sub-pixel smoothing procedure of Sec.~\ref{sec:smooth-algo}
  is implemented. All the vertices of the skeleton segments  are considered
  one by one together with  the value of the probability distribution
  in the center of the surrounding pixels. According to the sub-pixel
  algorithm, the extremities are moved in order to obtain a
  differentiable skeleton. 
\item Configurations that are identified as  problematic are corrected for
  following the method described in section \ref{sec:artifact}, and
  the resulting skeleton is smoothed over a few pixels (usually $d$ of
  them) while fixing either bifurcation points or maxima/minima.  
\item Eventually, individual filaments can be extracted (and tagged) following the
  method of section \ref{sec:filaments}.
\end{enumerate}
Figures~\ref{fig_3dskl_proj} and \ref{fig_3dskl_simu} show a 3D skeleton computed from a simulated density field at $z=0$, sampled over only $128^3$ pixels.\\

Note that in this paper, we did not address the issue of shot noise
that has for long been known to be a problem for most segmentation
algorithms, and in particular for   Watershed techniques (see {\it
  e.g.} \cite{jos01} for a review on the subject). In fact, shot noise
often leads to over segmentation, and numerous complex techniques have
been developed to try and compensate for it.  Instead, we chose here
to follow the approach  used  in \cite{2dskel}, \cite{3dskel} and
\cite{sklsdss}, that involve simply filtering the sampled fields using
a Gaussian kernel on large enough scales (in terms of number of
sampled pixels) so that  it is possible to consider that the sampled
field is a smooth enough representation of the underlying field. A
clear disadvantage of this method is that it introduces a particular
smoothing scale and thus adds one parameter (the smoothing scale) to
take into account when considering sets of critical lines an surfaces
computed on a field. A short study of the robustness of the skeleton extraction in the case 
of a smoothed scalar field is presented in appendix C. Improvements over this shortcoming are postponed
to further investigations.\\

Regarding performance, the computing time and memory consumption for the extraction of the skeleton
mainly depends on three parameters: the number of pixels $N_p$, the smoothing
length $L$ in units of pixel size and the number of dimensions $N_d$. Most of
the computational power is spent during the first step of the process: the
propagation of the  probabilities to compute the patches. For a constant value
of $L$ and $N_d$, the algorithm speed is linear in $N_p$, and so is the memory
consumption. A smaller
value of $L$ implies more smaller patches, which therefore have proportionally
more borders with each other thus increasing the number of different
probabilities to propagate. Indeed, for very small values
of $L$,  memory consumption is largely increased as well as the computational
time; it seems reasonable to keep $L$ above a minimal threshold of $L\ge 5$
pixels (which in any case is also necessary to ensure sufficient
differentiability of the sampled field). Finally, the value of $N_d$ is most
critical to  memory consumption and speed, not only because $N_p$ should
increase with $N_d$ to keep a constant sampling resolution, but also because
the number of neighbours for each pixels scales as  $3^{N_d}$ for a Cartesian
grid. The computational time and the memory
consumption follows, as the number of different
probabilities to keep track of is also much increased (each pixels having many
more neighbours, the ratio of patches interface surface to their volume
increases and so does the number of different probabilities to propagate, on
larger distances). For the different skeletons presented in this paper,
to give and order of magnitude, for a single modern CPU, $2D$ skeletons
of $1024^2$ pixels smoothed over $l\approx 10$ pixels are computed in a matter
of few seconds and the memory needed is of order $\approx 10$
MBytes. Computing a $3D$ skeleton on a $128^3$ pixels grid with $L\approx 6$
takes approximately $1$ minute and a hundred of MegaBytes of memory, while for
a $512^3$ grid, it takes about an hour and around $14$ GBytes of memory are
used. While $4D$ skeletons are still tractable at a descent resolution 
, higher
dimensionality seems difficult to reach with present facilities without
implementing a fully parallel version of the code.

\section{An application: validating the Zel'dovich mapping }
\label{sec:applications}
The scope of application of the algorithm presented in Sec.~\ref{sec:algo} is vast 
(see Sec.~\ref{sec:conclusion} for a discussion). Here we shall focus on a simple example which makes use 
of 
one of the clear virtues of the above implementation: it allows us 
to identify  as physical objects the filaments present in the matter
distribution on cosmological scales, and see how these objects evolve with time. 

Specifically, we intend to show, using the skeleton as a diagnostic tool, that a relatively
simple but powerful model, namely the  truncated Zel'dovich
approximation mapping \citep{zeldo70}, can capture
the  main features  of the {\sl cosmic evolution} of the web.
Indeed predicting
the evolution  of matter distribution from the point of view of the
topology  and the geometry of the cosmic web has been a recurrent issue in cosmology
(e.g. \cite{bbks}) and is becoming critical as the geometry of the cosmic environment is now believed to play a key role in 
shaping galaxies (see, e.g. \cite{ocvirk}).

Being able to carry such an extrapolation from the initial condition to the present day distribution
of filaments
  should  lead to a simplified and
broader understanding of large scale structures in the Universe, in
the same way the concept of clusters as important physical objects
gave birth to the hierarchical model of structure formation. 
The fully connected skeleton encompasses both the geometry and the topology
of the cosmic web: it is therefore the ideal tool to validate this mapping
between the initial condition  and the present day distribution of filaments.
Understanding and partially correcting for  the distorsion induced by the proper motions
of the structures is also of prime importance when dealing with observationnal data sets
(see e.g. \cite{pichon2001}).

The principle of the Zel'dovich
approximation (ZA hereafter) is to make a first order approximation, in
Lagrangian coordinates,  of the motion of the collisionless dark
matter (DM) particles. The motion of these particles from the  initial mass
distribution in Lagrangian coordinates $\vq$ to their Eulerian
coordinates ${\vx}$ can therefore be described as:
\begin{equation}
\vx\left(z,t\right) = \vq + D\left(z\right)/D\left(z_i\right){\bf
  \Psi}_i\left(\vq\right),
\end{equation}
where $z$ is the redshift, $D\left(z\right)$ the growth factor, and
${\bf \Psi}_i\left(\vq\right)$ the displacement field, computed in
the initial matter distribution as:
\begin{equation}
{\bf \Psi}_i\left(\vq\right) = \frac{-2D}{3{H_{\rm in}}^2\Omega_{\rm in}}{\bf \nabla_{\vq}}\phi
\end{equation}
where $H$ is the Hubble constant, $\Omega$ the quantity of energy in the Universe, $\phi$ the gravitational potential and the subscript ``in'' stands for initial conditions.
The truncated Zel'dovich
approximation simply consists
in filtering short scale modes  of the initial power spectrum before
computing the displacement field in order to prevent shell crossing
effects.  It has been shown to improve the precision of the
approximation \citep{coles93}. As we are mainly  interested in the
large scale behavior of the cosmic web,  the smoothing scale,
$L=L_{\rm NL}\approx 3.94$ Mpc, that we use hereafter to compute ${\bf \Psi}_i$
roughly corresponds to the scale of non linearity at $z=0$, as the  so called {\it truncated } Zel'dovich
approximation  has been shown 
to work best above this scale \citep{kofman92}. It was
computed as the scale at which, in the simulation, the smoothed
density field, $\rho(L)$, is such that
$\sigma^2(L_{\rm NL})=\left<(\rho(L_{\rm NL})-\bar\rho(L_{\rm NL}))^2\right>=1$ at $z=0$.
\\

\subsection{Simulation and skeletons}

The numerical simulation that we use in this section was computed with
the publicly available  N-body code GADGET2 \citep{gadget2}. It corresponds to
a dark matter only cosmological simulation of $512^3$ particles
within a  $250 h^{-1}$ Mpc cubic box, considering a $\Lambda$CDM
concordant model ($H_{0}=70$, $\Omega_{b}=0.05$, $\sigma_{8}=0.92$,
$\Omega_{\Lambda}=0.7$ \& $\Omega_{0}=0.3$). In order to study the
evolution of the cosmic web, a set of reference skeletons, ${\cal
  S}_{\rm simu}(z,L)$, were computed from different snapshots, at
redshift $z=\lbrace{0,0.15,0.3,0.5,0.66,1.15,3,5,10\rbrace}$, where
$z=z_i=10$ corresponds to the redshift of the  initial conditions of
the simulation. These skeletons were computed on density fields
generated by sampling  the particle distribution of the respective
snapshots on a $512^3$ grid and after smoothing with a Gaussian kernel
of size  $L=L_{\rm NL}\approx 3.94$. In order to understand if the
truncated Zel'dovich
approximation is able to capture the essential features of cosmic web,
these skeleton are compared to different sets of skeletons, generated
using the truncated Zel'dovich
approximation in different ways:
\begin{itemize}
\item{${\cal S}_{\rm ZA}(z,L_{\rm NL})$}: This set of skeletons is
  generated by applying the Zel'dovich
approximation to the DM particles of the simulation
  in the initial conditions. The displacement field is computed after
  smoothing over the scale $L_{\rm NL}$ and the resulting distribution
  is sampled and smoothed over the same scale to generate the
  skeletons.\\
\item{${\cal S}_{\rm SZA}(z,L_{\rm NL})$}: these skeletons are
  generated by applying the Zel'dovich
approximation directly to the skeleton of the initial
  conditions. The initial condition simulation ($z_i=10$) is sampled
  and smoothed over the scale $L_{\rm NL}$ to compute its
  skeleton. The displacement field is computed on the same field, but
  smoothed over a scale $L_{\rm l}\approx 8.81$ Mpc (such that
  $\sigma^2(L_{\rm l})=0.5$ at $z=0$) and the Zel'dovich
approximation is applied to each
  segment of the initial condition skeleton. We use a larger
  truncation scale for the Zel'dovich
approximation here in order to prevent shell crossing,
  which can be tolerated when applied to particles but would result in
  a very fuzzy displaced skeleton.\\
\item{${\cal S}_{\rm ZAL}(z,L_{\rm NL})$}: same as ${\cal S}_{\rm
  ZA}(z,L_{\rm NL})$, but with a displacement field smoothed over the
  scale $L_{\rm l}$, in order to check the influence of this choice on
  ${\cal S}_{\rm SZA}(z,L_{\rm NL})$.\\
\item{${\cal S}_{\rm ZA}(z,L_{\rm cor})$}: same as  ${\cal S}_{\rm
  ZA}(z,L_{\rm NL})$, but the sampled field is smoothed on a scale
  $L_{\rm cor}$ instead of $L_{\rm NL}$ to take into account that the Zel'dovich
approximation
  introduces an artificial additional smoothing scale (see below).
\end{itemize}

\subsection{Skeleton length}

\begin{figure}
\centering  
\includegraphics[width=9cm]{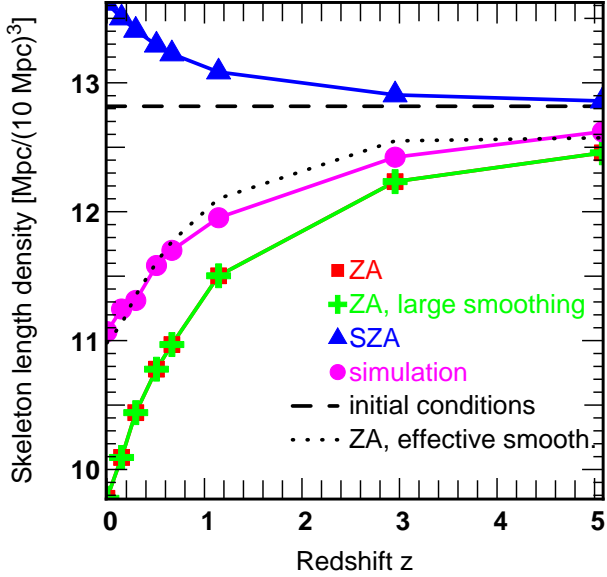}
\caption{ 
 Measured length of the skeleton per unit volume as a
  function of redshift $z$. The length density was measured  on the
  simulation ({\sl purple discs}), its truncated Zel'dovich
approximation whose displacement field
  was computed using a smoothing length $L\approx 3.94$ such that
  $\sigma(L,z=0)=1$ ({\sl red squares}), or $L_l\approx 8.81$ such that
  $\sigma(L,z=0)=0.5$ ({\sl green crosses}), and finally using the
  displacement field of the Zel'dovich
approximation at scale $L_l$, applied directly to the
  skeleton of the initial condition, at $z=10$ ({\sl blue triangles}). The
  black dashed line stands for the length of the skeleton in the
  initial conditions (at $z=10$), while the dotted line represents the
  length measured using the Zel'dovich
approximation on the initial condition while taking
  into account the effective smoothing introduced by using the
  Zel'dovich
approximation. This recipe yields the best match with the simulation.
 Except for this last case, the skeletons where computed after
  smoothing the density field with a Gaussian kernel of width $L$.
 \label{fig_skl_length}}
\end{figure}

There exist many different ways to compare one dimensional sets of
lines within a $3D$ space,  but one of the simplest certainly involves
comparing their lengths. Figure~\ref{fig_skl_length} presents the
measured  length per unit volume of the different sets of skeletons
(described above) as a function of redshift. Let us first consider the
length of ${\cal S}_{\rm simu}(z,L_{\rm NL})$ ({\sl purple curve with
  discs symbols}). It was shown in  \cite{3dskel} and \cite{sklsdss}
that, whereas for scale invariant fields such as the initial
conditions of the simulation, the length of the skeleton is expected
to grow as $L^{-2}$ ($L$ being the smoothing length), it grows in fact
as $\approx L^{-1.75}$ around $z=0$ for $\Lambda$CDM simulation.  Note
that the fact that the length of ${\cal S}_{\rm simu}(z,L_{\rm NL})$
decreases with time seems consistent with the expected evolution of
matter distribution  in the case a cosmological constant, where the
expansion accelerates around $z\approx 1$. In that case, matter in
fact tends to form separate distant heavy halos: more numerous small
filaments on smaller scales shrink and melt into each other as dark
matter halos merge, while larger scale filaments tend to stretch: the
net result is a total length decrease. 

This process seems to be well captured by the Zel'dovich
approximation as the
length of ${\cal S}_{\rm ZA}(z,L_{\rm NL})$ ({\sl red curve, square
markers}) exhibits  the same time evolution as the length of
${\cal S}_{\rm simu}(z,L)$. The discrepancy between the measured
length in the simulation and with Zel'dovich's
approximation is  nonetheless of  the  order of
$\approx 10\%$ at $z=0$. This disagreement should be explained in part
by the fact that the Zel'dovich
approximation uses a displacement field computed from a smoothed
version of the initial condition density field, thus introducing an
additional smoothing that one should take into account when computing
${\cal S}_{\rm ZA}$.\\

\begin{figure}
\centering \includegraphics[width=8cm]{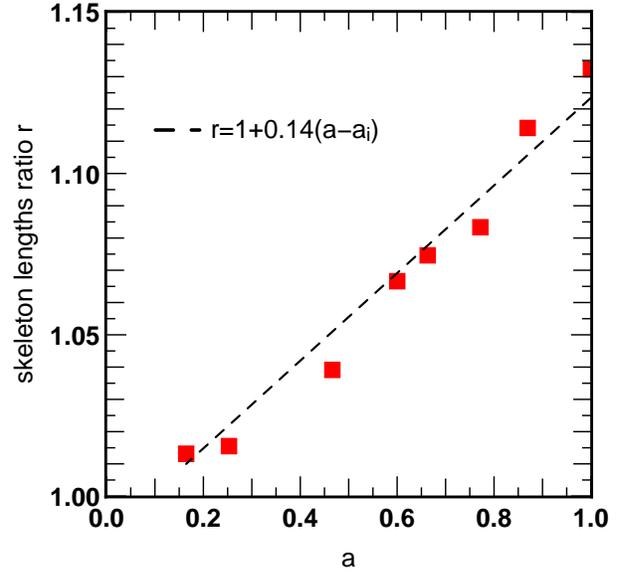}
\caption{ Ratio of the length of the skeleton measured in the
  simulation to the length of the skeleton  of the Zel'dovich
approximation as a function of
  time, a. The dashed line represents the best fit of the data (red
  squares).\label{fig_fit_r_a}}
\end{figure}

The measure of the ratio, $r$, of the length of ${\cal S}_{\rm
  simu}(z,L_{\rm NL})$ to the length of ${\cal S}_{\rm ZA}(z,L_{\rm
  NL})$ as a function of time, $a$, is displayed on Figure~\ref{fig_fit_r_a}. It appears that $r$  is approximately a linear
function of time, $a$, and can thus be fitted as 
\begin{equation}
\label{eq_rfit}
r=1.00+0.14\left(a-a_i\right),
\end{equation} 
where $a_i=1/\left(1+z_i\right)\approx 0.09$ is the time of the
initial conditions from which the Zel'dovich
approximation was computed. Moreover, the fact
that the value of $r$ is relatively close to unity confirms that the
artificial smoothing introduced by the Zel'dovich
approximation is  small;  we chose to model
it as a convolution with a Gaussian kernel of size $L_{\rm ZA}$. The
effective Gaussian smoothing used on Zel'dovich's
approximation  has scale $L_{\rm eff}$ and
is thus the result of the composition of two Gaussian smoothing of
scale $L_{\rm ZA}$ and $L_{\rm NL}$:
\begin{equation}
\label{eq_leff}
L_{\rm eff} = \sqrt{{L_{\rm ZA}}^2 + {L_{\rm NL}}^2}. 
\end{equation}
Using equations (\ref{eq_rfit}) and (\ref{eq_leff}), and the fact that
the skeleton length grows with smoothing scale as $\approx L^{-1.75}$ 
\citep{sklsdss}, the value of $L_{\rm ZA}$ one should chose to get the best match
with $\Lambda$CDM simulations is thus
\begin{equation}
L_{\rm ZA}\approx L_{\rm NL}\left(\frac{2\cdot 0.14}{1.75}\left(a-a_i\right)\right)^{1/2}=0.4L_{\rm NL}\sqrt{a-a_i}.
\end{equation}
In order to compute a skeleton that is comparable to ${\cal S}_{\rm simu}(z,L_{\rm NL})$, 
one should therefore smooth the distribution
obtained using the Zel'dovich
approximation on a scale $L_{\rm cor}$ such that
\begin{equation}
L_{\rm cor}=\sqrt{{L_{\rm NL}}^2-{L_{\rm ZA}}^2}=L_{\rm NL}\sqrt{1.00-0.16\left(a-a_i\right)}. \label{eq:defLcor}
\end{equation}
On Figure~\ref{fig_skl_length}, the dotted black curve represents the
measure of the length of ${\cal S}_{\rm ZA}(z,L_{\rm cor})$, when the
effective smoothing introduced by the Zel'dovich
approximation is taken into account. The
agreement with the  measurements in the simulation is significantly improved
compared to the naive approach; this suggest that the Zel'dovich
approximation can be used to
predict the shape of the evolved cosmic web from the initial
conditions distribution only. 

Of course, the length is only a global
characteristic of the skeleton and it certainly does not fully
constraint its shape. Higher order estimators that can compare  the
relative position and shapes of skeletons are needed to
quantify how good an approximation the skeleton obtained by Zel'dovich's
approximation is.\\

Before doing so, let us consider an alternative form of the Zel'dovich
approximation, where, instead of displacing  the particles from the
initial conditions of the simulation to derive the evolved density
field, we directly  use the displacement field to evolve the skeleton
of the initial conditions. This method will be called here the
skeleton Zel'dovich approximation (SZA hereafter), and the resulting
skeleton ${\cal S}_{\rm SZA}$. Studying the properties of ${\cal
  S}_{\rm SZA}$ is interesting as it should make it possible to
distinguish between two different processes that affect the properties
of the cosmic web: the simple deformation of the initial cosmic web on
the one hand and the creation or annihilation of filaments on the
other hand. Indeed, ${\cal S}_{\rm SZA}$ reflects only the
modification of the skeleton due to its deformation while ${\cal
  S}_{\rm ZA}$ also takes into account the merging and annihilation of
filaments. Note nonetheless that by definition, the locus of the
skeleton for the SZA is biased toward higher density regions; in these
regions, non-linear effects inducing shell-crossings in the Zel'dovich
approximation are more likely. To be conservative, we thus use a
larger smoothing length than $L_{\rm NL}$ to compute the displacement
field. This smoothing length, $L_{\rm l}\approx 8.81h^{-1}$ Mpc, was
chosen such that $\sigma(L_{\rm l},z=0)=0.5$; the green curve ({\sl
  cross markers}) of Figure~\ref{fig_skl_length} shows that using
$L_{\rm l}$ or $L_{\rm NL}$ does not make any difference regarding the
length of the skeleton. On this figure, the blue curve ({\sl triangle
  markers}) depicts the evolution of the length of  ${\cal S}_{\rm
  SZA}(z,L_{\rm NL})$: its behavior is clearly opposite to the ${\cal
  S}_{\rm ZA}$ case, as  the length rises with time. Although
surprising at first sight, this result only confirms our previous
interpretation of the evolution of the cosmic web. In fact, if  the
SZA can nicely capture the large scale evolution of long filaments,
the smaller ones cannot melt into each other, which induces several
small scale filaments to be located at the same loci, where only one
piece of filaments should have been measured. The disappearance of the
smaller scale filaments does not compensate anymore for the expansion
of  large scale filament: the net result is thus an increase  of the
total measured length of ${\cal S}_{\rm SZA}$ with time.

\begin{figure*}
\centering   \subfigure[ZA and
  simulation]{\includegraphics[width=7.5cm]{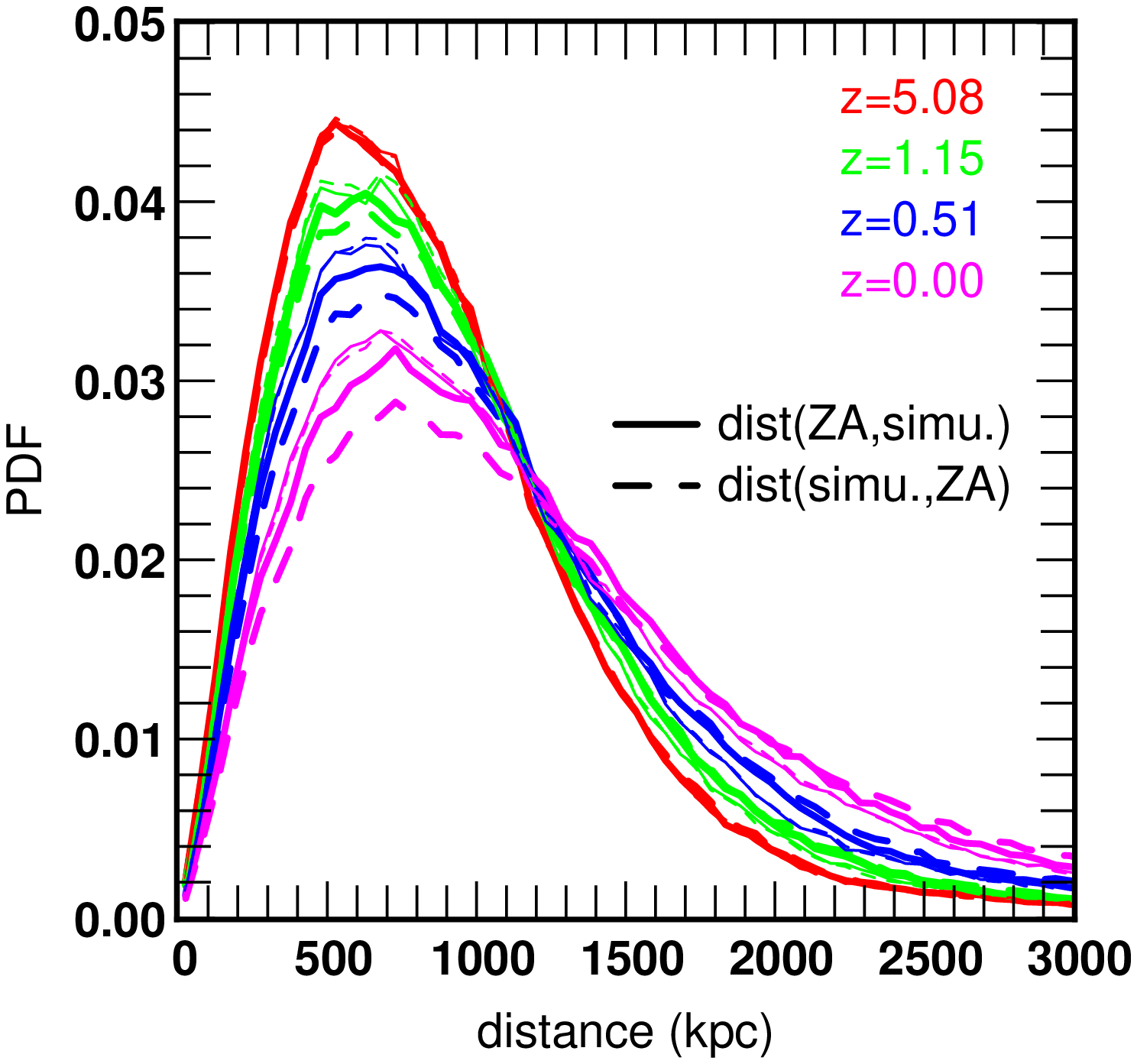}\label{fig_dist_z_t}} \subfigure[ZA
  applied to the skeleton (SZA) and
  simulation]{\includegraphics[width=7.5cm]{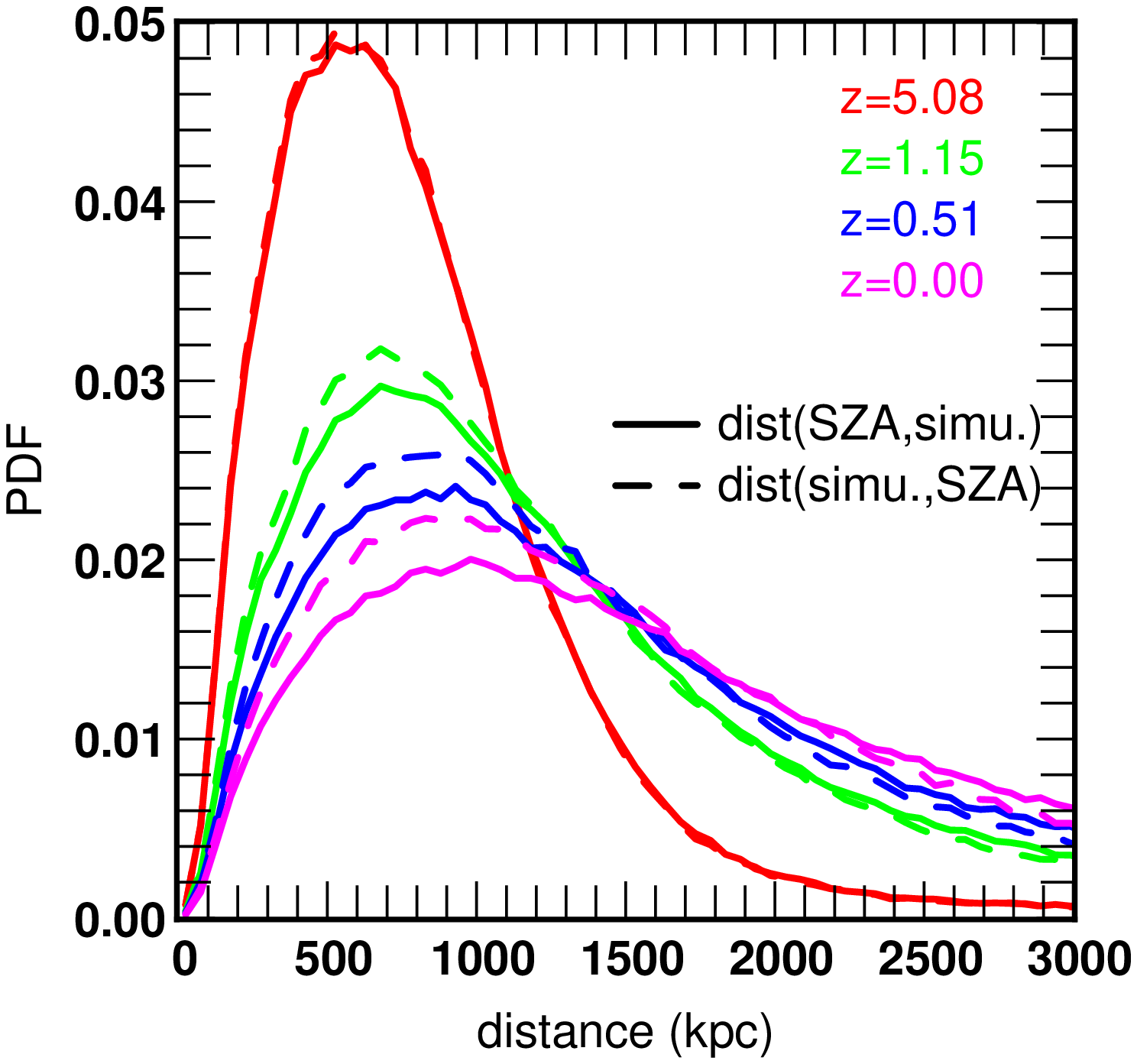}\label{fig_dist_zd_t}}
\caption{ The inter-skeleton distance as defined in the main text and
  appendix B, applied to the skeletons of the simulation and the
  Zel'dovich approximation (Figure~\ref{fig_dist_z_t}) and the
  skeletons of the simulation and displaced initial conditions
  skeleton, SZA (Figure~\ref{fig_dist_zd_t}). The displacement fields
  and skeletons are computed  after smoothing the field on a scale
  $L\approx 3.94$ such that $\sigma(L,z=0)=1$, except for SZA, where
  the displacement field was obtained after smoothing over $L_l\approx
  8.81$ Mpc, such that $\sigma(L,z=0)=0.5$. The full lines represent
  the distance from the simulation's skeleton to the other, while the
  dotted lines represent the reciprocal distance.  The thin lines on
  Figure~\ref{fig_dist_z_t} stand for the case where the effective
  smoothing introduced by the Zel'dovich approximation is taken into
  account. Note that SZA PDF is more skewed as the merging/annihilation of filaments is then
not taken into account.
\label{fig_skl_dist}}
\end{figure*}
\subsection{Inter-skeleton pseudo-distance}

Let us now 
define a way to compute a pseudo-distance between two different
skeletons (see also \cite{caucci}). 
In practice, a skeleton
${\cal S}$ is always computed from a sampled density and thus has a
maximal resolution $R_s$.  It can therefore be described, without loss
of information, as the union of a set of $N$ straight segments ${\cal
  S}^i$ of size $R_s$.  We define a pseudo-distance from a skeleton
${\cal S}_a$ to a skeleton ${\cal S}_b$, $\distance{a}{b}$,
as the probability distribution function (PDF) of the minimal distance
from the $N^a$ segments ${\cal S}_a^i$ to any of the $N^b$ segments
${\cal S}_b^j$. In practice, our algorithm  applied to a
density field sampled on a Cartesian grid naturally leads to a
skeleton described as a set of segments of size the order of the
sampling resolution. Hence we directly use these segments to compute
inter-skeleton distances.

Note that there is no reason, in
general, for $\distance{a}{b}$ to be identical to
$\distance{b}{a}$; this discrepancy, together with the value of the
different modes of the PDFs, do in fact quantify the differences
between ${\cal S}_a$ and ${\cal S}_b$ (see appendix B for details on
how to interpret pseudo-distances PDFs). The upper and lower panels of
Figure~\ref{fig_skl_dist} present the pseudo distance measurements
obtained by comparing ${\cal S}_{\rm simu}$ to ${\cal S}_{\rm ZA}$ and
${\cal S}_{\rm SZA}$ respectively. A close examination of Figure~\ref{fig_dist_z_t} confirms  the hypothesis we made in previous
subsection. First, the high correlation of ${\cal S}_{\rm ZA}$ and
${\cal S}_{\rm simu}$ ({\sl bold curves}) for any redshift, is demonstrated
by the localization of the mode around $d\approx 600h^{-1}$ kpc, well
below the smoothing length  $L_{\rm NL}=3.94$ Mpc. Second, the
asymmetry between the PDFs of $\distance{ZA}{simu}$ and
$\distance{simu}{ZA}$ follows from the fact that ${\cal S}_{\rm
  simu}$ has smaller scale filaments that have no counterpart in ${\cal
  S}_{\rm ZA}$ (the mode intensity is higher for $\distance{ZA}{simu}$
than $\distance{simu}{ZA}$). This is exactly what should happen if
${\cal S}_{\rm ZA}$ was effectively smoothed on a scale larger than
${\cal S}_{\rm simu}$. The thin curves, for which the effective Zel'dovich
approximation
smoothing was taken into account, confirms this,  as the asymmetry is
completely removed in that case.\\

\begin{figure*}
\centering  
\includegraphics[width=16cm]{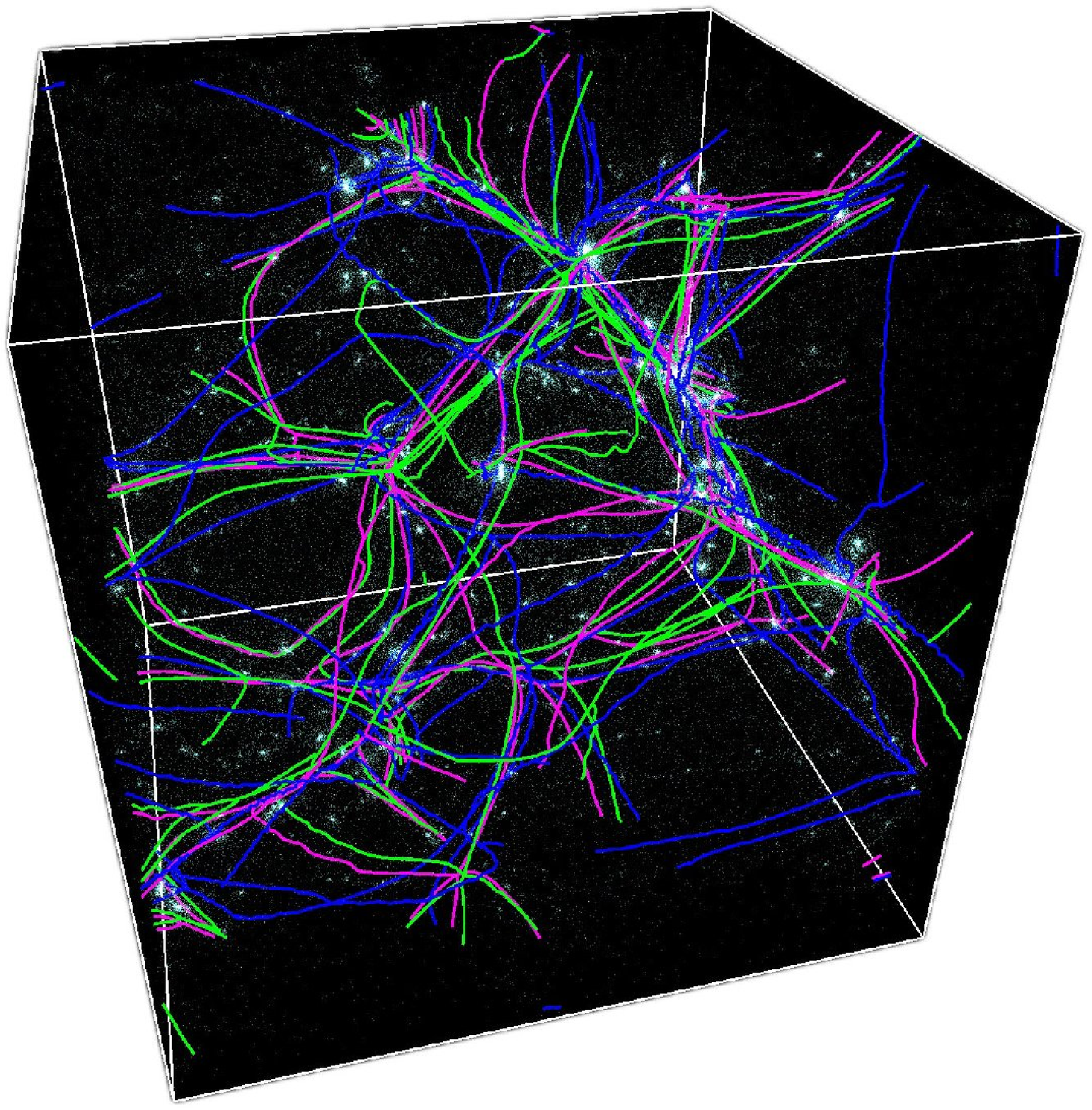}
\caption{ A $(50 h^{-1})^3$ ${\rm Mpc}^3$ section of the $512^3$
  particles simulation of a $250 h^{-1}$ Mpc large box (only $1$
  particle every $8$ is displayed). The purple skeleton is ${\cal
    S}_{\rm simu}(z,L)$ (computed from the simulation), the green one
  ${\cal S}_{\rm ZA}(z,L_{\rm cor})$ (computed on the Zel'dovich
  approximation using an effective smoothing length) and the blue one
  ${\cal S}_{\rm SZA}(z,L_{\rm NL})$ (computed by displacing the
  skeleton of the initial conditions). The simulation and corrected
  Zel'dovich approximation skeletons appear to be relatively close to
  each other and every individual filament has a counterpart in the
  other skeleton. The blue skeleton, computed from the skeleton
  Zel'dovich approximation, is more wiggly which reflects the small
  scale perturbation of the displacement field. Moreover, while many
  of its filaments have counterpart in the two other skeletons, others
  do not, as displacing the initial skeleton prevents the merging or
  disappearance of filaments. This results can be
  quantitatively measured, as shown on Figure~\ref{fig_skl_dist} and explained in appendix B.\label{fig_cmp_zeldo3d}}
\end{figure*}

 It is also interesting to  look at the distance PDFs between
 ${\cal S}_{\rm SZA}$ and ${\cal S}_{\rm simu}$ (see Figure~\ref{fig_dist_zd_t}). Except for high redshifts ($z=5$),  the general
 intensity of the modes are lower for $\distance{SZA}{simu}$ than for
 $\distance{ZA}{simu}$, suggesting that the Zel'dovich
approximation is a better description
 of the evolution of the filaments on large scales, and that filaments
 mergers and creation are important processes. The general position of
 the modes is still comparable, which means that SZA is nonetheless
 successful in describing the evolution of the general shape of the
 cosmic web. Also, the asymmetry between $\distance{SZA}{simu}$ and
 $\distance{simu}{SZA}$ suggests that ${\cal S}_{\rm SZA}$  has
 more small scale filaments than ${\cal S}_{\rm simu}$. These
 observations confirm our  previous assumption that although
 the cosmic web evolves in a simple inertial way on larger scales (a
 process captured by SZA),  the shrinking and fusion of the more
 numerous smaller scale filaments is the cause of the general
 length decrease of the cosmic web (as suggested by a simple visual
 examination of a $(50 h^{-1})^3$ ${\rm Mpc}^3$ subregion of ${\cal S}_{\rm
   simu}$, ${\cal S}_{\rm SZA}$ and ${\cal S}_{\rm ZA}$ on Figure~\ref{fig_cmp_zeldo3d}).\\

The above investigation opens the prospect of correcting for the
peculiar velocities of galaxies  induced by gravitational clustering,
and carry an Alcock-Paczynski (AP) test \citep{aptest} on the skeleton
of the large scale structures of the universe.  In short, the AP test
compares  observed transverse and longitudinal distances  to constrain
the global geometry of the universe.  Galaxy positions are usually
observed in redshift space which induces an important distortion
between the distances measured along and orthogonally to the line of
sight, which plagues the regular AP-test.  Our analysis suggests  that
it is in fact possible to correct  through the Zel'dovich
approximation for the distortions induced on the cosmic web.  Having
carried such a correction, we expect that the measure of the
anisotropy of the observed skeleton length ratio could yield a good
constraint on the value of the cosmological parameters.
Note finally that The Zel'dovich mapping smoothed with $L_{\rm cor}$ (see Equation~(\ref{eq:defLcor})) 
 could be used to generate synthetically sets of extremely large cosmic skeletons probing
 exotic cosmologies using codes such as 
  {\tt mpgrafic} \citep{prunet}  to generate the initial conditions and their Zel'dovich displacement.
  This construction could then be populated with halos and substructure using semi analytical models. Note finally that the total length and the skeleton's distance are two probes  amongst many on how to characterize
the difference between two skeletons. 
Moreover, there are other means to quantify the evolution of the cosmic web.
For instance, an interesting statistics would be to find out
how often the reconnection of the skeleton occurs as a function of redshift. 

\section{Discussion and prospects}
\label{sec:conclusion}
We have presented a method, based on an improved watershed
technique, to efficicently compute the full hierarchy of critical subsets from a
density field within spaces of arbitrary dimensions. Our algorithm
uses a fast one pass probability propagation scheme that is able to improve significantly
the quality of the segmentation by circumventing  the discreteness  of
the sampling. We showed that, following Morse theory, a recursive
segmentation of space yields, for  a $d$-dimensional
space, a succession of  $d-1$ $n$-dimensional subspaces that
characterize the topology of the density field. In 3D for
cosmological matter density distribution, we particularly focused on
the 3D subspaces which are the peak and void patches of the field
({\rm i.e. the attraction/repulsion pools}) and the 1D critical lines
which trace the filaments as well as the whole primary cosmic web structure
({\it i.e.} a  {\sl fully-connected}, non-local skeleton as
defined in \cite{2dskel}). For  the primary critical lines, we
also demonstrated that it is possible to use the probabilities
distribution from our algorithm to derive a smooth and differentiable
skeleton with a sub-pixel level resolution. Thus this method  allows
us to consider the cosmic web as a precise physical
object  and makes it possible to compute any of its properties such as
length, curvature, halos connectivity etc...\\

As an application, we used our algorithm to study the evolution of the
cosmic web, while comparing the time evolution of the skeleton (a proxy to the cosmic web) of a
simulation, to those corresponding to different versions of its Zel'dovich
approximation.  We first compared the 
evolution of the respective lengths of the different skeletons and then introduced a method to
compute pseudo-distances between different skeletons. This pseudo distance
makes it possible to compare different features of the skeleton such as the
{size} of their filaments and the similarity of their locations. Using
these measurements, we showed that two effects were competing, with
net result a decrease of the cosmic length with time: a general
dilation of the larger scales filaments that could be captured by a
simple deformation of the skeleton of the initial conditions on the one hand,  and the
shrinking, fusion and disappearance of the more numerous smaller
scales filaments on the other hand. 
We also showed that a simple Zel'dovich
approximation could accurately capture most features of the evolution of the cosmic
web  for all scales larger than a few megaparsecs  (provided an effective
smoothing introduced by the approximation  is taken into account).
Hence in this context, 
the skeleton has proven to be a useful tool both for insight and as a quantitative
probe and diagnostic. Conversely, the match between the simulated and the mapped skeleton 
confirms and extends geometrically the (point process elliptical)  peak patch theory \citep{bbks} since both the peaks and their frontiers 
(the skeleton in 2D and the peak patch volumes in 3D) are well mapped by the  Zel'dovich
approximation.
 \\

The domain of interest of the skeleton is quite vast and offer the prospect 
of  a number of applications. 

From a theoretical
point of view, using the points presented in this paper and in
\cite{3dskel}, we are presently developing a general theory of the
skeleton and its statistical properties \citep{pogo}
that aims to understand the properties of the critical lines of scale invariant
Gaussian random fields
as mathematical objects.  In particular, this companion paper
 provides quantitative analytic predictions for the 
length per unit volume (resp. curvature) of the critical lines and its scaling with the shape parameter of the field, 
 and checks successfully the current algorithm against these.
In this paper we focussed on the skeleton.  
One could clearly investigate on the rest of the peak patch hierarchy and measure, say, the 
surface or volume of the (hyper)-surfaces of the recursion (whose last intersection is 
given by the primary critical lines).  
Another interesting issue would be to 
estimate the fraction of special (degenerate) points which do not satisfy the Morse condition,
where the fields behaves pathologically  \citep{pogo}.

\begin{figure*}
\centering   \includegraphics[width=5.5cm]{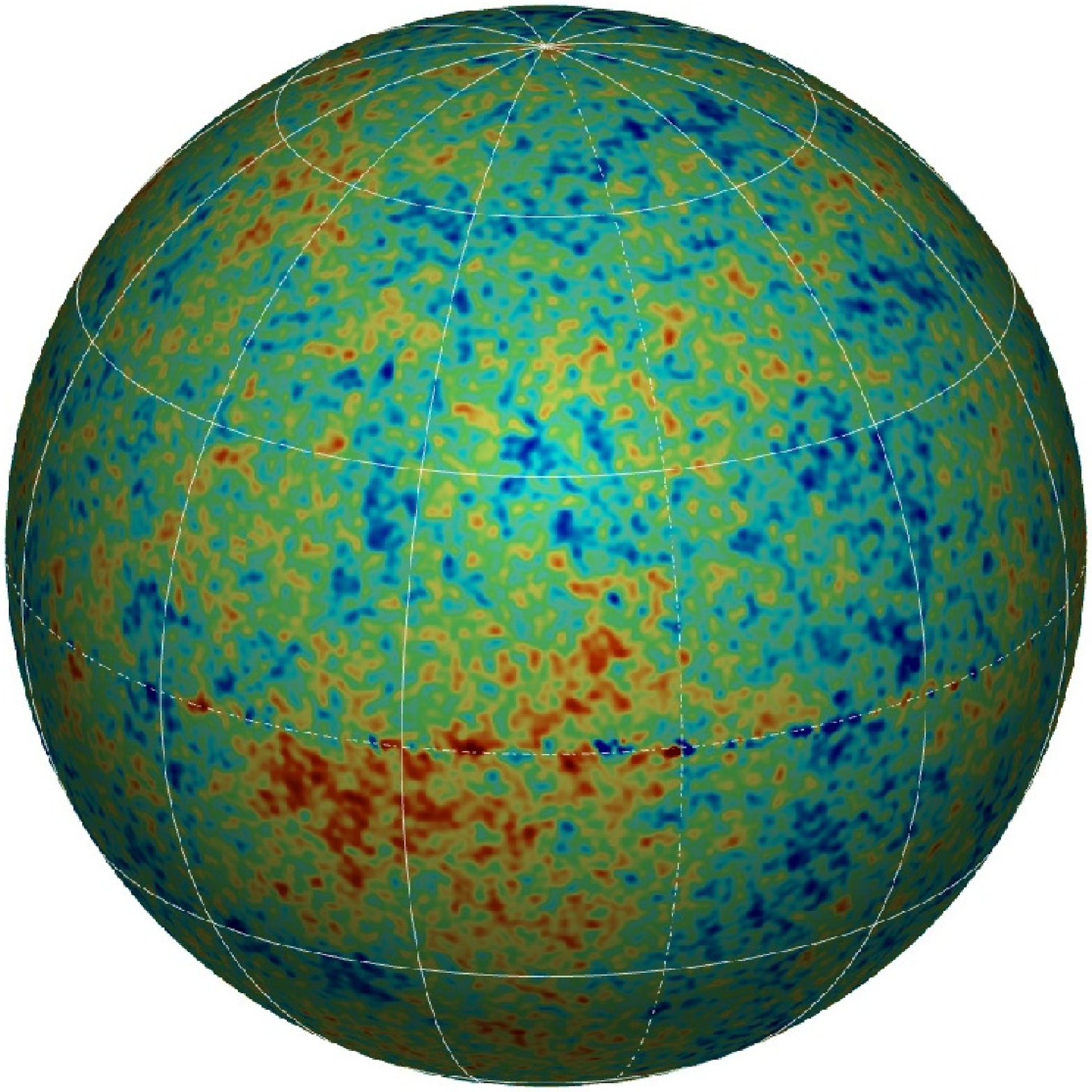}  \includegraphics[width=5.5cm]{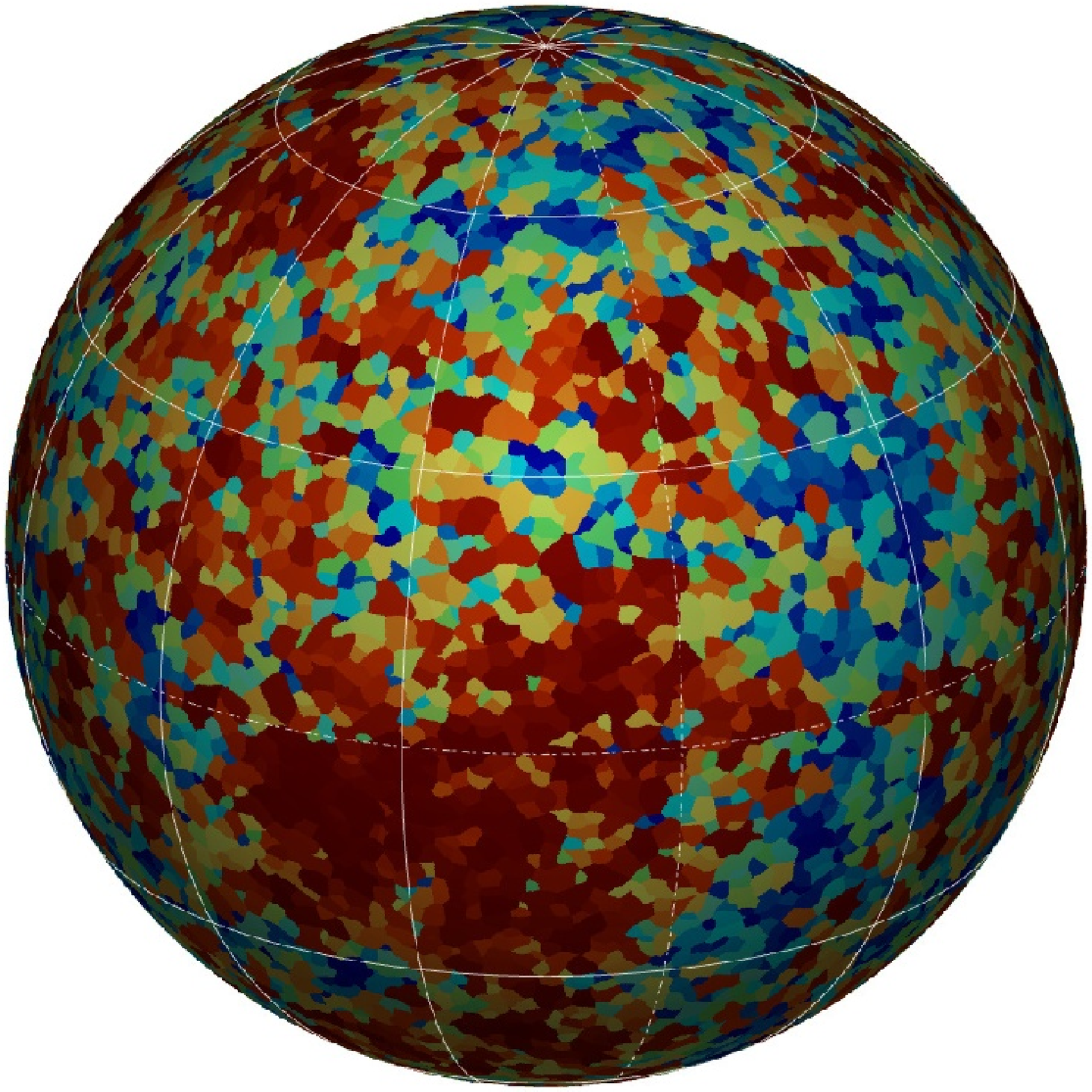}  \includegraphics[width=5.5cm]{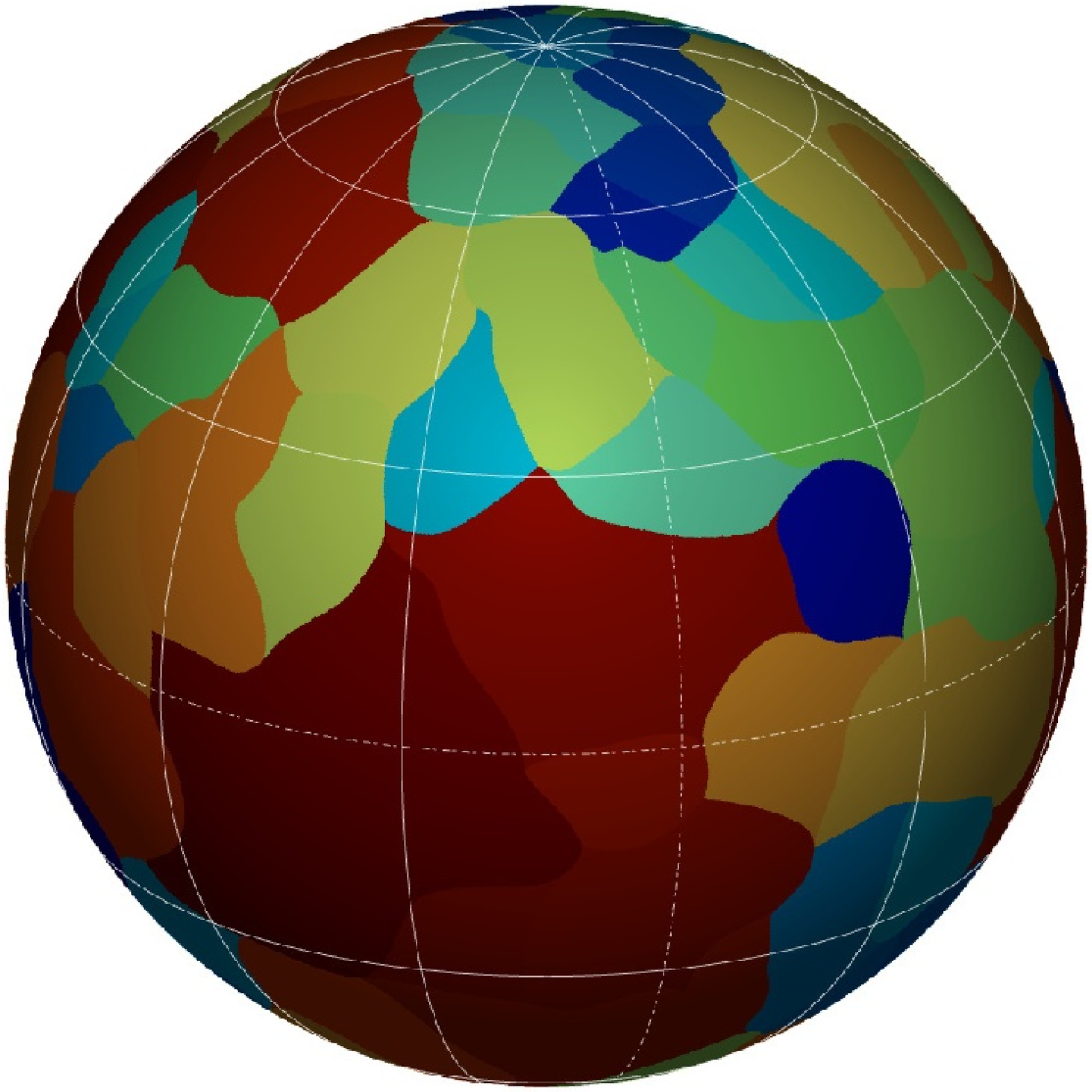}
\caption{{\sl from left to right}: the 5 year WMAP release of the CMB temperature map, the corresponding peak patches  and the peak patches of the same field smoothed over a FHWM of 420 arcmin.
Different colours represent different patches. The algorithm described in section~\ref{sec:algo}  is implemented here on 
the {\tt healpix} pixelisation.
 \label{healpix}}
\end{figure*}

For instance, one of the shortcoming of the present algorithm concerns special fields
where critical lines disappear, a situation which occurs, say, in the context of tracing dendrites 
in a neural network, or blood vessels within a liver. Note also that 
there exist other sets of (geometrical) critical lines that are not topological invariants such as the 
lines of steepest ascent connecting directly minima to maxima which are not accounted for by the present formalism.
In contrast, the algorithm is well suited to identify bifurcation points, and the connectivity
of the network. In particular, in an astrophysical context, it would be worthwhile to make use of this feature
and
study statistically how the skeleton connects onto dark matter halos as a function of, say, 
they mass or spin, and investigate the details of local spin accretion in the context of the 
cosmic web superhighways, hence completing the spin alignment 
measurements of  \cite{3dskel}  on smaller scales.
 More generally, the algorithm provides 
a neat bridge, via the provided connectivity,
 between the theory of continuous fields on the one hand, and 
graph theory for discrete networks on the other.
This could prove to be of importance in the context of percolation theory. For instance,
the
percolation threshold can be explained in terms of the properties of the connectivity of the relevant nodes (see {\it e.g.} \cite{colombi00}).\\
Here, as argued in section~2.4 we deliberately chose not to consider the issue of
shot noise and its consequences on segmentation, for which no
definitive  solution yet exists, though many improvements have been proposed in the literature (see {\it e.g.} \cite{jos01}). Instead, we  followed  the approach of \cite{3dskel},
 that simply  involves convolving the sampled
density field with a large enough (in terms of sampling scale)
Gaussian kernel so that the field can be considered smooth and
differentiable; the probabilistic algorithm allows for the removal of
sampling effects and small intensity residual shot noise. In appendix~C we show that the corresponding fully connected
skeleton is nonetheless quite robust (the core of the network remains quasi unchanged), so long as the SNR is above one. 
A possible
drawback of this method is that it introduces a smoothing scale
attached to the skeleton. This is not necessarily a
problem in cosmology as the scaling of the skeleton properties with
scale yields information on the distribution over these scales.
Moreover, one is usually interested  by the properties of the skeleton
on a given scale (typically larger than the halo scale, a few
megaparsecs).  Nonetheless, there exist more complex multi-scale
sampling and smoothing techniques such as the one presented in \cite{platen07} or \cite{colberg07}
 that could  straightforwardly  be adapted to our
implementation. All the algorithm requires is  a structured
sampling grid  where one can recover a one to one pixel neighbourhood
({\it i.e.} one needs to be able to find the neighbouring pixels of any
pixel and these pixels must have the former as neighbour as
well).  For instance, we already implemented the algorithm for an {\tt Healpix} \citep{hivon} pixelisation
of the sphere (see Figure~\ref{healpix}),  while a direct implementation on a
delaunay tesselation network is clearly an option\footnote{for instance to segment regions on the surface of skull}.\\
A natural extension of the theoretical component of this work would be to investigate numerically the properties of the bifurcation points in  abstract space or anisotropic settings (see \cite{pogo} for a theoretical discussion for isotropic Gaussian random fields). For instance, in  the context of cosmic structure formation,
\cite{hanami01}, 
relied  on the parallel between the skeleton of the density field in position-time 4D space 
and in position-scale 4D space to relate the two. In the former, the skeleton is a natural way of computing what is known as a halos 
merger tree, commonly used in semi-analytical galaxy formation models (see \cite{hatton} for instance): the 
skeleton traces the evolution of the critical points of the density field in time. The peak theory \citep{bbks}
tells us that the smoothing scale can be linked to time evolution on scales where gravitational effects remain weakly 
non-linear. A worthwhile goal 
is to establish the parallel between the properties of 4D skeleton in this position-smoothing scale space 
(which can be computed from the Gaussian initial conditions only) and the halo merger tree.
Finally, note that classical bifurcation theory is concerned  with the
evolution of a critical point as a function of a control parameter. 
In the language of the skeleton, this evolution may correspond to the skeleton in the extended ``phase space''.

From the physical and observational point of view,
 an other interesting venue would be to apply the skeleton to actual galaxy catalogs such as the SDSS
\citep{sdss}  to characterize the (universal) statistics of 
filaments as physical objects, like halos or voids,
and describe them in terms of their thickness, length, curvature and
environmental properties (galaxies types, halo proximity, color and morphology gradient...), both
in virtual and observed catalogs. It could also be used as a diagnosis tool for inverse methods 
which aim at reconstructing the three dimensional distribution of the IGM from say QSO bundles \citep{caucci}
or upcoming radio surveys (LOFAR, SKA etc...)
Clearly the peak patch segmentation
 developed in this paper will also be useful in the context of the upcoming surveys such as 
the LSST, or the the  SDSS-3 BAO surveys, for instance to identify rare events
such as large walls or voids and study their shape. 
Its application to CMB related full sky data, such as WMAP or Planck 
should provide insight into, e.g. the level of non Gaussianity in these maps. 
Similarly, upcoming large scale weak lensing surveys (Dune, SNAP...)
 could be analysed in terms 
of these tools (see \cite{pichon} for the validation of a reconstruction method in this context). 
Using the skeleton, the geometry of 
cold gas accretion that fuel stellar formation in the core of galaxies could be probed. 
The properties of the distribution of metals on smaller scales could be  also investigated using peak patches, to see 
 how they influence galactic properties; one could  compare these statistical results to those obtained 
through WHIM detection by Oxygen emission lines (\cite{aracil} \cite{caucci}). Indeed it has 
been claimed (see  e.g. \cite{ocvirk} \cite{dekel}) 
that the geometry of the cosmic inflow on a galaxy (its mass, 
temperature and entropy distribution,  the connectivity of the local filaments network etc. ) is strongly correlated to its history and nature.\\

In closing, let us emphasize  again that the scope of application of the algorithm presented in this paper 
extends well beyond the context of the large scale structure of the universe: it could be used in 
any scientific of engineering context (medical tomography, geophysics, drilling ...) where the geometrical structure of a given field needs to be characterized.


\subsection*{Acknowledgements}
{ \sl We thank  D. Pogosyan, D.~Aubert, J.~Devriendt, J.~Blaizot,
  S.~Peirani and S.~Prunet for fruitful comments during the course of this work, and
  D.~Munro for freely distributing his Yorick programming language and
  opengl interface (available at {\em\tt
    http://yorick.sourceforge.net/}).  This work was carried within
  the framework of the Horizon project,
  \texttt{www.projet-horizon.fr}.   }


\bibliographystyle{mn2e}


\begin{thebibliography}{}
\bibitem[Adelman-McCarthy et al.(2008)]{sdss} 
Adelman-McCarthy, J.~K., et al.\ 2008, \apjs, 175, 297 
\bibitem[Alcock \& Paczynski(1979)]{aptest} Alcock, C., \& Paczynski, B.\ 1979, \nat, 281, 358 
\bibitem[\protect\citeauthoryear{{Aracil}, {Petitjean}, {Pichon} \&
  {Bergeron}}{{Aracil} et~al.}{2004}]{aracil}
{Aracil} B.,  {Petitjean} P.,  {Pichon} C.,    {Bergeron} J.,  2004, \aap, 419,
  811
\bibitem[Arag{\'o}n-Calvo et 
al.(2007)]{aragon-calvo07a}  Arag{\'o}n-Calvo, M.~A., Jones, B.~J.~T., van de Weygaert, R., \& van der Hulst, J.~M.\ 2007, \aap, 474, 315 
\bibitem[Arag{\'o}n-Calvo et al.(2007)]{aragon-calvo07} 
Arag{\'o}n-Calvo, M.~A., van de Weygaert, R., Jones, B.~J.~T., 
\& van der Hulst, J.~M.\ 2007, \apjl, 655, L5 
\bibitem[Aubert \& Pichon(2006)]{aubert06} Aubert, D., \& Pichon, C.\ 2006, EAS Publications Series, 20, 37 
\bibitem[Aubert et al.(2004)]{ADHOP} Aubert, D., Pichon, C., 
\& Colombi, S.\ 2004, \mnras, 352, 376 
\bibitem[Barrow et al.(1985)]{barrow85} Barrow, J.~D., Bhavsar, 
S.~P., \& Sonoda, D.~H.\ 1985, \mnras, 216, 17 
\bibitem[Bertschinger(1985)]{bertschinger85} Bertschinger, E.\ 1985, 
\apjs, 58, 1 
\bibitem[Beucher \& Lantuejoul(1979)]{beucher79}
  {Beucher} S., {Lantuejoul} C., 1979, in Proceedings International
  Workshop on Image Processing, CCETT/IRISA, Rennes, France
\bibitem[Beucher \&  Meyer(1993)]{beucher93}
  {Beucher} S., {Meyer} F., 1993, Mathematical Morphology in Image Processing
  , ed. M. Dekker, New York, Ch. 12, 433
\bibitem[Bharadwaj et al.(2000)]{bharadwaj00} Bharadwaj, S., Sahni, 
V., Sathyaprakash, B.~S., Shandarin, S.~F., 
\& Yess, C.\ 2000, \apj, 528, 21 
\bibitem[Bond \& Myers(1996)]{bbks} Bond, J.~R., \& Myers, S.~T.\ 1996, \apjs, 103, 1 
\bibitem[Bond et al.(1991)]{bond91} Bond, J.~R., Cole, S., 
Efstathiou, G., \& Kaiser, N.\ 1991, \apj, 379, 440 
\bibitem[Colberg(2007)]{colberg07} Colberg, J.~M.\ 2007, \mnras, 
375, 337
\bibitem[Coles et al.(1993)]{coles93} Coles, P., Melott, A.~L., 
\& Shandarin, S.~F.\ 1993, \mnras, 260, 765 
\bibitem[Colless et al.(2003)]{2df} Colless, M., et al.\ 
2003, ArXiv Astrophysics e-prints, arXiv:astro-ph/0306581 
\bibitem[Colombi et al.(2000)]{colombi00} Colombi, S., Pogosyan, 
D., \& Souradeep, T.\ 2000, Physical Review Letters, 85, 5515 
\bibitem[Caucci et~al.(2008)]{caucci}
{Caucci} S.,  {Colombi} S.,  {Pichon} C.,  {Rollinde} E.,  {Petitjean} P.,
  {Sousbie} T.,  2008, \mnras \, in press, pp 000--000
\bibitem[Dekel  et~al. (2008)]{dekel}
{Dekel} A.,  {Birnboim} Y.,  {Engel} G.,  {Freundlich} J.,  {Goerdt} T.,
  {Mumcuoglu} M.,  {Neistein} E.,  {Pichon} C.,  {Teyssier} R.,    {Zinger} E.,
   2008, ArXiv e-prints, 808
\bibitem[de Lapparent et al.(1986)]{delap86} de Lapparent, V., 
Geller, M.~J., \& Huchra, J.~P.\ 1986, \apjl, 302, L1 
\bibitem[El-Ad 
\& Piran(1997)]{el-ad97} El-Ad, H., \& Piran, T.\ 1997, \apj, 491, 421 
\bibitem[Forero-Romero et al.(2008)]{forero08} Forero-Romero, 
J.~E., Hoffman, Y., Gottloeber, S., Klypin, A., 
\& Yepes, G.\ 2008, arXiv:0809.4135 
\bibitem[Gottloeber(1998)]{gottlober98} Gottloeber, S.\ 1998, Large 
Scale Structure:  Tracks and Traces, 43 
\bibitem[\protect\citeauthoryear{{G{\'o}rski}, {Hivon}, {Banday}, {Wandelt},
  {Hansen}, {Reinecke} \& {Bartelmann}}{{G{\'o}rski} et~al.}{2005}]{hivon}
{G{\'o}rski} K.~M.,  {Hivon} E.,  {Banday} A.~J.,  {Wandelt} B.~D.,  {Hansen}
  F.~K.,  {Reinecke} M.,    {Bartelmann} M.,  2005, \apj, 622, 759
\bibitem[Graham et al.(1995)]{graham95} Graham, M.~J., Clowes, 
R.~G., \& Campusano, L.~E.\ 1995, \mnras, 275, 790 
\bibitem[Hahn et al.(2007)]{hahn07} Hahn, O., Carollo, C.~M., 
Porciani, C., \& Dekel, A.\ 2007, \mnras, 381, 41 
\bibitem[Hanami(2001)]{hanami01} Hanami, H.\ 2001, \mnras, 327, 
721
\bibitem[Harker et al.(2006)]{harker06} Harker, G., Cole, S., 
Helly, J., Frenk, C., \& Jenkins, A.\ 2006, \mnras, 367, 1039 
\bibitem[\protect\citeauthoryear{{Hatton}, {Devriendt}, {Ninin}, {Bouchet},
  {Guiderdoni} \& {Vibert}}{{Hatton} et~al.}{2003}]{hatton}
{Hatton} S.,  {Devriendt} J.~E.~G.,  {Ninin} S.,  {Bouchet} F.~R.,
  {Guiderdoni} B.,    {Vibert} D.,  2003, \mnras, 343, 75
\bibitem[Hoffman 
\& Shaham(1982)]{hoffman82} Hoffman, Y., \& Shaham, J.\ 1982, \apjl, 262, L23 
\bibitem[Huchra 
\& Geller(1982)]{FOF} Huchra, J.~P., \& Geller, M.~J.\ 1982, \apj, 257, 423 
\bibitem[Icke(1984)]{icke84} Icke, V.\ 1984, \mnras, 206, 1P
\bibitem[Jost(1995)]{jost} {Jost }, Jorgen., {Riemannian Geometry and Geometric Analysis, Fourth Edition}, 1995, Springer
\bibitem[Kirshner et al.(1981)]{kirshner81} Kirshner, R.~P., 
Oemler, A., Jr., Schechter, P.~L., 
\& Shectman, S.~A.\ 1981, \apjl, 248, L57 
\bibitem[Kofman et al.(1992)]{kofman92} Kofman, L., Pogosyan, 
D., Shandarin, S.~F., \& Melott0, A.~L.\ 1992, \apj, 393, 437 
\bibitem[Lacey 
\& Cole(1993)]{lacey93} Lacey, C., \& Cole, S.\ 1993, \mnras, 262, 627 
\bibitem[Martinez 
\& Saar(2002)]{martinez02} Martinez, V., \& Saar, E.\ 2002, \procspie, 4847, 86 
\bibitem[Merritt et al.(2006)]{merritt06} Merritt, D., Graham, 
A.~W., Moore, B., Diemand, J., \& Terzi{\'c}, B.\ 2006, \aj, 132, 2685 
\bibitem[Milnor (1963)]{milnor63} Milnor, J., 1963, Morse Theory (Princeton University,  
Princeton,  NJ)
\bibitem[Navarro et al.(1997)]{NFW97} Navarro, J.~F., Frenk, 
C.~S., \& White, S.~D.~M.\ 1997, \apj, 490, 493 
\bibitem[Neyrinck(2008)]{neyrinck08} Neyrinck, M.~C.\ 2008, 
\mnras, 386, 2101 
\bibitem[Neyrinck et al.(2005)]{neyrinck05} Neyrinck, M.~C., 
Gnedin, N.~Y., \& Hamilton, A.~J.~S.\ 2005, \mnras, 356, 1222 
\bibitem[Novikov et al.(2006)]{2dskel} Novikov, D., Colombi, 
S., \& Dor{\'e}, O.\ 2006, \mnras, 366, 1201 
\bibitem[Ocvirk et~al. (2008)]{ocvirk}
 Ocvirk P.,  Pichon C.,    Teyssier R.,  2008, ArXiv e-prints, 803
\bibitem[Peebles(1980)]{peebles80} Peebles, P.~J.~E.\ 1980, 
Research supported by the National Science Foundation.~Princeton, N.J., 
Princeton University Press, 1980.~435 p.,
\bibitem[Peebles(1993)]{peebles93} Peebles, P.~J.~E.\ 1993, 
Princeton Series in Physics, Princeton, NJ: Princeton University Press, 
|c1993,
\bibitem[\protect\citeauthoryear{{Pichon}, {Vergely}, {Rollinde}, {Colombi} \&
  {Petitjean}}{{Pichon} et~al.}{2001}]{pichon2001}
{Pichon} C.,  {Vergely} J.~L.,  {Rollinde} E.,  {Colombi} S.,    {Petitjean}
  P.,  2001, \mnras, 326, 597
\bibitem[Pichon et al.(in prep.)]{pichon}  Pichon C., Thi{\'e}baut E., Prunet S., Benabed K., Sousbie,T., Teyssier R., Colombi S. {\sl in prep.}
\bibitem[Pogosyan et al.(in prep.)]{pogo}  Pogosyan D.,  Pichon C., Prunet S.,  Gay C., Sousbie,T., Colombi S. {\sl in prep.}
\bibitem[Platen et al.(2007)]{platen07} Platen, E., van de 
Weygaert, R., \& Jones, B.~J.~T.\ 2007, \mnras, 380, 551 
\bibitem[\protect\citeauthoryear{{Prunet}, {Pichon}, {Aubert}, {Pogosyan},
  {Teyssier} \& {Gottloeber}}{{Prunet} et~al.}{2008}]{prunet}
{Prunet} S.,  {Pichon} C.,  {Aubert} D.,  {Pogosyan} D.,  {Teyssier} R.,
  {Gottloeber} S.,  2008, ArXiv e-prints, 804
\bibitem[Roerdink(1995)]{jos01} Jos, B.~T.~M., Roerdink \& Arnold Meijster, \ 2001, {\it Fundamenta informaticae}, 41, 187-228
\bibitem[Sahni et al.(1998)]{sahni98} Sahni, V., Sathyaprakash, 
B.~S., \& Shandarin, S.~F.\ 1998, \apjl, 495, L5 
\bibitem[Sathyaprakash et al.(1996)]{sathyaprakash96} Sathyaprakash, 
B.~S., Sahni, V., \& Shandarin, S.~F.\ 1996, \apjl, 462, L5 
\bibitem[Sheth(1998)]{sheth98} Sheth, R.~K.\ 1998, \mnras, 300, 
1057
\bibitem[Sousbie et al.(2008a)]{3dskel} Sousbie, T., Pichon, 
C., Colombi, S., Novikov, D., \& Pogosyan, D.\ 2008a, \mnras, 383, 1655 
\bibitem[Sousbie et al.(2008b)]{sklsdss} Sousbie, T., Pichon, 
C., Courtois, H., Colombi, S., \& Novikov, D.\ 2008b, \apjl, 672, L1 
\bibitem[Springel(2005)]{gadget2} Springel, V.\ 2005, \mnras, 
364, 1105 
\bibitem[Springel et al.(2001)]{springel01} Springel, V., White, 
S.~D.~M., Tormen, G., \& Kauffmann, G.\ 2001, \mnras, 328, 726 
\bibitem[Stoica et 
al.(2005)]{stoica05} Stoica, R.~S., Mart{\'{\i}}nez, V.~J., Mateu, J., \& Saar, E.\ 2005, \aap, 434, 423 
\bibitem[Teyssier et al. (2008)]{teyssier} Teyssier, R, Pires, S, Prunet, S,
  Aubert, D. Pichon, C Prunet, Amara, A Benabed, K
Colombi, S Refregier, A. \& Starck, J.L. 2008, submitted.
\bibitem[van de Weygaert 
\& Schaap(2007)]{rien07} van de Weygaert, R., \& Schaap, W.\ 2007, ArXiv e-prints, 708, arXiv:0708.1441
 \bibitem[Wang et al.(2007)]{wang07} Wang, H.~Y., Mo, H.~J., 
\& Jing, Y.~P.\ 2007, \mnras, 375, 633 
\bibitem[Zel'Dovich(1970)]{zeldo70} Zel'Dovich, Y.~B.\ 1970, A\&A, 5, 84 
\end{thebibliography}

%

\onecolumn

\appendix
%
\section{A generic minimization algorithm}
In this appendix, we present a generic algorithm that aims at
minimizing a multi-linear scalar function $f(x_1,..,x_d)$ of $d$
variables within a polygonal volume, in a $d$-dimensional space, by
reducing the problem to finding the respective minima of a set of
polynomials of order $d$. It takes as input the location of the
minima, $M_i^0$, of $f(x_1,..,x_d)$ on the edges of the square and
simply consists in recursively minimizing the value of $f(x_1,..,x_d)$
along the lines joining them. 

\begin{figure}
\centering
\includegraphics[width=8.5cm]{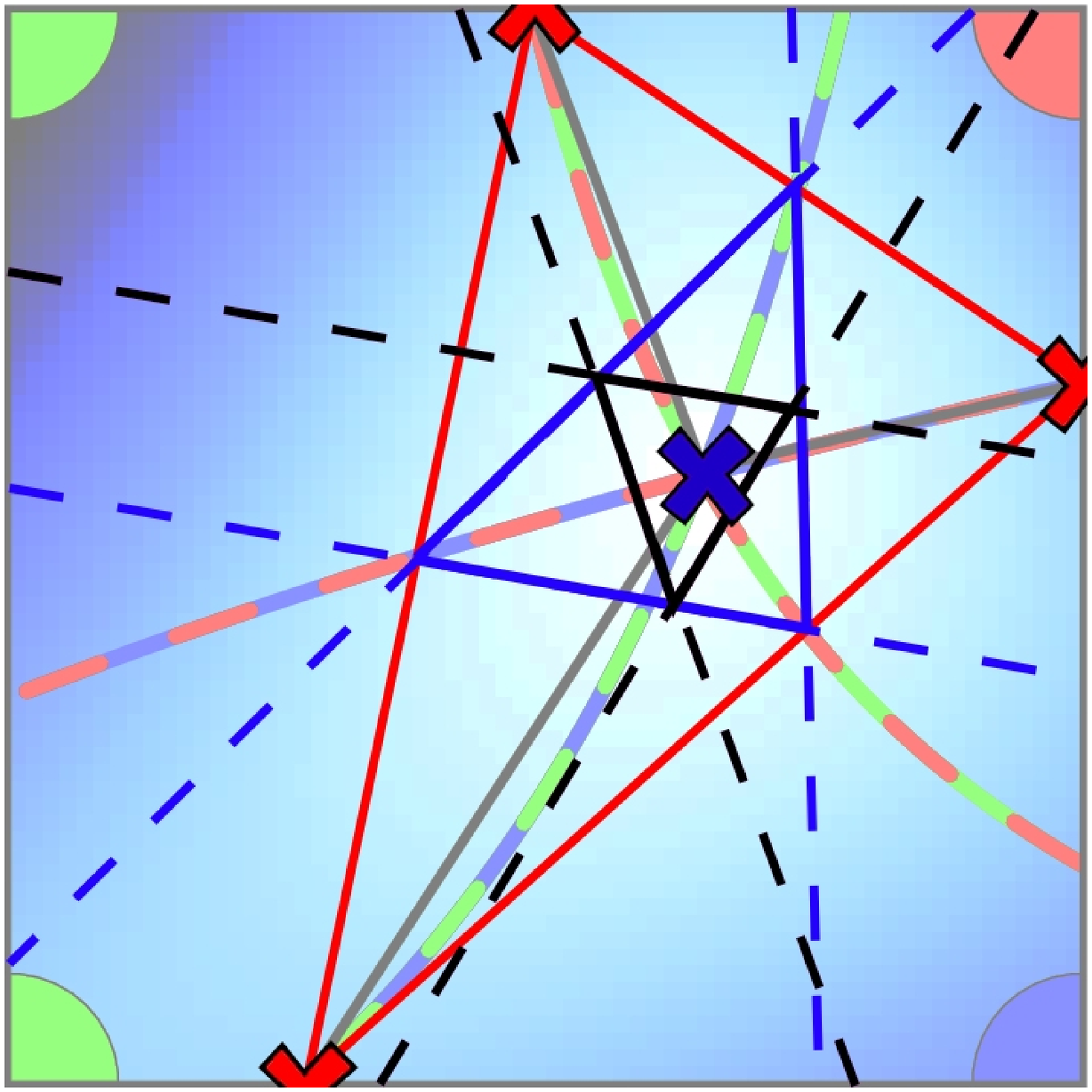}
\caption{ Illustration in the 2D case of the recursive minimization
  algorithm, applied to the case of Figure~\ref{fig_smooth3p}. The
  reader can refer to the legend of Figure~\ref{fig_propersmoothing} for more
  details. The scalar field to minimize is represented by the blue
  shading in the background while its minimum is located at the
  intersection of the 3 quadrics. The red crosses locate the field
  minima along the edges while the red, blue and black sets of lines
  result from the first three recursion
  steps.\label{fig_minimization}}
\end{figure}

Let us first consider the 2D case illustrated by Figure~\ref{fig_minimization}, where the cell is a square. In this case,
three minima, $M_1^0$, $M_2^0$ and $M_3^0$ (represented by red
crosses) can be easily found on the edges of the square from the
linearly interpolated value of $f$ along them. One can then compute
the location of the minima along the three lines linking them (the red
triangle), noting that because of the multi-linearity of $f$, its
value along a line can be expressed as a second order polynomial. One
thus obtains $3$ new points, $M_1^1$, $M_2^1$ and $M_3^1$, and the
process can be repeated, as represented by the blue and black sets of
lines, until convergence to the solution, represented by the blue
cross ({\it i.e.} when the three points are close enough to each
other).

This algorithm can be generalized to the case of the $p$-face of an
$n$-cubic cell, $p\leq n$, thus providing the solution over the
$p$-face from the $k$ solutions, $M_i^0\; i\in\lbrace1,..,k\rbrace$ ,
over the sets of $(p-1)$-faces that are its edges. As explained in
section \ref{sec:smooth-algo}, this algorithm is thus recursively
applied to the edges of the cell, starting from the $1$-faces, in the
order of their increasing dimensionality. The $j^{th}$ step of the
algorithm thus goes as follows:
\begin{enumerate}
\item Compute the equations of the $(k)(k-1)/2$ lines joining pairs of
  $M_i^{j-1}$.
\item Evaluate the value of $f(x_1,..,x_d)$ at $p+1$ points along
  these lines using multi-linear interpolation, and fit a polynomial
  of order $p$.
\item Find the minima of these polynomials that belong to the cell and
  keep the $k$ lowest among them, with coordinates $M_i^{j}$.
\item If these points are all contained in a sphere of radius a given
  fraction of the cell, stop, else start over.
\end{enumerate}

Note that although only the case of a Cartesian sampling grid was
presented here, the algorithm is easily transposable to any type of
grid, such as the one produced by Voronoi tessellation on a manifold, which is
composed of simplex shape cells. 
\section{Inter-skeleton pseudo-distance}

The inter-skeleton pseudo-distance from one skeleton ${\cal S}_a$ to
another skeleton ${\cal S}_b$ was defined in the main text by the
probability distribution function (PDF) of the minimum of the distance
from each segment of ${\cal S}_a$ to any segments of ${\cal S}_b$. In
this appendix, we  show how this measure can be interpreted
using realizations of scale invariant Gaussian random fields (GRFs)
with different power spectrum index $n$ (such that
$P\left(k\right)\propto k^{-n}$) and different smoothing lengths
$L$. All the skeletons that we use were computed from $512^3$ pixels
realizations of GRFs, smoothed over a scale $L=8$ pixels or $L_L=16$
pixels. These  scales are defined as the width of the Gaussian kernel
that we used to smooth the fields and the value of $L$ roughly
corresponds, in number of pixels, to the smoothing scale we used in
the main text, $L_{\rm NL}$. A total of six different skeletons were computed:
\begin{itemize}
\item{${\cal S}_{\rm GRF0}$ and ${\cal S}_{\rm GRF0^\prime}$}: skeletons computed from two realizations (GRF0 and ${\rm GRF0}^\prime$) of GRFs with spectral index $n=0$, smoothed over a scale $L=8$ pixels.\\
\item{${\cal S}_{\rm GRF3}$ and ${\cal S}_{\rm GRF3^\prime}$}: skeletons computed from two realizations (GRF3 and ${\rm GRF3}^\prime$) of GRFs with spectral index $n=3$, smoothed over a scale $L=8$ pixels.\\
\item{${\cal S}_{\rm GRF0_T}$}: this skeleton was computed from the field GRF0, smoothed on scale $L$. The resulting skeleton was then translated by ${\bf v}=\left(L/2,0,0\right)$.\\
\item{${\cal S}_{\rm GRF0_L}$}: this skeleton was computed from the field GRF0, smoothed on scale $L_L=2L=16$ pixels.\\
\end{itemize}

\begin{figure*}
\centering  
\subfigure[$n=0$]{\includegraphics[width=5.5cm]{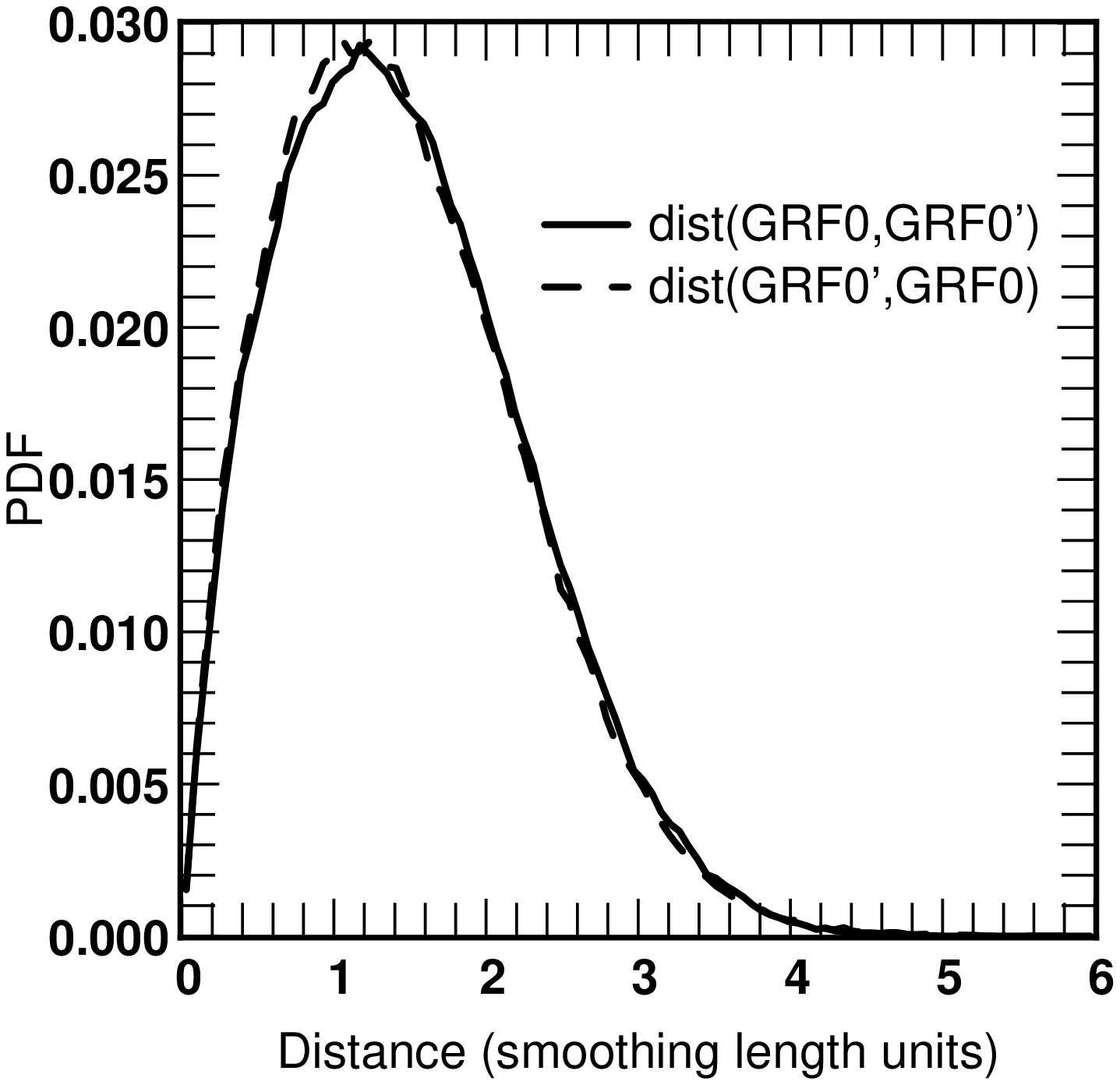}\label{fig_grf00}}
\hfill  \subfigure[$n=3$]{\includegraphics[width=5.5cm]{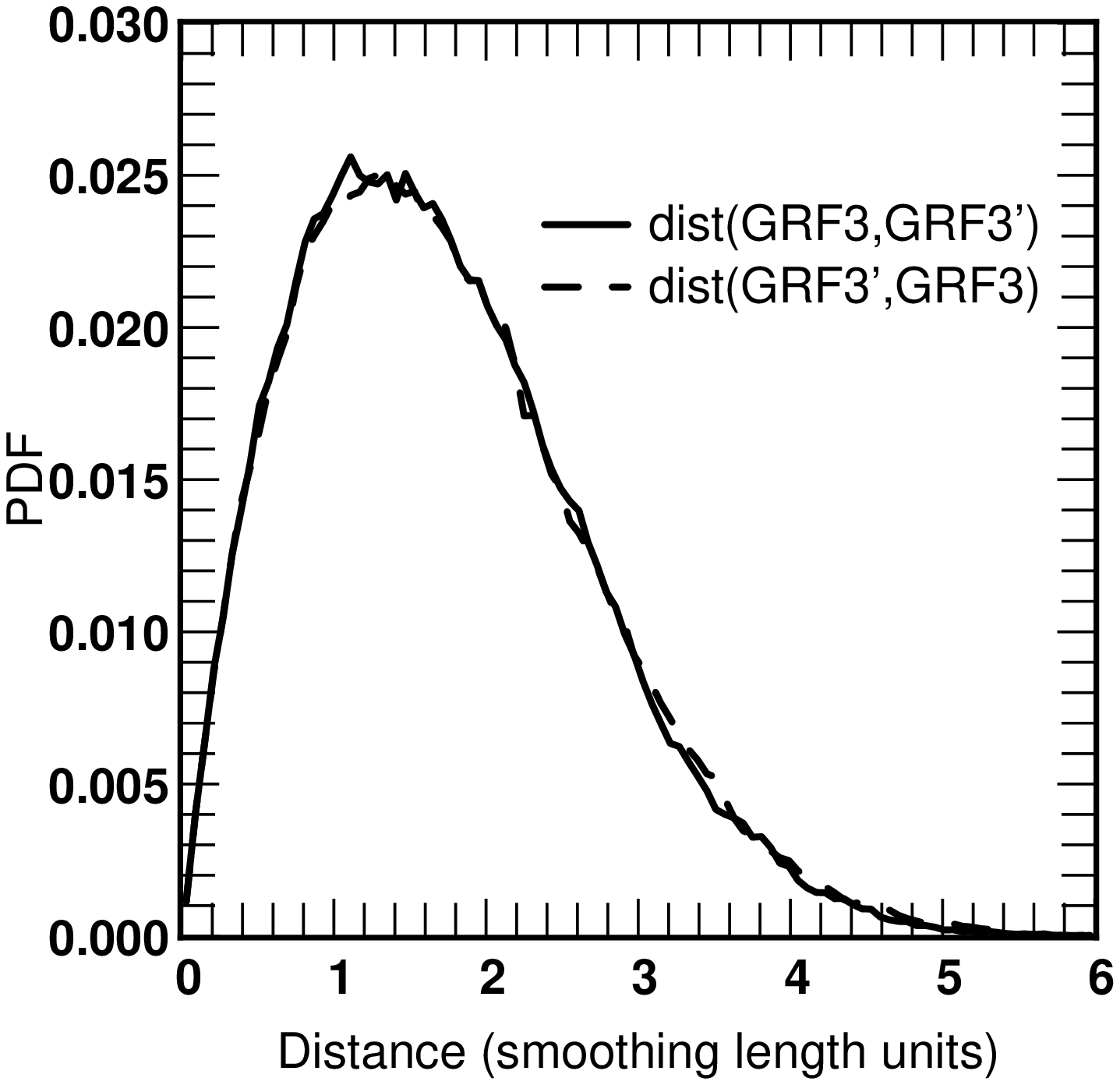}\label{fig_grf33}}
\hfill  \subfigure[$n=0$ and $n=3$]{\includegraphics[width=5.5cm]{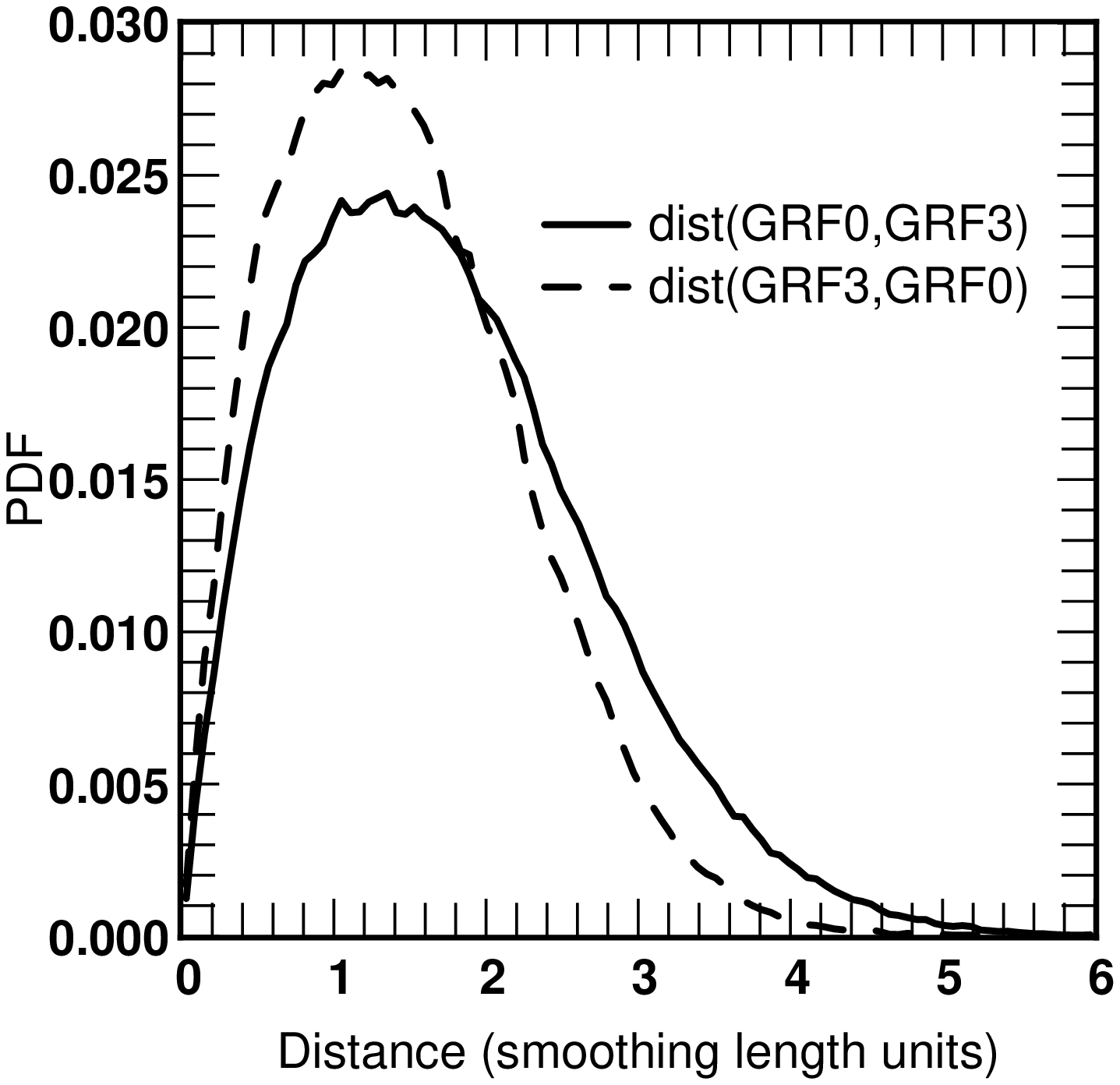}\label{fig_grf03}}\\
 \subfigure[$n=0$ and its translation]{\includegraphics[width=5.5cm]{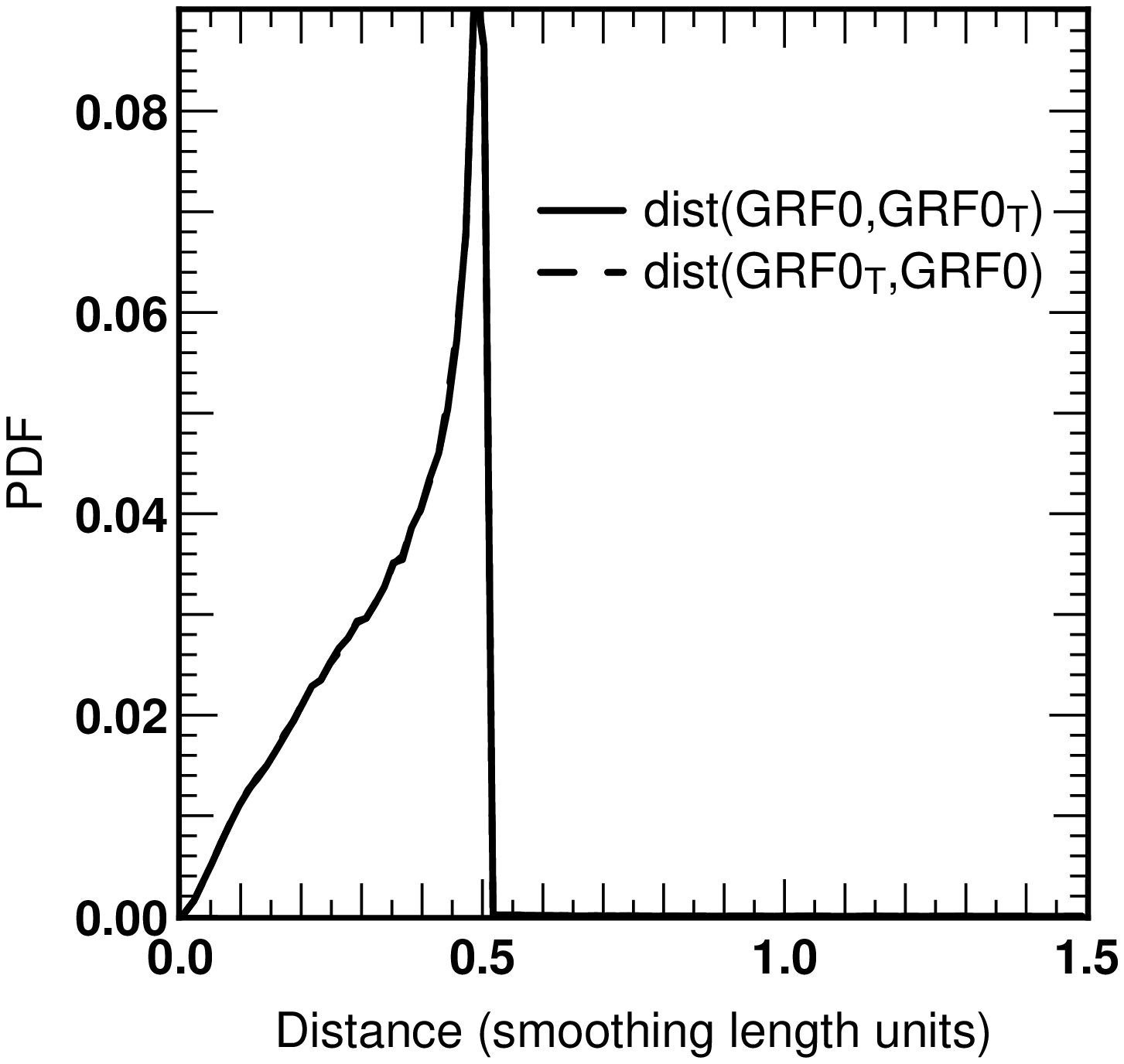}\label{fig_grf00t}}
\hfill  \subfigure[$n=0$, different smoothing lengths]{\includegraphics[width=5.5cm]{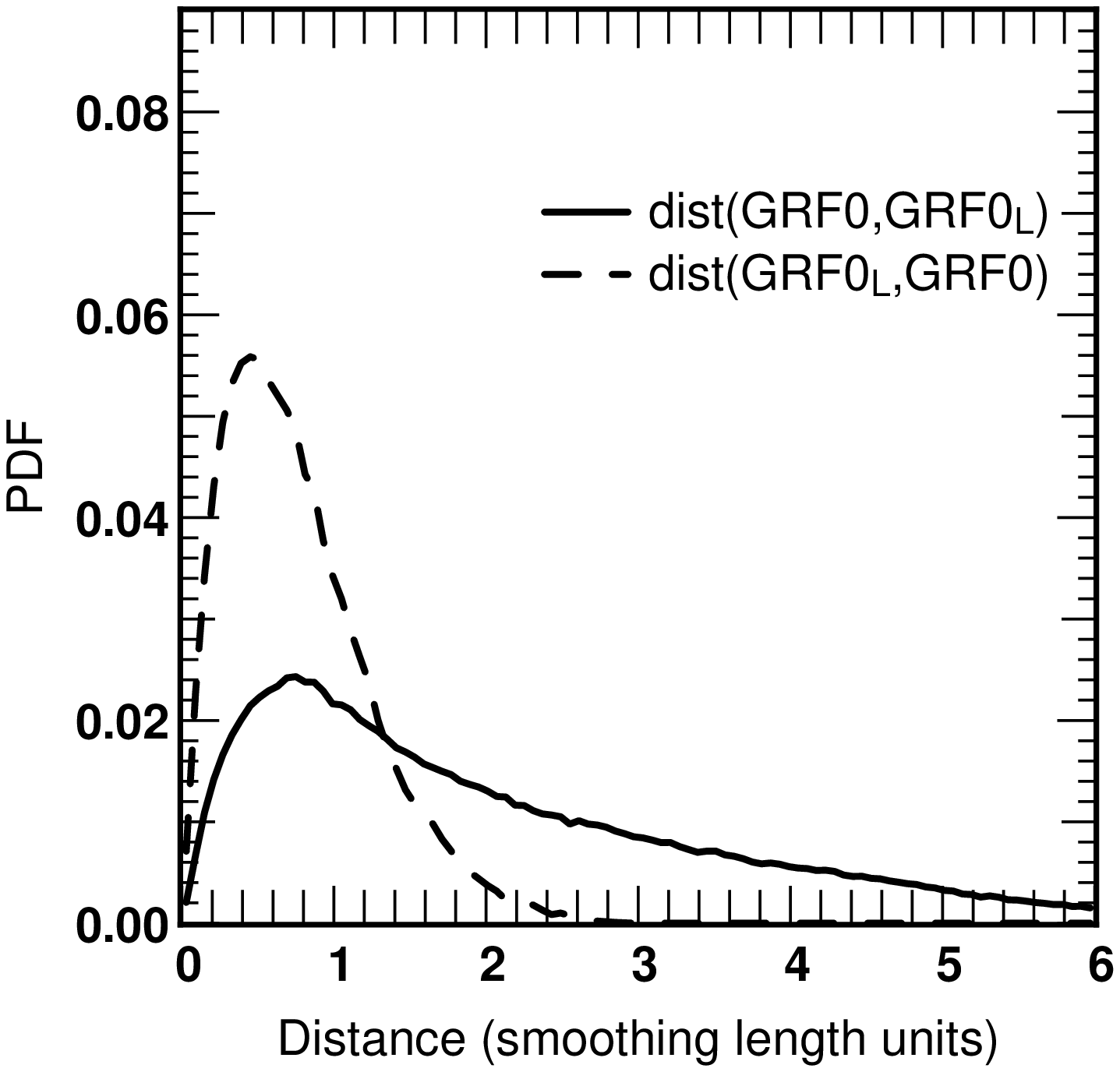}\label{fig_grf00l}}
\hfill  \subfigure[$n=0$, translation  and different smoothing lengths]{\includegraphics[width=5.5cm]{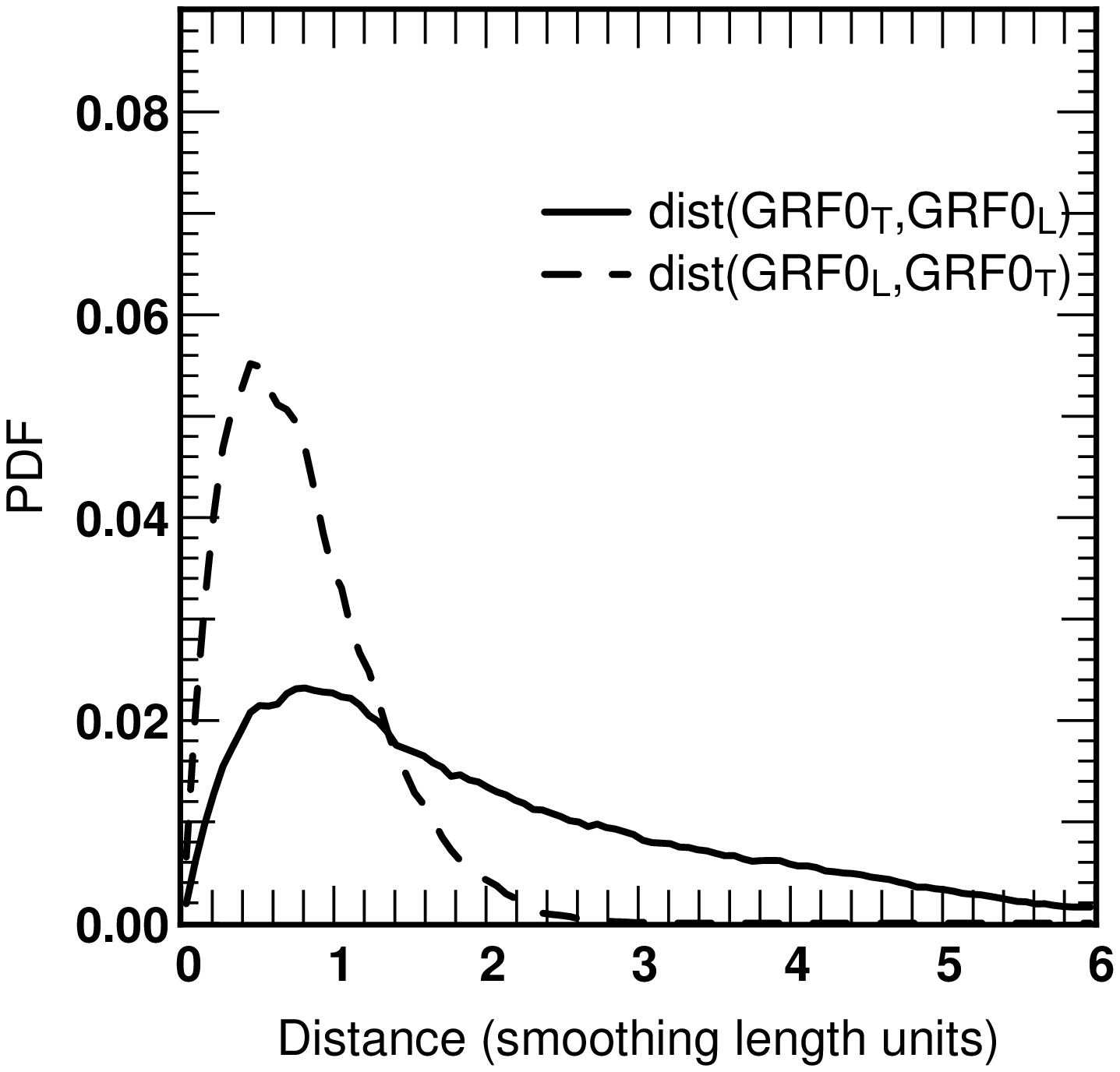}\label{fig_grf0l0t}}\\
\caption{Measures of the inter-skeleton pseudo distances for Gaussian random fields with different power spectrum index $n$ and smoothing length $L$. These plots show how the pseudo-distances measurements can be used to assess the discrepancies between two skeletons. 
\label{fig_GRF}}
\end{figure*}

Figure~\ref{fig_GRF} presents the different pseudo-distances
between these skeletons, 
$\distance{a}{b}$. Figures~\ref{fig_grf00} and~\ref{fig_grf33} present the results obtained when comparing
uncorrelated fields (i.e. different realizations of GRFs). As expected
in that case, $\distance{GRF0}{GRF0^\prime}=\distance{GRF0^\prime}{GRF0}$ and 
$\distance{GRF3}{GRF3^\prime}=\distance{GRF3^\prime}{GRF3}$ and the position of the mode  is 
about the smoothing length. One should also note that the mode
intensity differs between $n=0$ and $n=3$, which can be explained by the
fact that in the latter case, small scale fluctuations are suppressed
together with smaller scale filaments, thus making it less probable
for a segment of one realization to be very close to one of the other
realization. Figure~\ref{fig_grf03} shows that these pseudo distance
measurements  make it possible to distinguish the different nature of
two skeletons. In fact, whereas ${\cal S}_{\rm GRF0}$ has filaments on
any scales, only  the larger scales are present in ${\cal S}_{\rm
  GRF3}$, which translates into an asymmetry between $\distance{GRF0}{GRF3}$  and $\distance{GRF3}{GRF0}$. Whereas in the first case, there is no
reason why every segment of ${\cal S}_{\rm GRF0}$ should be close to a
segment of ${\cal S}_{\rm GRF3}$, the reciprocal is not true : ${\cal
  S}_{\rm GRF0}$ spreads on all scales and every segment of ${\cal
  S}_{\rm GRF3}$ should be as close as any other from a segment of
${\cal S}_{\rm GRF0}$ (hence the higher intensity of the mode for $\distance{GRF3}{GRF0}$). When comparing a skeleton ${\cal S}_{\rm a}$
with less filaments to a skeleton ${\cal S}_{\rm b}$ with more
filaments, the intensity of the mode is thus expected to be higher for
$\distance{a}{b}$ than for $\distance{b}{a}$.\\

This observation is confirmed by Figure~\ref{fig_grf00l} where ${\cal
  S}_{\rm GRF0_L}$ is compared to ${\cal S}_{\rm GRF0}$, which has
small scale filaments that ${\cal S}_{\rm GRF0_L}$ does not have. But
in that case, the two skeletons are correlated  as only the smoothing
length changes. This results in a higher intensity of the mode of
$\distance{GRF0_L}{GRF0}$: the larger scale
filaments are present in both skeletons. It also results in a shift in
the position of the mode, located at a distance smaller than the
smoothing length. Figure~\ref{fig_grf00t} illustrates the case of a
simple translation of length half the smoothing length $L$: in that
case,  both PDFs are identical and a very asymmetric and high
intensity mode is present at distance $L/2$. Finally, it is also
interesting to note that the comparison of  ${\cal S}_{\rm GRF0_L}$ to
${\cal S}_{\rm GRF0_T}$ almost gives the exact same result as the one
for ${\cal S}_{\rm GRF0_L}$ to ${\cal S}_{\rm GRF0}$ and it is
difficult to distinguish one from the other.

\section{Robustness of Fully connected skeleton}

\begin{figure}
\centering     \includegraphics[width=8.5cm]{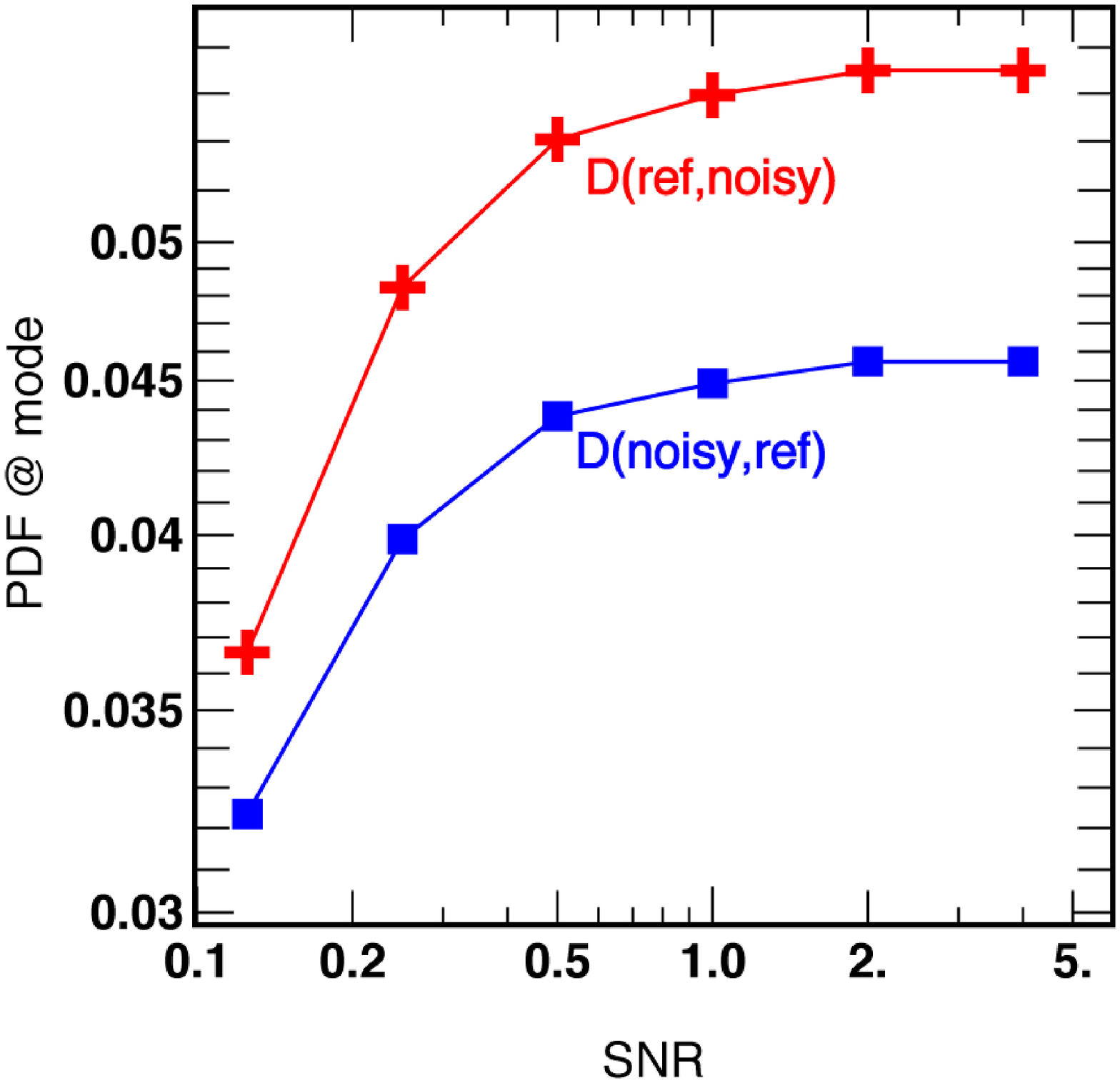} 
\caption{ The evolution of the PDF of distances at the mode as a function of the SNR of a noisy  
field. Here the distance is computed between the reference skeleton and its noisy counter part.
 For SNR above one, only small differences between weak filaments account for the difference between the 
 two distances. Conversely, for more noisy fields, the fraction of match between the two skeleton drops. 
 \label{sklnoise}}
\end{figure}

In order to investigate the robustness of the fully connected skeleton with respect to 
small changes in the underlying field, the following experiment is carried.
A given 2D white random field of size $4096^2$ is generated. It is then smoothed over 10 pixels, and 
its  reference skeleton, ${\cal S}_{\rm ref}$ is computed. 
A white random field of amplitude SNR is added to the reference field, and the 
corresponding skeleton, ${\cal S}_{\rm SNR}$, is computed after smoothing over 10 pixels.
The PDF of the pseudo-distances $\distance{{\cal S}_{\rm ref}}{{\cal S}_{SNR} }$ and 
$\distance{{\cal S}_{\rm SNR}}{{\cal S}_{ref} }$ is then calculated (see Appendix B).
The distance at the maximum (its mode) of both PDF remains unchanged for all the SNR considered ($1/8,1/4,1/2,1,2,4$),
which demonstrates that the core of the skeleton is quite robust: the reference skeleton is always shadowed by
its noisy counter part.
The amplitude of the PDF at its maximum is plotted in Figure~{\ref{sklnoise}}. 
This amplitude is sensitive to the high distance tail of mismatch between the two skeleton since 
the PDF is normalized.
In short, within the network there is a small subset of filaments which  are sensitive to any small 
variation of the field. For the vast majority of the network, the skeleton is globally only weakly affected
by changes of the underlying field so long as the amplitude of the change has a SNR above one.
When the SNR drops bellow one, spurious filaments occur more and more.
 The  discrepancy between the two plateaux at larger SNR reflects the fact that weaker filaments will occur somewhat
 randomly from one realisation to another, depending on very small details in the field.





\label{lastpage}

\end{document}